\documentclass[
reprint,
nofootinbib,
 amsmath,amssymb,
 aps,
]{revtex4-2}

\usepackage{graphicx}
\usepackage{dcolumn}
\usepackage{bm}
\usepackage{color}
\usepackage{amsmath}
\usepackage{wasysym}
\usepackage{float}
\usepackage{array,url,kantlipsum}
\usepackage{verbatim}
\usepackage[dvipsnames]{xcolor}
\usepackage{CJK}
\usepackage{textcomp}
\usepackage{tabularx}
\newcolumntype{P}[1]{>{\centering\arraybackslash}p{#1}}
\usepackage{array}
\usepackage{upgreek}
\usepackage{scrextend}
\usepackage{hyperref}

\renewcommand{\thesection}{\arabic{section}}
\renewcommand{\thesubsection}{\thesection.\arabic{subsection}}
\renewcommand{\thesubsubsection}{\thesubsection.\arabic{subsubsection}}

\makeatletter
\renewcommand{\p@subsection}{}
\renewcommand{\p@subsubsection}{}
\makeatother

\usepackage{hyperref}
\hypersetup{
    colorlinks,
    citecolor=blue,
    filecolor=blue,
    linkcolor=blue,
    urlcolor=blue
}

\newcolumntype{x}[1]{>{\centering\arraybackslash\hspace{0pt}}p{#1}}

\newcommand{\YAL}[1]{\textcolor{violet}{#1}}

\begin{document}



\title{Expanding Nuclear Physics Horizons with the Gamma Factory}

\author{Dmitry Budker}
\affiliation{Johannes Gutenberg-Universit{\"a}t Mainz, 55128 Mainz, Germany}
 \affiliation{Helmholtz-Institut, GSI Helmholtzzentrum f{\"u}r Schwerionenforschung, 55128 Mainz, Germany}
\affiliation{Department of Physics, University of California, Berkeley, California 94720, USA}

\author{Julian C. Berengut}
\affiliation{School of Physics, University of New South Wales,  Sydney 2052,  Australia}
\affiliation{ Max-Planck-Institut f\"ur Kernphysik, Saupfercheckweg 1, 69117 Heidelberg, Germany}

\author{Victor V.\,Flambaum} 
\affiliation{School of Physics, University of New South Wales,  Sydney 2052,  Australia}
\affiliation{Johannes Gutenberg-Universit{\"a}t Mainz, 55128 Mainz, Germany}
 \affiliation{Helmholtz-Institut, GSI Helmholtzzentrum f{\"u}r Schwerionenforschung, 55128 Mainz, Germany}
\affiliation{The New Zealand Institute for Advanced Study, Massey University Auckland, 0632 Auckland, New Zealand}

\author{Mikhail Gorchtein}
\affiliation{Johannes Gutenberg-Universit\"at Mainz, 55128 Mainz, Germany}

\author{Junlan Jin}
\affiliation{Department of Modern Physics, University of Science and Technology of China, Hefei 230026, China}

\author{Felix Karbstein}
\affiliation{Helmholtz-Institut Jena, GSI Helmholtzzentrum f{\"u}r Schwerionenforschung, 07743 Jena, Germany}
\affiliation{Theoretisch-Physikalisches Institut, Abbe Center of Photonics, Friedrich-Schiller-Universit\"at Jena, Max-Wien-Platz 1, 07743 Jena, Germany}

\author{Mieczyslaw Witold Krasny}
\affiliation{LPNHE, Sorbonne Universit\'{e}, CNRS/IN2P3, Paris; France}
\affiliation{CERN, Geneva, Switzerland}

\author{Yuri A. Litvinov}
 \affiliation{GSI Helmholtzzentrum f{\"u}r Schwerionenforschung, Planckstrasse 1, 64291 Darmstadt, Germany}

\author{Adriana P\'alffy}
\affiliation{ Department of Physics, Friedrich-Alexander-Universit\"at Erlangen-N\"urnberg, 91058 Erlangen, Germany}

\author{Vladimir Pascalutsa}
\affiliation{Institut f\"{u}r Kernphysik, Johannes Gutenberg-Universit{\"a}t Mainz, 55128 Mainz, Germany}

\author{Alexey Petrenko}
\affiliation{Budker Institute of Nuclear Physics, Novosibirsk, Russia}
\affiliation{Novosibirsk State University}

\author{Andrey Surzhykov}
\affiliation{Physikalisch--Technische Bundesanstalt, D--38116 Braunschweig, Germany}
\affiliation{Institut f\"ur Mathematische Physik, Technische Universit\"at Braunschweig, D--38106 Braunschweig, Germany}
\affiliation{Laboratory for Emerging Nanometrology Braunschweig, D-38106 Braunschweig, Germany}

\author{Peter G. Thirolf}
\affiliation{Fakult\"at f\"ur Physik, Ludwig-Maximilians-Universit\"at M\"unchen, 85748 Garching, Germany}

\author{Marc Vanderhaeghen}
\affiliation{Institut f\"{u}r Kernphysik, Johannes Gutenberg-Universit{\"a}t Mainz, 55128 Mainz, Germany}

\author{Hans A. Weidenm\"uller}
\affiliation{ Max-Planck-Institut f\"ur Kernphysik, Saupfercheckweg 1, 69117 Heidelberg, Germany}

\author{Vladimir Zelevinsky}
\affiliation{Department of Physics and Astronomy and National Superconducting Cyclotron Laboratiory/Facility for Rare Isotope Beams, Michigan State
University, 640 S. Shaw Lane, East Lansing, MI 48824, USA}

\date{\today}

\begin{abstract}
The Gamma Factory (GF) is an ambitious proposal, currently explored within the CERN Physics Beyond Colliders program, for a source of photons with energies up to $\approx 400\,$MeV and photon fluxes (up to $\approx 10^{17}$ photons per second) exceeding those of the currently available gamma sources by orders of magnitude. The high-energy (secondary) photons are produced via resonant scattering of the primary laser photons by highly relativistic partially-stripped ions circulating in the accelerator. The secondary photons are emitted in a narrow cone and the energy of the beam can be monochromatized, eventually down to the $\approx1$\,ppm level, via collimation, at the expense of the photon flux. This paper surveys the new opportunities that may be afforded by the GF in nuclear physics and related fields. 

\end{abstract}

\maketitle
\tableofcontents

\begin{figure*}[!htpb]\centering
    \includegraphics[width=1.0\textwidth]{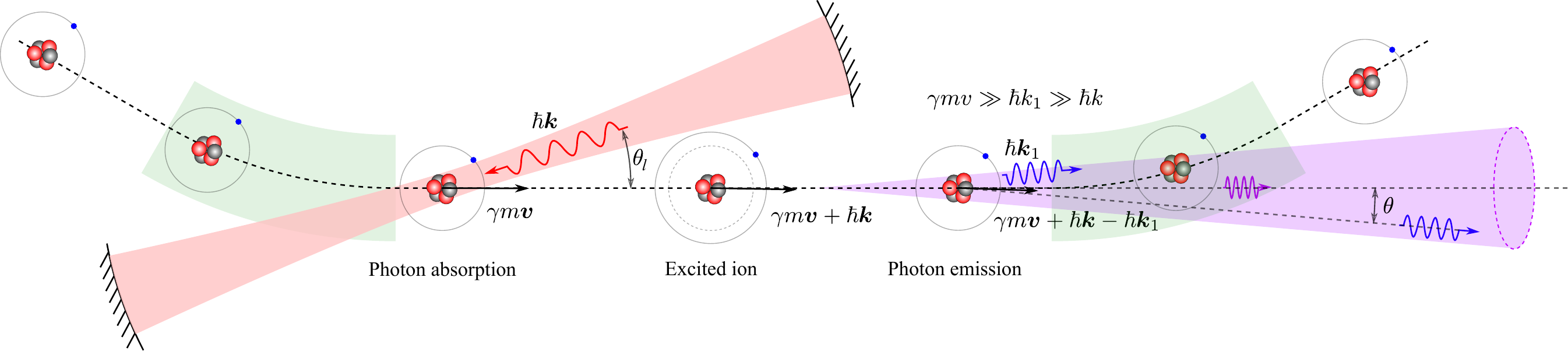}
    \caption{The Gamma Factory concept.  Laser photons with momentum $\hbar k$ (the primary photon beam) impinge onto ultrarelativistic ions (relativistic factor $\gamma$, mass $m$, velocity $v$) circulating in a storage ring. Resonantly scattered photons, as seen in the laboratory frame, are emitted in a narrow cone with an opening angle $\approx 1/\gamma$ in the direction of the motion of the ions. The energy of these secondary photons is boosted by a factor of up to $4\gamma^2$ with respect to the energy of the initial photons.}
    \label{Fig:Gamma_Factory_concept}
\end{figure*}
\section{Introduction}
\label{sec:Introduction}

The Gamma Factory (GF) is a novel research tool proposed in Ref.~\cite{Krasny:2015ffb} and subsequently developed at CERN as part of its Physics Beyond Colliders studies \cite{jaeckel2020quest}. The main aim of the GF is the generation of high-intensity gamma-ray beams  of tunable energy and relatively small energy spread. At the same time, the GF facility offers unique scientific opportunities by far not limited to the various uses of the produced gamma rays (the secondary beam), but also includes spectroscopy of stored relativistic ions, including radioactive ions, with the (primary, energy boosted) laser photons that are used to produce the gamma rays, as well as the production and use of various tertiary beams.

As a unique and unprecedented research tool, the GF opens new possibilities across several fields of physics, from atomic physics (with the opportunities surveyed in \cite{Budker2020_AdP_GF}), to nuclear physics and related fields (the subject of the present paper), to elementary-particle physics and searches for physics beyond the standard model (SM). While the GF today is still a proposal, some of its key components such as the production and storage of partially stripped ions (PSI), for instance hydrogen- and helium-like Pb and phosphorus-like Xe \cite{Hirlaender:2018rvt,Schaumann:2019evk,Gorzawski:2020dgx}, have been already demonstrated experimentally in the Super Proton Synchrotron (SPS) and the Large Hadron Collider (LHC) at CERN. Moreover, the proof-of-principle experiment \cite{Krasny2019PoP}, further discussed in Sec.\,\ref{Section:NP_PoP}, will demonstrate the full GF concept at SPS. The design of the experiments is complete,  awaiting  approval of the SPS Committee. 
 
\subsection{The Gamma Factory}
\label{subsec:The Gamma Factory} 

The principle of the GF is illustrated in Fig.\,\ref{Fig:Gamma_Factory_concept}. Highly charged  partially-stripped ions circulate in a storage ring at ultrarelativistic speeds with a Lorentz factor $\gamma=(1-\beta^2)^{-1/2}$, with $\beta=v/c\approx 1$ being the ion speed normalized by the speed of light. The electrons bound to these ions interact with a (primary) laser beam and undergo transitions between the atomic shells. The resonance fluorescence photons, as seen in the laboratory frame, are emitted in a narrow cone with opening angle of $\approx 1/\gamma$, and form the secondary beam. The photon energy is boosted by a factor of $\approx 4\gamma^2$ with respect to the original laser photons used for the electronic excitation. The  anticipated secondary beam parameters are listed in Table\,\ref{tab:GF_parameters}. In many respects, the GF will be a qualitative leap compared to the existing gamma sources, as well as   sources currently under construction. These are briefly discussed in Appendix\,\ref{Appendix:gamma_facilities}.

Successful excitation of an atomic shell relies on a resonance condition involving both the relativistic factor $\gamma$ of the PSI as well as the primary (optical) laser frequency $\omega$. Thus, the tunability of the secondary beam energy is achieved by  simultaneously  tuning  $\gamma$ and  adjusting the wavelength
or the incident angle $\theta_l$
of the primary laser beam to maintain the resonance with an atomic transition in the PSI ($\hbar\omega'$):
\begin{equation}
    \hbar\omega' = \hbar \omega \gamma (1 + \beta\cos\theta_l).
\end{equation}
The ion beam energy variation is a routine procedure in a storage ring. Indeed, at the LHC, the ions are injected with an initial relativistic factor of around $\gamma=220$, and are subsequently accelerated up to $\gamma=2900$. Tunable-laser technologies are also well developed.

The secondary gamma-ray beam will have a strong correlation between the emission angle $\theta$ (see Fig.~\ref{Fig:Gamma_Factory_concept}) and the gamma-ray energy $E$ given by the expression
\begin{equation}
    E(\theta) = \frac{E_{max}}{2\gamma^2(1-\beta\cos\theta)},
\end{equation}
where $E_{max} = (1+\beta)\gamma\hbar\omega'$.
For small angles this can be approximated as
\begin{equation}
    E(\theta) \approx \frac{E_{max}}{1+\gamma^2\theta^2} \approx \frac{4\gamma^2\hbar\omega}{1+\gamma^2\theta^2}.
\end{equation}
A small fraction of photons are emitted with large angles, $\theta \sim 1$, and in this case the emitted photon energy is comparable to the incident laser photon energy. For instance,
\begin{equation}
    E\left(\frac{\pi}{2}\right) = \frac{E_{max}}{2\gamma^2} \approx 2\hbar\omega.
\end{equation}

\begin{figure}[!htpb]\centering
    \includegraphics[width=\columnwidth]{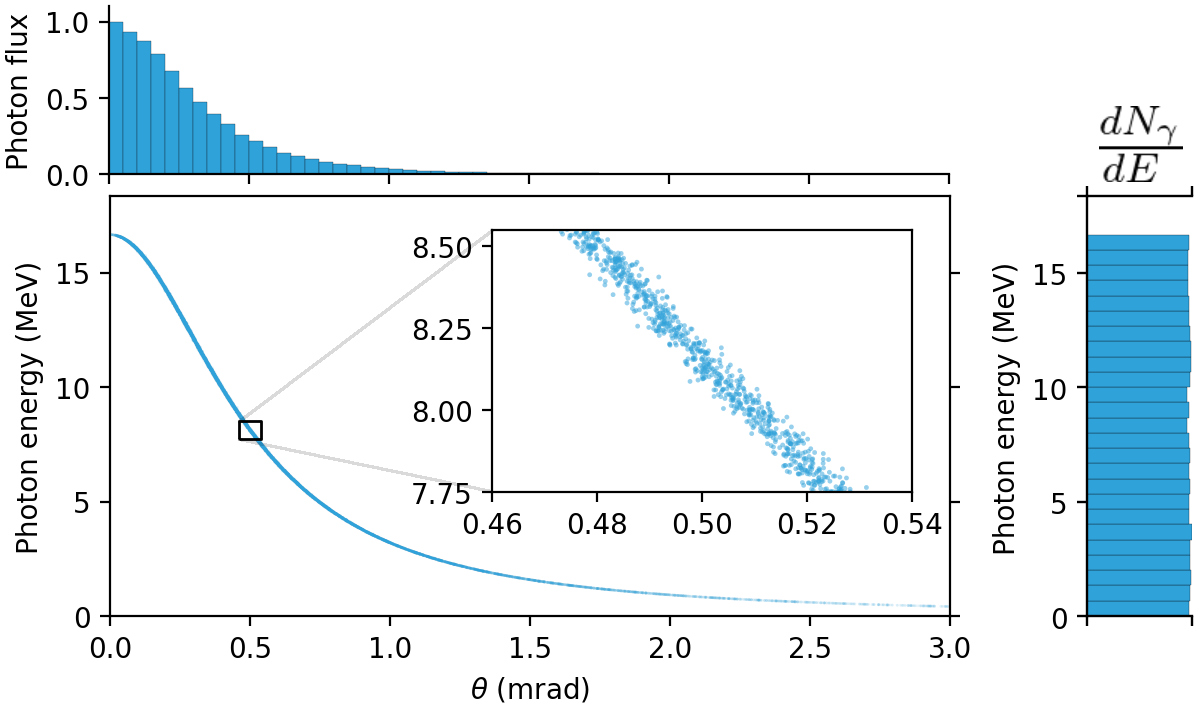}
    \caption{Correlation between the energy and angle of emitted photon in one particular case of the Gamma Factory based on hydrogen-like calcium ion beam in the LHC. The top histogram shows the corresponding photon flux density (photons per unit area per unit time) into a small solid angle. The right histogram shows the flat energy spectrum. The ion beam parameters in this example are: $\gamma = 2048$, $\sigma_x = 0.44$~mm, $\sigma_{x'} = 2.2$~$\mu$rad (normalized emittance $\epsilon_{n}=\gamma \sigma_x\sigma_{x'}=2$~mm$\cdot$mrad) \cite{Petrenko2021_EnergyAngleNotebook}.}
    \label{Fig:angle_energy}
\end{figure}

For a given emission angle, the residual photon energy spread is mainly determined by the angular spread of the PSI beam and its angular size as seen from the observation point (the location of the target or detector). A typical normalized transverse emittance of the ion beam in the LHC is $\epsilon_{n} = \gamma\sigma_x\sigma_{x'} \approx 1$\,mm$\cdot$mrad. The transverse beam size in the LHC can be varied from $\sim$1\,mm down to $\sim$10\,$\mu$m (in the collider interaction points). With a mm-wide ion beam, $\sigma_x \approx 1$\,mm and for $\gamma \approx 10^3$, this corresponds to angular spread on the order of a microradian $\sigma_{x'}=\epsilon_n/\gamma\sigma_x \approx 10^{-6}$. Therefore, the uncorrelated energy spread of the secondary gamma-photon beam will be on the order of $\delta\theta/\theta \approx \sigma_{x'}\gamma \approx 10^{-3}$
(see Fig.~\ref{Fig:angle_energy}). However, for the visible angular size of the mm-wide source of photons to drop below 1\,$\mu$rad, the distance to the observation point should be more than 1\,km. 

Laser cooling can potentially dramatically reduce the ion-beam emittance and energy spread. The transverse-only laser cooling \cite{Krasny2020_HL_LHC} can be applied to reduce the transverse beam size and angular spread by one order of magnitude at which point the energy spread in the ion beam will become important, since the typical relative energy spread in the LHC ion beam is on the order of $10^{-4}$. To improve the energy-angle correlation in the resulting gamma-photon beam to better than $10^{-4}$, both transverse and longitudinal laser cooling should be applied \cite{Bessonov1996,Zolotorev1997,Krasny2020_HL_LHC}. Deep laser cooling will be limited by the intrabeam scattering and other collective effects.

In summary, an optimal  trade-off between the gamma beam intensity and its spectral resolution will have to be made for each of the usage cases of the GF beams, including  case-by-case  choice of the number of ions per bunch, interaction-point collision optics, beam emittance and the distance of the gamma beam target to its production point.

Beside their energy and emission pattern, the secondary photons are characterized by their polarization. The polarization depends not only on the emission angle $\theta$, but also on the polarization state (e.g., magnetic sublevel population) of the incident PSI and the polarization of incident laser photons. A proper choice of these parameters results in emission of secondary gamma rays with high degrees of circular or linear, and generally, arbitrary polarization. A detailed analysis of the polarization of scattered photons, as performed using the density-matrix approach, is presented in Ref.~\cite{Serbo:2021AdP}. 

\subsection{The goals and main results of this paper}\label{subsec:Intro_goals_results}

Realization of the Gamma Factory project will result in a new facility based on the significant progress achieved in accelerator physics, laser technology,
detectors, control devices, computer control, and data-analysis techniques. The state-of-the art modern instrumentation and measurement techniques will enable the next step in our understanding of deep problems in nuclear, atomic, and hadron physics. 


   
  

Historically, new tools have led to new discoveries. Therefore, as we survey the science opportunities offered by the GF, we attempt to identify novel, potentially breakthrough directions. Here are some of these: 

\begin{itemize}
    \item Physics opportunities with primary (Sec.\,\ref{Sec:Nuclear_spectrosc_stored_ions}) and   secondary (Sec.\,\ref{Sec:Nucl_photophys_fixed_target})  beams with previously unattainable parameters;
    
    \item Spectroscopy of nuclear gamma transitions on par with laser spectroscopy of atoms (Sec.\,\ref{Sec:Nuclear_spectrosc_stored_ions});
    
    \item Investigation of the physics of exotic nuclei and of the mechanism of their stability along the road to the drip \YAL{lines} 
    (Sec.\,\ref{Sec:Radioisotope_Storage_Ring});
    
    \item Direct measurements of astrophysical S-factors at relevant energies (Sec.\,\ref{subsec:AstroSfactors});

    \item  Resolving narrow resonances in the photofission cross section of actinides via state-selective
       high-resolution spectroscopy (Sec.\,\ref{Subsec:Photofission});

    \item Pionic (Sec.\,\ref{Subsec:Pion_photoproduction}) and Delta-resonance (Sec.\,\ref{subsec:Delta}) physics on a qualitatively and quantitatively new level of precision;
    
    \item Precision measurement of parity violation (PV) in hadronic and nuclear systems at previously inaccessible levels (Sec.\,\ref{sec:P-violation});
    

    \item Investigation of exotic radioactive nuclei in conjunction with the CERN ISOLDE facility (Sec.\,\ref{Sec:Nucl_photophys_fixed_target}) and/or a dedicated storage ring (Sec.\,\ref{Sec:Radioisotope_Storage_Ring});
    
    \item Production of copious amounts of isotopes and nuclear isomers for medicine, dark matter search and gamma lasers via photonuclear reactions
    (Sec.\,\ref{Sec:Medical_Isotopes});
 
    \item Gamma polarimetry at the $<10^{-6}$\,rad level (Sec.\,\ref{subsec:Polarimetry_with_narrow_res});
    
    \item Production of and physics opportunities with high-intensity, monoenergetic and small-emittance tertiary beams  (neutrons, muons, neutrinos, etc.; Sec.\,\ref{Sec:Nucl_phys_tertiary_beams}).

\end{itemize}

\begin{table}[htpb]
    \centering
    \begin{tabular*}{\linewidth}{@{\extracolsep{\fill}} cc}
    \hline 
    \hline
    Parameter & Value\\
    \hline \\[-0.2cm]
    Ion $\gamma$ factor     &  10\footnote{The lowest $\gamma$ factors are for the SPS. At the LHC, they are $\approx200$.} -- 2900 \\[0.1cm]
    Ion species             & Pb$^{q+}$ as an example \\
    Transverse beam radius  & 16\,$\mu$m \\[0.1cm]
    Number of ions in a bunch & 10$^{8}$\footnote{A larger number can be expected for lighter PSI \cite{Citron2018future}, for example, $3\cdot10^9$ for He-like Ca.} \\[0.1cm]
    Number of bunches in the ring & 592 -- 1232 \\[0.1cm]
    Effective repetition rate         & 10\,MHz\footnote{20\,MHz planned after the LHC injector upgrade} \\[0.1cm] 
    Ion energy spread       & 10$^{-4}$ \\[0.1cm] 
    RMS bunch length        & 7.9\,cm  \\[0.1cm]
    Normalized emittance    & 1.6\,$\mu$m \\ [0.1cm] 
    Circumference of the LHC    & 26.7\,km \\
    \hline
    \hline
    \end{tabular*}
    \caption{Representative parameters of the Gamma Factory at CERN with currently available ion beams. $q$ denotes the charge state of the ions. The numbers are presented for Pb ions. Note that the current optimization for the LHC is for the collider mode. A dedicated optimization for the GF may lead to improvements.}
    \label{tab:GF_parameters}
\end{table}

In this paper, we outline a possible nuclear physics program for the GF, successively addressing the opportunities afforded by using the primary photon beams (starting with spectroscopy of stored ions with primary photons in Sec.\,\ref{Sec:Nuclear_spectrosc_stored_ions}), the secondary photon beams (starting with a discussion of fixed-target experiments in Sec.\,\ref{Sec:Nucl_photophys_fixed_target}), followed with a discussion of production and use of tertiary beams of various particles in Sec.\,\ref{Sec:Nucl_phys_tertiary_beams}.  

\section{Nuclear spectroscopy in the ion beam}
\label{Sec:Nuclear_spectrosc_stored_ions}

Laser spectroscopy of atomic transitions is well established in ion storage rings, see, for example, Refs. \cite{Nortershauser-2015,Ullmann-2017}. Precision spectroscopy of nuclear transitions may also be possible with the GF.

With the initial laser-photon energy of up to 10\,eV and the relativistic factor at the LHC of up to $\gamma\simeq 3000$, in their reference frame, the circulating ions ``see'' the photon energies up to 60\,keV, high enough to excite low-lying nuclear states. This opens up possibilities of nuclear spectroscopy or even nuclear quantum optics already with the primary GF beam. For this purpose, it is not necessary to use PSI and the experiments can be performed with bare ions. The prospects of combining relativistically accelerated nuclei as a target with presently available coherent light sources for first nuclear quantum optics experiments were originally discussed in Ref.\,\cite{Buervenich2006}. At that time,  a combination of X-ray free electron lasers and more moderate target acceleration was envisaged. This in turn  posed severe challenges due to the geographical distance between the required large-scale facilities, with table-top laser or accelerator alternatives still lacking in performance \cite{Liao2011,Liao2013}. The GF envisaged at CERN is in  a unique position of achieving high photon energies even with (table-top) lasers operating in the visible and ultraviolet (UV) ranges, thus opening new possibilities for nuclear quantum optics experiments.

In this Section, we give examples of low-lying nuclear transitions that can be driven by the primary laser light of the GF, discuss the opportunities related to nuclear isomers and possible couplings between the nuclear and atomic degrees of freedom, and put forward further prospects for using nuclear transitions for laser cooling or obtaining even higher gamma-ray energies at the GF. 

One needs to point out that direct spectroscopy of nuclear transitions at the GF would require overcoming a number of specific challenges, including ensuring the availability of ion sources with stable or radioactive isotopes capable of producing beams of sufficient intensity, operation of the full accelerator chain with radioactive species, as well as low excitation rates and long decay times of the narrow excited states. None of these appear as a-priori ``show stoppers'', but would need to be carefully addressed on a case-by-case basis.

\subsection{Low-energy nuclear transitions}
\label{sbsec:Low_En_Nucl_Trans}
Examples of low-energy gamma transitions starting from a stable or long-lived nuclear ground state are listed in Table\,\ref{tab:Low_En_Gamma_Trans}.
Typically, the low-lying nuclear spectrum can be described by collective rotations and vibrations of the nuclear surface \cite{Ring1980}, and such excited states frequently connect to the ground state  via electric-dipole-forbidden  transitions \cite{Palffy2008}. 
Direct spectroscopy on these transitions can provide detailed information, for example, on the relative role of different transition multipoles (e.g., $M1$ and $E2$ mixing), or provide accurate transition probabilities. In the context of nuclear quantum optics, many of these transitions were theoretically investigated for coherent driving with X-ray free electron laser sources \cite{Buervenich2006,Palffy2008,Junker2012}.

We note that some of the transitions to the ground state listed in Table\,\ref{tab:Low_En_Gamma_Trans} have not been directly observed, indicating their strong degree of suppression. For instance, due to its low energy, the radiative decay channel of the 8\,eV first excited state  in $^{229}$Th has not been observed so far, despite many experimental attempts of direct photoexcitation \cite{Jeet2015}. This peculiar ``isomeric'' transition is discussed in more detail in the context of the GF in the next section. In addition, the 13.034 and 51.697\,keV excited states \YAL{in} $^{235}$U have only been observed to connect to the 76\,eV long-lived first excited state \cite{HOJEBERG1970249}.   This is attributed to selection rules for the $K$-quantum number, i.e., the nuclear spin projection on the symmetry axis. In Table~\ref{tab:Low_En_Gamma_Trans} we therefore list instead the available experimental data on the two transitions to the isomeric state.   Detailed investigations of  strongly suppressed transitions at the GF, for instance the ones connecting the 13.034 and 51.697\,keV levels of $^{235}$U to the ground state,  would contribute to a better understanding of nuclear structure and the interplay between nuclear collective and single-particle degrees of freedom.

\begin{figure}[!hpb]\centering
    \includegraphics[width=\linewidth]{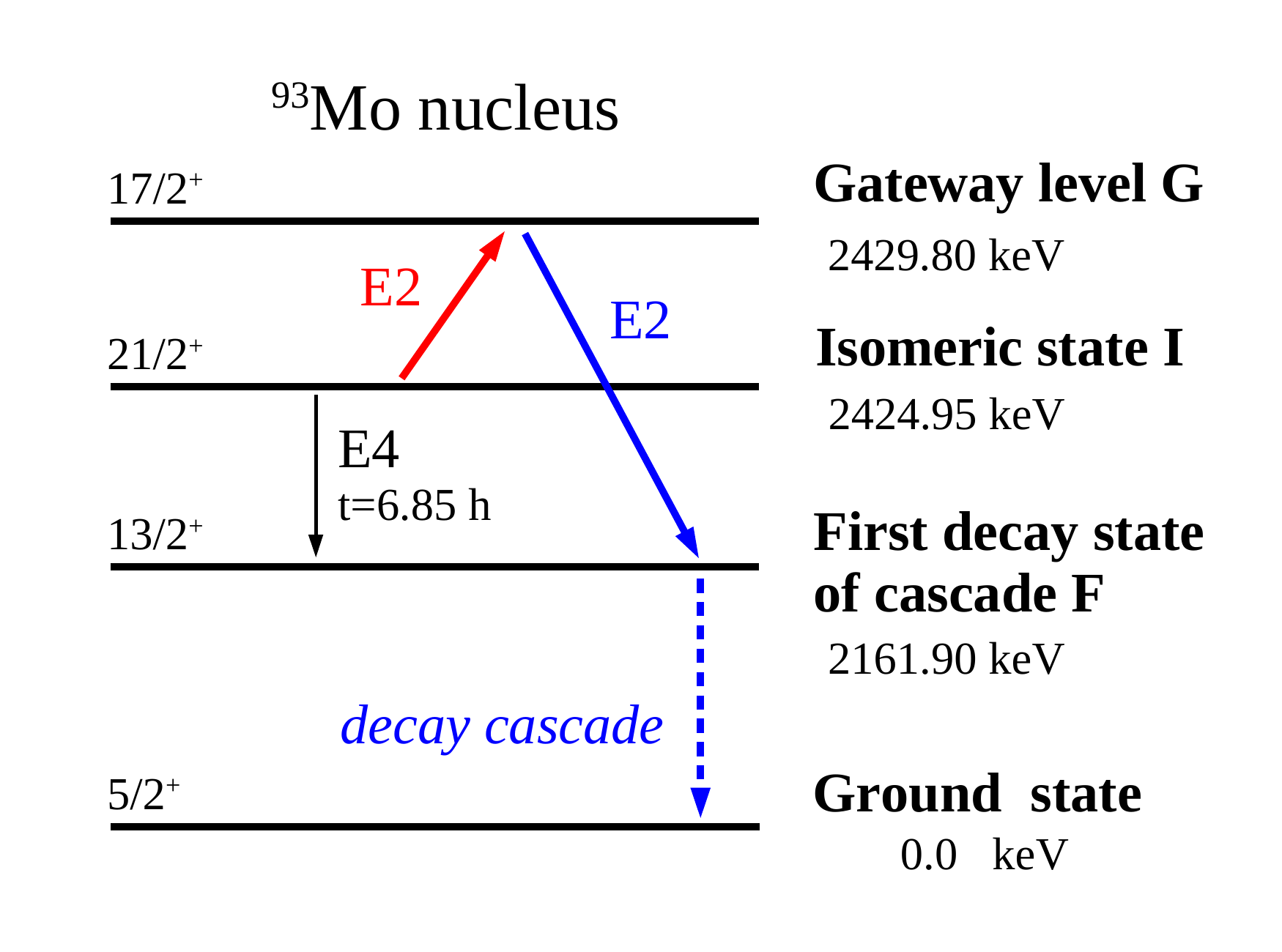}
    \caption{ Partial level scheme of $^{93}_{42}\mathrm{Mo}$. The isomeric state (I) can be excited to the gateway  level (G) which subsequently decays back to I or to a level F, initiating a cascade via different intermediate states (dashed line) to the ground state. The direct I$\to$F decay is a strongly hindered $E4$ transition, while I$\to $G  and G$\to $F are prompt $E2$ transitions.}      
    \label{fig:93Mo}
\end{figure}

In addition to transitions connecting the nuclear ground state with a low-lying excited state, one can envisage the use of nuclei in metastable states as PSI in the ion beam. Such nuclear metastable states are also known as isomers and  can store large amounts of energy over long period of time. Their existence is typically attributed to large differences in spin, shape or $K$ quantum number, between the isomer and the lower-lying levels \cite{Walker1999, Walker2020}.    In an advantageous configuration of the nuclear excited levels, once the excitation occurs from the isomeric state
to an upper gateway level, the subsequent nuclear decay may  proceed directly to a state below the isomer, thus reaching the ground state in a fast cascade. Such a process is called isomer  depletion, since it allows for the depopulation of the isomeric state and thus a controlled release of the energy stored in the metastable state. A typical example is the case of the 2.4\,MeV $^{93m}$Mo isomer, for which we present the relevant partial level scheme in Fig.~\ref{fig:93Mo}. An excitation scheme involving such a $\Lambda$-like three-level system (of the states $I$, $G$ and $F$ in the Fig.\,\ref{fig:93Mo} example), known as Raman-type excitation in  atomic and molecular physics, is advantageous since the excitation and decay photons have different energies and can therefore be more easily distinguished in experiments. In the context of isomers,  an additional fascinating potential application is related to the controlled release of nuclear energy on demand that would allow the design of a clean nuclear energy storage solution without involving fission or fusion \cite{Walker1999}. A list of nuclear isomer parameters useful for potential isomer depletion as well as other applications such as  medical applications or gamma-ray lasers are presented in Tables~\ref{tab:Isomer_(g,g')}, \ref{tab:Isomers_medicine}, \ref{tab:Isomer_dark_matter} and \ref{tab:Isomer_graser}. An idea on how to use the GF for spectroscopy of isomers and possible depleting levels above them is sketched in Sec.\,\ref{subsec:isomer-spectroscopy}.

\begin{table*}[t]
    \centering
\resizebox{0.95\textwidth}{!}{
\begin{tabular*}{\textwidth}{@{\extracolsep{\fill}} rcccccc c c }
    \hline 
    \hline
      Isotope  &$T^g_{1/2}$ &$E_e$\,(keV) &$I_g$   &$I_e$    &$\lambda L$ &$T^{rad}_{1/2}$ (s) & $\alpha(K)$ & $\alpha (L)$\\
    \hline \\[-0.3cm]
     ${}^{229}$Th  & 7880\,y   & 0.008\footnote{transition energy from Refs.\,\cite{Seiferle_Nature_2019,Sikorsky2020}, and radiative transition rate deduced using theoretical predictions in Ref.\,\cite{Minkov_Palffy_PRL_2017}}  &5/2$^+$   & 3/2$^+$  & $M1$  & $ 5.19\times 10^3$  & - & -\\
     ${}^{235}$U   &$7\times10^8$\,y    &0.076  &7/2$^-$     &1/2$^+$    &$E3$   &$7.03\times10^{23}$\footnote{using theoretical predictions from Ref.~\cite{berengut18prl0}} & - & - \\
     ${}^{201}$Hg   & stable  &1.565  &3/2$^-$      &1/2$^-$      & $M1$  & $3.76\times 10^{-3}$ & - & - \\
     ${}^{205}$Pb  &$1.7\times10^7$\,y  &2.329  &5/2$^-$   &1/2$^-$  & $E2$  & $9.07\times 10^2$ & - & - \\
     ${}^{181}$Ta  & stable  &6.238  &7/2$^+$ &9/2$^-$ & $E1$     & $4.34\times 10^{-4}$ & - &  -\\
     $^{239}$Pu &$2.4\times10^4$\,y  &7.861  &$1/2^+$  &$3/2^+$  &$M1$  &$2.04\times10^{-7}$   &-  &-\\
     ${}^{169}$Tm  & stable &8.410  &1/2$^+$ &3/2$^+$    & $M1$    & $1.07\times 10^{-6}$ & - & - \\
     $^{83}$Kr  & stable  &9.406  &9/2$^+$  &7/2$^+$  & $M1$  &$2.80\times10^{-6}$  &- & 14  \\
     ${}^{187}$Os  & stable &9.756 &1/2$^-$   &3/2$^-$  & $M1$   & $9.01\times 10^{-7}$ & - & - \\
     ${}^{137}$La  &$6\times10^4$\,y  &10.560  &7/2$^+$   &5/2$^+$    & $M1$   & $1.04\times 10^{-5}$ & - & 93.2\\
    ${}^{45}$Sc   &stable  &12.400  &7/2$^-$  &3/2$^+$ & ($M2$)    & $1.96\times 10^2$ & 362 & 54\\
    ${}^{235}$U   &-\footnote{\label{u235-ground}The isomeric state at 0.076\,keV is considered as the ground state for this transition.  In PSI the IC channel of the isomeric decay is closed and the  radiative half-life would be much longer than the ground-state $\alpha$ decay half-life. }  &13.034
    &1/2$^+$\footnote{\label{u235-spin}isomeric state spin}  &3/2$^+$   &M1   & $2.43\times 10^{-7}$\footnote{\label{u235-halflife}Radiative decay rate is given for the transition to the isomeric state at 0.076 keV. } & - & - \\
    ${}^{73}$Ge   &stable  &13.284  &9/2$^+$  &5/2$^+$  &$E2$      &$3.1\times 10^{-3}$ & 299 & 666 \\
    ${}^{57}$Fe   &stable &14.413   &1/2$^-$  &3/2$^-$  &$M1$      & $9.32\times 10^{-7}$ & 7.35 & 0.78\\
    ${}^{151}$Eu  &$\geq1.7\times10^{18}$\,y &21.541 &5/2$^+$   &7/2$^+$  & $M1$   &$2.62\times 10^{-7}$ & - & 21.7\\
    ${}^{149}$Sm  &stable &22.507  &7/2$^-$   &5/2$^-$  & $M1$   & $2.24\times 10^{-7}$&- &22.2\footnote{\label{footref:BrIcc_coef}theoretical values derived using the \textit{BrIcc} database \cite{BrIcc} based on Ref.\,\cite{Kibedi2008IC_coeff} with corresponding mixing-ratio data from the ENSDF database \cite{ENSDF}} \\       
    ${}^{119}$Sn  &stable  &23.871   &1/2$^+$  &3/2$^+$ &$M1$    & $1.07\times 10^{-7}$ & - & 4.1\\
    ${}^{161}$Dy  &stable  &25.651  &5/2$^+$   &5/2$^-$ & $E1$   & $9.59\times 10^{-8}$  & - &1.79\footref{footref:BrIcc_coef}\\
    $^{201}$Hg  & stable  &26.272              &$3/2^-$  &$5/2^-$  &$M1$  &$4.61\times10^{-8}$   &-  &55.9\footref{footref:BrIcc_coef}\\
    $^{129}$I   &$1.6\times10^7$\,y  &27.793 &$7/2^+$  &$5/2^+$  &$M1$  &$1.02\times10^{-7}$   &-  &4.06\\
    ${}^{229}$Th  &7880\,y   &29.190 &5/2$^+$   &5/2$^+$ &$M1$   & $3.26\times 10^{-8}$\footnote{using theoretical predictions from Ref.~\cite{Barci2003}} & -  &168\footref{footref:BrIcc_coef} \\   
    ${}^{40}$K  &$1.2\times 10^9$\,y & 29.830 &4$^-$  &3$^-$   &$M1$   & $5.47\times 10^{-9}$  &0.26\footref{footref:BrIcc_coef}   &0.023\footref{footref:BrIcc_coef} \\   
    $^{201}$Hg  &stable  &32.145               &$3/2^-$  &$3/2^-$  &$M1$  &$5.04\times10^{-9}$\footnote{calculated for the direct decay to the ground state only\label{gs-only}}   &-  &
     30.8\footref{footref:BrIcc_coef}
     \\
     $^{237}$Np  &$2.1\times10^6$\,y &33.196 &$5/2^+$  &$7/2^+$  &$M1$  &$9.92\times10^{-9}$   &-  &131\footref{footref:BrIcc_coef}\\
     $^{125}$Te  &stable  &35.492              &$1/2^+$  &$3/2^+$  &$M1$  &$2.15\times10^{-8}$   &11.69  &1.602\\
    ${}^{189}$Os  & stable  &36.200   &3/2$^-$  &1/2$^-$  &$M1$  & $1.09\times 10^{-8}$ & - & 15.6\\  
         $^{121}$Sb  &stable  &37.129              &$5/2^+$  &$7/2^+$  &$M1$  &$4.06\times10^{-8}$   &9.36  &1.227\\
    ${}^{129}$Xe  &stable &39.578  &1/2$^+$  &3/2$^+$   &$M1$    &$1.25\times 10^{-8}$ & 10.27 &1.41\\
         $^{233}$U   &$1.6\times10^5$\,y &40.351 &$5/2^+$  &$7/2^+$  &$M1$\footnote{This transition has $M1+E2$ multipolarity with a large mixing ratio $\delta\approx0.93$. $M1$ is the dominant radiative  channel by a small marge. }  &$1.03\times10^{-7}$   & -  &374\footref{footref:BrIcc_coef}\\
     
    $^{243}$Am  &7364\,y  &42.20                &$5/2^-$  &$7/2^-$  &$M1$  &$6.43\times10^{-9}$   &-  &110\\
    ${}^{229}$Th  &7880\,y  &42.435 &5/2$^+$ &7/2$^+$    &$M1$   & $2.59\times 10^{-8}$ & - &99.3\footref{footref:BrIcc_coef} \\   
    $^{240}$Pu  &6561\,y  &42.824               &$0^+$    &$2^+$  &$E2$   &$1.55\times10^{-7}$   &-  &658\\
     $^{246}$Cm  &4706\,y  &42.852               &$0^+$    &$2^+$  &$E2$   &$1.31\times10^{-7}$   &-  &770\footref{footref:BrIcc_coef}\\
    ${}^{248}$Cm  &$3.5\times10^5$\,y & 43.400  &$0^+$  &2$^+$  &$E2$    & $1.22\times 10^{-7}$ & - &724\footref{footref:BrIcc_coef}\\   
    $^{234}$U   &$2.5\times10^5$\,y  &43.498  &$0^+$    &$2^+$  &$E2$   &$1.80\times10^{-7}$   &-  &520\\
     $^{244}$Pu  &$8.1\times10^7$\,y   &44.2    &$0^+$    &$2^+$  &$(E2)$   &$1.23\times10^{-7}$   &-  &560\\
    $^{242}$Pu  &$3.7\times10^5$\,y   &44.54   &$0^+$    &$2^+$  &$E2$   &$1.20\times10^{-7}$   &-  &543\footref{footref:BrIcc_coef}\\
    ${}^{238}$U   &$4.5\times10^9$\,y &44.916  &$0^+$ &$2^+$   &$E2$   &$1.26\times 10^{-7}$ & - & 444\\
    ${}^{236}$U   &$2.3\times10^7$\,y &45.242  &0$^+$   &$2^+$         &$E2$   & $1.38\times 10^{-7}$& - & 429\\
    ${}^{235}$U   &$7\times10^8$\,y   &46.103  &7/2$^-$    &9/2$^-$      &$M1$   & $7.15\times 10^{-10}$ & - & 40 \\
    ${}^{183}$W   &$\ge 6.7\times10^{20}$\,y  &46.484 &1/2$^-$       &3/2$^-$     &$M1$  & $1.73\times 10^{-9}$ & - &6.46\footref{footref:BrIcc_coef}\\
    ${}^{232}$Th  &$1.4\times 10^{10}$\,y  & 49.369 &0$^+$    &$2^+$  &$E2$  &$1.15\times 10^{-7}$ & - &244 \\
        $^{81}$Kr   &$2.3\times10^5$\,y   &49.57   &$7/2^+$  &$9/2^+$  &$M1$  &$9.41\times10^{-9}$   &1.117\footref{footref:BrIcc_coef}  &0.169\footref{footref:BrIcc_coef}\\
    ${}^{235}$U   &-\footref{u235-ground}   &51.697
    &1/2$^+$\footref{u235-spin}   &5/2$^+$  &$E2$  &  $8.34\times 10^{-8}$\footref{u235-halflife} & - & 226 \\
    ${}^{230}$Th  &$7.5\times 10^{4}$\,y  & 53.227   &0$^+$  &$2^+$  &$E2$   &$8.08\times 10^{-8}$  & - & 166.8\\ 
      ${}^{157}$Gd  & stable    &54.536  &3/2$^-$  &5/2$^-$  & $M1$  & $1.74\times 10^{-9}$  & 9.50\footnote{This value corresponds to neutral atoms. For He- or H-like ions, the ionization potential increases closing the IC channel.} & 2\\       
    $^{239}$Pu &$2.4\times10^4$\,y  &57.275  &$1/2^+$  &$5/2^+$  &$E2$  &$3.58\times10^{-8}$   &-  &161.1\\
      $^{237}$Np  &$2.1\times10^6$\,y  &59.540  &$5/2^+$  &$5/2^-$  &$E1$  &$1.86\times10^{-7}$\footref{gs-only}  &-  &
     0.376\footref{footref:BrIcc_coef}\\
      ${}^{155}$Gd  & stable    &60.010  &3/2$^-$  &5/2$^-$  & $M1$  & $2.04\times 10^{-9}$ & 7.25 & 1.48\\
      
    \hline
    \hline
    \end{tabular*}}
    \caption{Examples of low-energy nuclear transitions, sorted by energy, with the lower state being a stable or long-lived ground (or isomeric) state with half-life $T^g_{1/2}$ and nuclear spin $I_g$. The  excited state energy $E_e$ and the  dominant multipolarity $\lambda L$ are given together with the excited state spin $I_e$.  $T^{rad}_{1/2}$ corresponds to the calculated radiative half-life of the excited state, and $\alpha(K/L)$ is the internal conversion coefficient 
    (of the transition) corresponding to the $K/L$-shell electrons.  Nuclear parameters were taken from the ENSDF database \cite{ENSDF} unless otherwise specified. 
     \label{tab:Low_En_Gamma_Trans}}
\end{table*}
\clearpage

\subsection{The lowest-lying nuclear isomer \texorpdfstring{\textsuperscript{229m}}{229m}Th}
\label{Subsec:Nuclear_Raman_Transitions}
Considerable interest in both theory and experiment has been paid in the past years to the lowest-known nuclear excited state, lying at only $\approx$8\,eV above the ground state of $^{229}$Th \cite{Seiferle_Nature_2019,Sikorsky2020}. $^{229}$Th belongs to the light actinide nuclear mass region known for the presence of enhanced collectivity and shape-dynamic properties \cite{MinkovPRC2021}. The single-particle states of the odd neutron determine the
$^{229}$Th ground state  with $K^{\pi}=5/2^{+}$ and the isomeric state with
$K^{\pi}=3/2^{+}$ based on the $5/2^+[633]$ and $3/2^+[631]$ single-particle orbitals. Here $K$
refers again to the projection of the total nuclear angular momentum on the
body-fixed principal symmetry axis of the system, $\pi$ is parity, and we use the usual Nilsson
notation $K^\pi[Nn_z\Lambda]$ with $N$, $n_z$ and $\Lambda$ being the asymptotic
Nilsson quantum numbers \cite{NR1995}. 

Because the isomeric transition energy is accessible with vacuum-ultraviolet lasers, $^{229m}$Th is the best candidate for a nuclear clock, i.e., a clock based on a nuclear rather than an electronic transition \cite{Peik_Clock_2003,Beeks2021,Peik2021}. At the GF, the low-energy isomeric transition can be  in principle accessed  with a primary optical laser beam directed at an obtuse angle to the ion beam.
In addition, one can also make use of the higher levels of $^{229}$Th to populate the isomer. Indeed, in a recent nuclear resonant scattering (NRS) experiment performed on a solid-state thorium-oxide target at the synchrotron-radiation source at SPring-8 in Japan,  the $^{229m}$Th isomer was excited indirectly via population of a higher-lying excited state \cite{Masuda2019ThXray}. Resonant X-rays were driving the transition between the ground state  $5/2^+[633]$ and the second excited state $5/2^+[631]$ at 29.19\,keV (see also Table\,\ref{tab:Low_En_Gamma_Trans}). This excited state decays with a branching ratio of $\approx$0.9 to the isomeric state $3/2^+[631]$. This is another example of Raman-type excitation in a nuclear $\Lambda$ three-level system similar to the isomer-depletion process discussed in Sec.\,\ref{sbsec:Low_En_Nucl_Trans}, however, now with the aim of populating instead of depleting the long-lived state. 

The NRS experiment  provided valuable information about the properties of the second excited state of ${}^{229}$Th, whose excitation energy of about 29.2\,keV has been determined with sub--eV accuracy. In addition, it also opened a new route for the determination of the ${}^{229m}$Th energy  \cite{Masuda2019ThXray, Yamaguchi2019}, though so far less precisely than the best prior measurements \cite{Seiferle_Nature_2019,Sikorsky2020}.
A different approach at the GF would be to  employ a $^{229}$Th beam in a dedicated storage ring, see Section~\ref{Sec:Radioisotope_Storage_Ring}. The intense GF photon flux would drive the 29 keV transition in the relativistic Th ions, allowing for precision studies of  excited state properties. Another interesting advantage at the GF is the possibility to study the properties of ${}^{229m}$Th in different atomic charge states. Here, due to strong hyperfine splitting one may expect sizeable differences between odd- and even charge states \cite{Brandau}.

One of the drawbacks for NRS experiments is the use of solid-state targets, in which the predominant nuclear decay is via internal conversion, quenching the isomeric state in microseconds \cite{Peik2021}. This is where the unique capabilities of the GF can come into play. Using $^{229}$Th$^{q+}$ ions with charge states $q=84-90$ in the LHC, the internal conversion channel  would be closed, since only  $2p_{3/2}$ orbitals and higher have ionization potentials smaller than 29\,keV. We may then envisage coherent nuclear driving with two 
narrowband UV primary beams that can excite simultaneously the two $M1$ transitions $5/2^+[633]\rightarrow 5/2^+[631]$ and $3/2^+[631]\rightarrow 5/2^+[631]$, both at $\approx29$\,keV. Varying the frequency of the second laser beam driving the  $3/2^+[631]\rightarrow 5/2^+[631]$ transition, would provide an independent  route to determine the energy of the isomeric nuclear state ${}^{229m}$Th with $10^{-2}$\,eV or perhaps even better accuracy. In addition, the two secondary X-ray beams would  open, for the first time, the possibility to exploit quantum-optics schemes as known from atomic or molecular $\Lambda$-systems directly with nuclear transitions.

In particular we can, in principle, envisage so-called $\pi$-pulses which swap the population between the lower and the upper state for each of the transitions, or stimulated Raman adiabatic passage (STIRAP) \cite{K.Bergmann1998,StirapRoadmap2019} between the ground state and the isomeric state. The latter would have the advantage that the population is coherently and efficiently transferred from the ground state   $5/2^+[633]$ to the isomeric state  $3/2^+[631]$ independently of the branching ratio of the excited state. While already proposed for nuclear coherent population transfer \cite{Liao2011}, the lack of suitable X-ray and gamma-ray facilities has prevented STIRAP from being observed for nuclear transitions so far. This could change with the advent of the GF. Laser phase fluctuations are known to reduce the performance of STIRAP \cite{Kuhn1992} as adiabaticity is disturbed during phase jumps. Nearly transform-limited pulses are best suited for experimental implementation of STIRAP \cite{Kuhn1992,K.Bergmann1998}. Such coherence properties  of the driving beams should  be available from the primary UV lasers. 


%
%
%
\subsection{Interaction of nuclear and atomic degrees of freedom}
\label{Subsec:Interaction_nucl_atom_DF}
The GF can be used to drive nuclear or atomic transitions in the ion beam with energies of up to $\approx$60\,keV. In this region, the coupling of the nuclear and atomic degrees of freedom can be  rather strong. 
Thus, the GF could become a powerful tool to explore the interplay between nuclear excitation or decay and electronic transitions. In the following, we outline the relevant processes and experimental prospects.

We address separately weak-interaction and electromagnetic-interaction processes involving atomic electrons.  In the first category fall the electron-capture (EC) and bound $\beta$ decay channels. EC cannot occur in bare ions because of the absence of $K$-shell electrons. Furthermore,  it  was shown that for the cases of $^{140}_{~59}$Pr and $^{142}_{~61}$Pm, the nuclear lifetime of the one-electron
Pr$^{58+}$ (Pm$^{60+}$) ion is shorter than the ones of the corresponding two- or many-electron cases \cite{Litvinov2007, Winckler-2009}. This could be explained by conservation of the total angular momentum, since only particular spin orientations   of the nucleus and of the captured electron can contribute to the allowed EC decay \cite{Patyk2008}. 
This selection rule leads to the acceleration of the allowed Gamow-Teller $1^+\rightarrow 0^+$ decay in one-electron Pr$^{58+}$ and Pm$^{60+}$ ions, while it stalls $1^+\rightarrow 2^+$ decays in one-electron $^{122}_{~53}$I ions \cite{Atanasov-2012}.
Bound $\beta$ decay, on the other hand, can only occur if vacancies in the atomic shell are available for the $\beta^-$ particle to occupy. The process and corresponding nuclear lifetime are sensitive to the electronic environment. For example, for bare  $^{163}$Dy and $^{187}$Re ions, the empty $K$ shells enable $\beta^-$  decay into these atomic orbits  with half-lives of 48$\pm 3$\,d and 32.9$\pm  2.0$\,y, respectively, while in the corresponding neutral atoms the nuclei are stable ($^{163}$Dy) or live $10^9$ times longer ($^{187}$Re) \cite{Jung1992,Bosch1996,Klepper1997}.  The study of such processes, however, does not require the primary GF photon beam to drive any nuclear transition, but rather $\beta$-unstable ions in the storage ring and  control of the electronic states. 

Studies of weak decays of highly-charged ions are routinely conducted at GSI in Darmstadt \cite{Litvinov-2011,Atanasov-2013}. However, there are numerous cases of forbidden decays of interest to astrophysics, for which such measurements are not presently feasible at GSI due to insufficient production rates \cite{Walker-2013}. 
To give just one example, $^{56}_{28}$Ni is among the most abundant products in supernova explosions. 
The half-life of a neutral atom is 6.075(10)\,d~\cite{NNDC} and the main decay branch is the Gamow-Teller $0^+\rightarrow1^+$ transition to the 1720.19\,keV state in $^{56}_{27}$Co.
In fully-ionized $^{56}$Ni$^{28+}$ nuclei, the EC decay channel is disabled. 
Experiments indicate a weak $\beta^+$ decay branch \cite{Zaerpoor-1999} rendering $^{56}$Ni a possible cosmo-chronometer, in contradiction to shell-model predictions \cite{Fisker-1999}.
The ISOL production method employed at ISOLDE can be superior in providing high-intensity beams of some elements \cite{ISOLDE_database}.
Furthermore, the production yields will be significantly improved within the EPIC project (Exploiting the Potential of ISOLDE at CERN) \cite{Catherall-2019}, which will facilitate conducting such experiments at the GF.

We now turn to the electromagnetic coupling between atomic and nuclear transitions for decay channels such as internal conversion, bound internal conversion or electronic bridge \cite{Palffy-CP}, which are suitable case studies that can take full advantage of the GF. Internal conversion is typically the stronger decay channel of low-lying excited nuclear states, in particular for excitation energies below 60\,keV. This in turn means that the lifetimes of such excited states strongly depend on the electronic configuration of the ion, both in terms of available electrons and spin couplings \cite{Palffy2008PLB}. In a pioneering study by Phillips and co-workers, the internal conversion rates for the 14-keV M\"ossbauer level in $^{57}$Fe were studied in F-like to H-like ions produced as secondary beams in Coulomb excitation. For H-like ions, effects  due to conservation of the total spin $F$ were revealed \cite{Phillips1989}.
The investigation of nuclear decays in highly ionized $^{125}_{~52}$Te$^{44+}$ up to $^{125}_{~52}$Te$^{48+}$ ions led to the discovery of a new decay mode, bound internal conversion (BIC) \cite{Attallah-1995}.

Studies of the influence of the electron shell on isomer lifetimes are particularly noteworthy since they might provide valuable information about energies and symmetry properties of nuclear states \cite{Karpeshin:2007, Karpeshin:2019}. Experimental information is presently limited to only a few cases \cite{Litvinov-2003, Sun-2010, Reed-2010, Reed-2012, Akber-2015, Bosch-2013}. In addition, a strong incentive comes from the field of astrophysics, since nucleosynthesis in stellar plasmas proceeds at high temperatures and, therefore, at a high degree of ionization \cite{Klepper1997,Wallerstein1997, Cowan-2021}.

With the unique capabilities of the GF, we can design, for the first time, scenarios in which both the nuclear excited states and electronic excited states are populated by the primary laser beam Lorentz-boosted in the reference frame of the accelerated ions. An additional degree of freedom for the experiment is the charge state of the ions. In order to investigate the effects of the atomic shell on nuclear decay lifetimes, PSI with few remaining electrons would be of interest. Nuclear transitions below 60\,keV such as those listed in Table\,\ref{tab:Low_En_Gamma_Trans} could be driven by one laser beam, while a second laser beam could be used to control the electronic shell. For convenience, we provide also the $K$-shell and $L$-shell internal conversion coefficients, i.e., the ratios between the internal conversion (for the respective electronic shell)  and the radiative decay rates, for the transitions listed in Table\,\ref{tab:Low_En_Gamma_Trans}. Due to low excitation energies and the rather high ionization potentials of the $K$ and $L$-shell electrons, for many of these transitions the respective internal conversion channels are closed; these are indicated with ``-'' in the last two columns of Table\,\ref{tab:Low_En_Gamma_Trans}. 

We also note that although the special cases of the low-lying $^{229m}$Th and $^{235m}$U isomers are well-known for having large internal conversion coefficients for the outer-shell electrons, the corresponding low-charge states are not accessible at the GF which will operate with beams of partially stripped but nevertheless highly charged ions. For example, the internal conversion coefficient for the 8\,eV isomeric transition in $^{229m}$Th is $\approx10^9$ for the neutral atom \cite{Karpeshin:2007}, but the channel is closed already for Th$^+$ ions. From Table\,\ref{tab:Low_En_Gamma_Trans} we identify 
interesting cases with large internal conversion coefficients for the $K$- and $L$-shell electrons, which are appealing candidates for future studies at the GF. Focusing on internal conversion coefficients larger than 100, we identify as promising candidates the transitions in $^{45}$Sc, $^{73}$Ge,  $^{237}$Np (the 33\,keV transition),  $^{243}$Am, together with almost all listed transitions in Th and U isotopes, and with all listed transitions in Pu and Cm isotopes.  For these nuclei, studies at the GF of  the dependence of the nuclear excited-state lifetime on the charge state and spin coupling of the electrons would be very interesting.

A different and promising prospect coupling the atomic and nuclear degrees of freedom at the GF involves the so-called electronic bridge process.  Thereby  the interaction of a nucleus with a photon is mediated by atomic electrons. In the electronic bridge process, the nuclear decay is accompanied by virtual excitation of an electron into a higher bound orbital. The electron then de-excites by photoemission. The process is third-order in QED, but can nevertheless be the dominant channel for the  decay of a nuclear isomer. Electronic bridge processes involving nuclear excitation at the expense of the initially excited electronic shell are sometimes called ``inverse electronic bridge'' \cite{tkalya90sov-phys-dokl}, however, many works use the same term for both excitation and decay channels.
The electronic bridge may dominate, for example, in low-energy transitions where the hyperfine interaction between the electron and the nucleus is less suppressed by selection rules than the direct gamma-decay, since the photon coupling will include a lower power of the photon energy. In this case the electrons can act as an effective ``bridge'' for the interaction, particularly if the resonance condition is met~\cite{krutov68ann-phys}.

The electronic bridge has been studied  for several nuclear transitions  including the 76\,eV $^{235m}$U isomer \cite{tkalya90sov-phys-dokl,hinneburg79zpa,hinneburg81zpa,berengut18prl0}, the 8\,eV $^{229m}$Th isomer \cite{tkalya92jetplett,tkalya92sjnp,kalman94prc,karpeshin99prl,kalman01prc,porsev10pra,porsev10pra0,nickerson20prl}, and the 3.05\,keV nuclear transition starting from the $6^-$ 464\,keV isomer   of $^{84}$Rb  \cite{tkalya14prc}. Laser-induced electronic bridge has been proposed to determine the excitation energy of the $^{229}$Th isomer in various charge states \cite{porsev10prl,karpeshin92plb,bilous18njp,bilous20prl}. However, to date, the electronic bridge process has not been experimentally observed \cite{tkalya04laser-phys}.

The Gamma Factory provides a new opportunity to explore the electronic bridge process, particularly as a means of driving nuclear excitation. Not only are there many suitable transitions (see Table\,\ref{tab:Low_En_Gamma_Trans}), but one can also tune the charge state of the PSI to optimize the resonance condition for the electronic excitation. As a concrete example we may consider the 2.329\,keV nuclear excited state in $^{205}$Pb, which in neutral atoms is an isomer with 24\,$\mu$s half-life \cite{ENSDF}. Understanding the decay properties of this state is important for nuclear astrophysics, where a much faster EC-decay is expected to $^{205}$Tl than the 17~My decay of the $^{205}$Pb ground state \cite{Pavicevic-2010, Pavicevic-2016}. This process affects the very end of s-process nucleosynthesis in the Tl-Bi region \cite{Tonchev-2018}.
Furthermore, since $^{205}$Pb is the only short-lived cosmic radioactivity produced exclusively in the s-process, the evidence for which was found in meteorites, it turns to be important for constraining cosmochemical simulations \cite{nucleosynthesis}.

Once the outer N-, O- and P-shell electrons are stripped off, the internal conversion channel is closed and the  decay can only proceed via an $E2$ radiative decay, leading to a half-life of approx.~3\,h. 
 In PSI, the $M1$ electronic fine-structure transition $2p_{1/2}$-$2p_{3/2}$ has approximately the same energy as the nuclear transition. Therefore, in suitably chosen PSI, electronic bridge decay paths will open where the $2p_{1/2}$-$2p_{3/2}$ transition mediates the nuclear transition. For example, in N-like Pb, the electronic bridge rate will exceed the radiative rate by several orders of magnitude, reducing the lifetime to minutes or seconds, depending on how well the resonance condition is met. 
This electronic bridge process may be directly observed in the GF by comparing excitation rates in different PSI where the resonant electronic transition is available (such as N-like) against those for which it is not (such as He-like).

\subsection{Further experimental prospects involving  nuclear transitions in the ion beam}
\label{subsec:Further_prospects_ion}

In addition to the aspects discussed so far, several proposals on how to use low-energy nuclear transitions for isomer spectroscopy, laser cooling or production of even higher gamma-ray energies in the secondary beams have been put forward. We shortly address these ideas here. 

\subsubsection{Spectroscopy of isomers} \label{subsec:isomer-spectroscopy}

The fact that the stored ions can be accelerated and decelerated in the storage ring opens the  possibility  to perform spectroscopy of appropriately long-lived isomeric states using two subsequent photoinduced transitions in a scenario reminiscent of pump-probe spectroscopy. A primary laser with a fixed photon energy is first used. The ion relativistic factor $\gamma$ is adjusted to excite nuclear transitions which populate a long-lived isomeric state.  This could be either a direct transition or excitation of the long-lived isomeric state via intermediate states.

After a sufficient amount of isomeric ions are stored, one can change the relativistic factor of the ions to a different value $\gamma'$. At this energy, with the same primary laser, transitions depleting the isomeric state would be driven,  and secondary photons detected. This provides a method for performing high-resolution spectroscopy of isomeric states and possible gateway levels around it in a broad range of frequencies, effectively driving (\texorpdfstring{$\gamma,\gamma'$}{g,g'}) reactions as the ones discussed in Sec.\,\ref{subsec:prod_isomers_g_gp} in the context of isomer production for medical applications. A limitation is that, with primary-beam photon energies of up to $\approx$10\,eV, the energy of the laser-driven gamma transitions needs to be below 60\,keV. One candidate with an adequate level scheme below this energy is $^{229m}$Th, for which transitions from the 29\,keV level connect to both the isomeric and ground states.
With the development of short-wave laser and mirror technologies and/or the advent of the LHC high-energy upgrade, the limitation of 60\,keV as maximum photon energy will be overcome, significantly enlarging the set of potential candidates.

\subsubsection{Laser cooling with nuclear transitions}
\label{subsubsec:laser_cooling_nuc_trans}
Cooling via scattering of laser photons near-resonant to gamma transitions in bare nuclei was suggested in Ref.\,\cite{Bessonov2002_Tkalya}, where the authors show that reasonable damping times (of less than an hour) can be achieved for several bare ions at the LHC. While laser cooling of PSI using electron transitions is generally much more efficient, laser  cooling of bare nuclei may be advantageous for colliding-beam applications to avoid beam losses due to mutual stripping of the colliding PSI beams \cite{Bessonov2002_Tkalya}. Candidate transitions for
laser cooling can be selected from those listed in Table~\ref{tab:Low_En_Gamma_Trans} or from the original proposal in Ref.\,\cite{Bessonov2002_Tkalya}.

\subsubsection{Production of higher-energy gamma rays}
\label{subsubsec:higher_en_gammas}
In principle, the GF concept can be extended to produce photons with much higher energies than 400\,MeV that can be generated by scattering conventional laser photons on relativistic PSI. This could be achieved by replacing the optical laser source\YAL{,} which excites the electronic shells of the primary beam\YAL{,} with an X-ray laser source driving nuclear transitions of the relativistic ions.  Suitable nuclear resonances include the so-called Giant Dipole Resonances (GDR) which are discussed in Sec.\,\ref{Subsect:GDR-multi}. We can envisage an efficient excitation up to nuclear energies of $\approx$15\,MeV, where the dipole response for stable medium to heavy nuclei exhausts almost 100$\%$ of the
Thomas-Reiche-Kuhn sum rule. Much higher energies would lead to particle loss and reduced radiative decay of the nuclear excitation. In turn, 15\,MeV corresponds to $\approx$2.6\,keV photon energy for the primary beam and $\approx$87\,GeV photon energy for the secondary beam  considering $\gamma=2900$.

X-ray free electron sources such as the LCLS \cite{LCLSBLs}, SACLA \cite{SACLABLs} or the European XFEL \cite{EuropeanXFELBLs} can easily cover the region of interest of \YAL{a} few keV. In addition, since the GDR is broad, the limited temporal coherence of the X-ray source should not play an important role. The difficulty when envisaging such a setup is that the GF and the X-ray facilities would need to be co-located, which is an obvious limiting factor. This has been so far a major impediment for other proposed combinations of relativistic acceleration and coherent X-ray sources in nuclear quantum optics \cite{Buervenich2006,Liao2011,Liao2013}. However, this impediment might be solved by using table-top plasma-driven sources \cite{Kneip2}.

Further opportunities for producing higher-energy gamma rays may be afforded by scattering off relativistic beams of the secondary photons produced by the GF itself. This is discussed in Sec.\,\ref{subsec:Production_ultrahigh_en_gammas_colliding}.

\section{\texorpdfstring{$P$}{P}- and \texorpdfstring{$CP$}{CP}-violating Compton scattering of primary photons from stored ions}
\label{sec:PCP_Compton}
    
    Searching for exotic signatures of violation of symmetry under spatial inversion $P$ and time reversal $T$ in Compton scattering,
    $\gamma(\vec q)+N(\vec p)\to\gamma(\vec q^{\,\prime})+N(\vec p^{\,\prime})$,
    is a natural task for the GF. A clear distinction of the GF (where photons are scattered from stored ion beams) from existing gamma sources based on laser-light backscattering from an electron beam is the possibility to probe specifically hadronic $P$-violating, $T$-conserving (PVTC) and $P$- and $T$-violating (PVTV) interactions \cite{deVries:2020iea}. The experiment will consist in counting the number of the generated secondary photons as a function of the circular polarization of the primary light.
    
    We start this discussion with the case of the proton beam at the LHC. For the laser-photon energy of $\sim10$\,eV and the relativistic factor $\gamma_p\approx7000$ (approximately double that of the ion beam), the photon energy in the photon-proton c.m. frame of $\sim140$\,keV can be reached. The theory of Compton scattering on a nucleon at low energies with PVTC was laid out in Refs.\,\cite{Bedaque:1999dh,Chen:2000mb}, and the PVTV case was considered in Ref.\,\cite{Gorchtein:2008cg}. Long-wavelength photons only interact with the bulk properties of the nucleon: its charge, mass, magnetic moment, and, in the case of a PVTV process, the electric dipole moment (EDM). Additionally, polarizabilities parametrize the response of the internal structure of the nucleon (which also depends on PVTC and PVTV interactions) to a quasistatic electromagnetic field. Peculiar for PVTC and PVTV Compton processes, the effect of polarizabilities dominates over the respective ground-state contribution, the opposite to the case of Compton scattering without symmetry violation~\cite{Gorchtein:2008cg}.
    This opens up a possibility to search for a new class of PVTV interactions other than those generating the EDM. Disregarding the effects of the nucleon spin for this discussion, the PVTC and PVTV signatures in the Compton scattering process with circularly polarized incoming laser photons are given by the $\vec S_\gamma\cdot(\vec q+\vec q^{\,\prime})$ and $\vec S_\gamma\cdot(\vec q-\vec q^{\,\prime})$ terms, respectively, where $\vec S_\gamma=i\vec \varepsilon_\lambda(\vec q)\times \vec\varepsilon^*_{\lambda\prime}(\vec q^{\,\prime})$ is the photon spin defined in terms of the initial and final photon polarization vectors $\vec\varepsilon_\lambda(\vec q)$ and $ \vec\varepsilon^*_{\lambda\prime}$, respectively. The presence of such PVTC and PVTV terms leads to a single-spin asymmetry~\cite{Gorchtein:2008cg} 
    \begin{equation}
     A_\gamma\equiv\frac{\sigma^+-\sigma^-}{\sigma^++\sigma^-}=A_\gamma^{\rm PVTC}\cos^4\frac{\theta}{2}+A_\gamma^{\rm PVTV}\sin^4\frac{\theta}{2},\label{eq:PVCompton}
    \end{equation}
    where $\sigma^\pm$ stands for the differential Compton cross section with left/right circular polarization of the incident photon. 
    Asymmetries $A_\gamma^{\rm PVTC}$ and $A_\gamma^{\rm PVTV}$ encode the respective polarizabilities. 
    Importantly, both PVTC and PVTV signals arise in the same observable, and the only difference is in its angular dependence. This feature follows from the fact that the $P$-even, $T$-odd vector $\vec S_\gamma$ projects onto either the $P$-odd, $T$-odd combination $(\vec q+\vec q^{\,\prime})$, or the $P$-odd, $T$-even combination $(\vec q-\vec q^{\,\prime})$, so that the single-spin asymmetry is purely PVTC in the forward direction, purely PVTV in the backward direction, and a mixture of the two in between.
    
    The size of the asymmetry $A_\gamma$ in Eq.\,\eqref{eq:PVCompton} depends on the model used to generate $A_\gamma^{\rm PVTC}$ and $A_\gamma^{\rm PVTV}$. For the nucleon case, the relevant mechanism is due to parity-violating couplings of lightest mesons, most notably the pions. 
\begin{figure}[!hpb]\centering
    \includegraphics[width=0.8\linewidth]{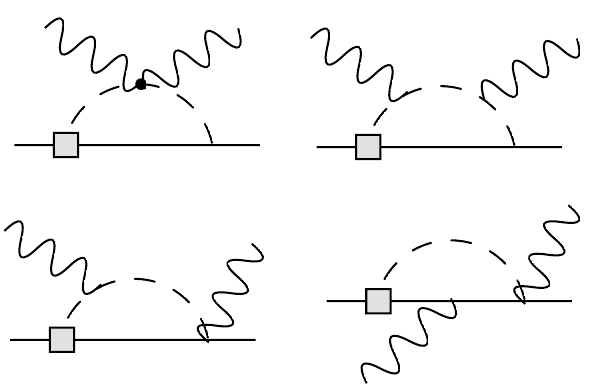}
    \caption{Representative Feynman diagrams responsible for generating PVTC polarizabilities. The square denotes the PVTC pion-nucleon coupling $h^1_\pi$, while the dotted vertex denotes the  two-photon coupling to the pion. }     
    \label{fig:PVTCpol}
\end{figure}

    The diagrams of  Fig.~\ref{fig:PVTCpol} yield \cite{Bedaque:1999dh,Chen:2000mb}:
    \begin{equation}
        A_\gamma^{\rm PVTC}\sim3\times10^{-8}\left(\frac{h^1_\pi}{5\times10^{-7}}\right)\left(\frac{\omega}{100\,\mathrm{MeV}}\right)^3,
    \end{equation}
     where $\omega$ stands for the c.m. photon energy. A recent measurement of the PV asymmetry in $\vec n+p\to d+\gamma$ by the NPDGamma collaboration~\cite{Blyth:2018aon} obtained
     \begin{equation}
         h^1_\pi=(2.6\pm1.2\pm0.2)\times10^{-7},
     \end{equation} 
     with the first and second uncertainties being statistical and systematic, respectively. This, together with the maximum energy of $140$\,keV in this GF setting, makes the PVTC asymmetry on a proton beam too small to be observed.
\begin{figure}[!hpb]\centering
    \includegraphics[width=0.3\linewidth]{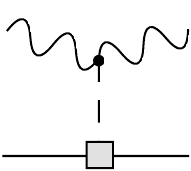}
    \caption{The $\pi^0$-pole contribution to the PVTV polarizabilities. The square denotes the PVTV pion-nucleon coupling $\bar g_0$.}      
    \label{fig:PVTVpol}
\end{figure}

    Similarly, the PVTV pion-nucleon coupling $\bar g_0$ generates PVTV polarizabilities, for example, via $\pi^0$ exchange as shown in Fig.~\ref{fig:PVTVpol}. This mechanism leads to an estimate~\cite{Gorchtein:2008cg} 
    \begin{equation}
        A_\gamma^{PVTV}\sim \bar g_0\left(\frac{\omega}{m_\pi}\right)^2.
    \end{equation} 
    The natural size of $\bar g_0$ is constrained by the neutron EDM~\cite{Crewther:1979pi}, $\bar g_0\lesssim10^{-11}$.
    
        To access the PVTC and PVTV signatures via laser backscattering at the GF, we therefore turn our attention to systems in which i) the characteristic energy scale is (ideally) comparable to the c.m. photon energy available,  i.e. $60$\,keV for the ion beam, and ii) the natural size of symmetry violation is enhanced by the presence of parity doublets, nearly degenerate pairs of states of opposite parity \cite{Sushkov_1982}. 
    We denote the average excitation energy of the parity doublet by $E_{\rm pd}\approx E_1\approx E_2$ and the energy splitting within the parity doublet by $\Delta E=E_2-E_1$. If the typical nuclear energy scale $E_N\approx10$\,MeV is much larger than this splitting, an enhancement factor $R\sim E_N/\Delta E\gg1$ arises, and 
    the expectation for the size of the symmetry-violating asymmetries in the ``polarizability regime" $\omega\leq E_{\rm pd}$ is
    \begin{align}
        A_\gamma^{\rm PVTC}&\sim10^{-8}
        R \left(\frac{\omega}{E_{\rm pd}}\right)^3,\label{eq:APVTC}\\
        A_\gamma^{\rm PVTV}&\lesssim10^{-11}
        R \left(\frac{\omega}{E_{\rm pd}}\right)^2.\label{eq:APVTV}
    \end{align}
    We list a few known examples of parity doublets in light nuclei in Table~\ref{tab:Parity_Doublets}.
    Given that the photon energy in the center-of-momentum of the laser photon and the stored beam is constrained to be below $60$\,keV, while the typical nuclear excitations reside at a few MeV, we can still operate in the polarizability regime. The enhancement due to the small energy splitting within a doublet comes in linearly, whereas energy suppression appears quadratically or cubically. 
    Because of this, one should aim at the smallest energy denominator in Eqs.\,\eqref{eq:APVTC}, \eqref{eq:APVTV} making ${}^{18,19}$F and ${}^{21}$Ne promising candidates.

Until now, these nuclei were only used for looking for PVTC signals. If a small PVTV component is present additionally to the parity mixing (PV is a prerequisite for PVTV effects), a backward-peaked component will arise. Note that due to a quadratic energy suppression rather than cubic for PVTC, the PVTV signal will have an additional enhancement with respect to the PVTC case, adding to the motivation for looking for such tiny asymmetries. 

\begin{table*}[t]
    \centering
\begin{tabular*}{\textwidth} {@{\extracolsep{\fill}} cccccccc c cc}
    \hline 
    \hline
      Isotope & $T_{1/2}$ &$E_1$\,(keV) & $I_1^P$   &$E_2$\,(keV) &$I_2^P$ & $\Delta E$\,(keV) & $R$ & 10$^8$ $A^{\rm PVTC}_\gamma$& FOM\,(s$^{-1}$) \\
    \hline \\[-0.2cm]
     ${}^{18}$F\,($1^+$) & 109.77(5) min &1042  &0$^+$  & 1081   &0$^-$ & 39 & 256 & 0.05 & $2.5\times10^{-8}$ \\
     ${}^{19}$F\,($\frac{1}{2}^+$) & stable &0  &$\frac{1}{2}^+$  & 110 &$\frac{1}{2}^-$ & 110 & 91 & 15 & 2$\times 10^{-3}$\\
     ${}^{20}$Ne\,($0^+$) & stable & 11255 &$1^-$  &  11258  &$1^+$ & 3.2 & 3125 & $5\times10^{-4}$& $2.5\times10^{-22}$ \\
     ${}^{21}$Ne\,($\frac{3}{2}^+$) & stable &2789  &$\frac{1}{2}^+$  & 2795   &$\frac{1}{2}^-$ & 5.7 & 1754 & 0.02 & $4\times10^{-9}$ \\
     
    \hline
    \hline
    \end{tabular*}
    \caption{Parameters of the low-lying parity doublets in isotopes of fluorine and neon, from the
     ENSDF database \cite{ENSDF}.}
    \label{tab:Parity_Doublets}
\end{table*}
    
    The longitudinal polarization of the relativistic ions (those that have non-zero spin) will allow to address the spin-dependent PVTC and PVTV polarizabilities. Following Ref.\,\cite{Bedaque:1999dh} for the PVTC 
    case worked out for the polarized proton, we expect the PV longitudinal single-spin asymmetry 
    \begin{align}
        A_{\rm Ion\;spin}^{\rm PVTC}&\sim10^{-8}
        R \left(\frac{\omega}{E_{\rm pd}}\right)^2 \label{eq:APVTC_ion}.
    \end{align}
The lower power of energy in this estimate suggests that if using a polarized ion beam is a viable option the PVTC effects Compton scattering might be easier to access than the photon asymmetry. In contrast, the contribution of PVTV spin polarizabilities to this observable is suppressed by an extra power of the photon energy, and involving the ion spin does not offer better sensitivity to PVTV.

The figure of merit (FOM) is defined as \cite{Horowitz:1999fk}
\begin{equation}
\textrm{FOM}=\mathrm{Rate}\times A^2,    
\end{equation}
and corresponds to the inverse time necessary for an observation of the asymmetry. 
Above, the rate is proportional to the cross section and fluxes of the colliding particles, and $A$ denotes the asymmetry. 
The cross section for low-energy Compton scattering (Thomson scattering) is
\begin{equation}
    \sigma=\frac{8\pi}{3}
    \left(
    \frac{Z^2\alpha \hbar}{Mc}
    \right)^2 .
\end{equation}
Here $Z$ and $M$ are the charge and mass of the ion. The cross-section is $\sigma\approx 3$\,$\mu$b for the case of $^{19}$F. 
The event rate is obtained as a product of the laser photon rate of $N_\gamma\approx 10^{25}$ per second, number of ions per bunch $N_{I/b}\approx10^{10}$ (see Table\,\ref{tab:GF_parameters}), 
and the ratio of the process cross section $\sigma$ to the cross section of the laser beam $S\approx(20\,\mu\mathrm{m})^2$:
\begin{equation}
    \mathrm{Rate}=N_{I/b}\times N_\gamma\times\frac{\sigma}{S}\approx 10^{11}\,\mathrm{s}^{-1}.
\end{equation}
Here we assume that the time structure of the laser pulses is matched to that of the ion bunches. Numerical estimates of the FOM for the four ions under consideration are summarized in Table\,\ref{tab:Parity_Doublets}. These results indicate that a promising candidate nucleus is ${}^{19}$F, for which a 10\% measurement of the asymmetry can be achieved with a day of statistics accumulation.   
 
 A final note in this section concerns the background not associated with symmetry-violating effects.
 Final-state interaction (FSI) due to electromagnetic rescattering will generate a non-zero phase of the Compton amplitude if even a tiny linear polarization component $P$ is present, leading to a false asymmetry due to a correlation $\vec S_\gamma \cdot[\vec q\times {\vec q\,}^\prime]$.
 In Ref.\,\cite{Gorchtein:2008cg} this background was estimated as $A_\gamma^{\rm FSI}\sim\alpha\frac{\omega^2}{M_N^2}\,P\,\sin^2\theta\cos2\phi$. The maximal degree of misalignment of circular polarization of laser photons can  reasonably be assumed to be $P\lesssim10^{-6}$, and for $\omega\leq60$\,keV the false asymmetry should be small. 
 The modulation with the azimuthal angle $\phi$ can be used to further suppress this undesired  background: when integrated over the full $2\pi$ range, this non-symmetry-violating background drops out of the asymmetry.
The above estimate relies on the assumption that the photon energy lies below the inelastic threshold. For $\omega\leq60\,$keV this requirement suggests that a fully stripped ion beam has to be used to avoid that atomic excitations and interferences thereof mimic the PVTC or PVTV signatures. An additional PVTV-type correlation may be generated by FSI on top of a PVTC signature. We expect this effect at the level $10^{-5}$ of PVTC or smaller.

The experiment at the GF will entail measuring the flux of secondary photons depending on the circular polarization of the laser photons. 
The detector will have to be $2\pi$-symmetric in order to eliminate the $\cos2\phi$-modulated electromagnetic background. 


\section{Nuclear physics with GF secondary-photon beam on fixed  targets} 
\label{Sec:Nucl_photophys_fixed_target}
This Section addresses the opportunities provided by the GF secondary beam for nuclear spectroscopy and hadron physics in the energy range from a few to hundreds of MeV.
A wide variety of targets can be used in conjunction with the GF secondary beam, benefiting from the vast experience and  infrastructure at CERN. Apart from stable or long-lived nuclear species, unique opportunities to use rare radioactive elements may be opened by the close physical proximity to the Isotope mass Separator On-Line facility (ISOLDE) \cite{Catherall2017_ISOLDE}, one of the world's leading sources of radioactive nuclides. Further opportunities may be rendered possible by a dedicated storage ring for rare isotopes discussed in Sec.\,\ref{Sec:Radioisotope_Storage_Ring}. 

\begin{figure}[!hpb]\centering
    \includegraphics[width=1.0\linewidth]{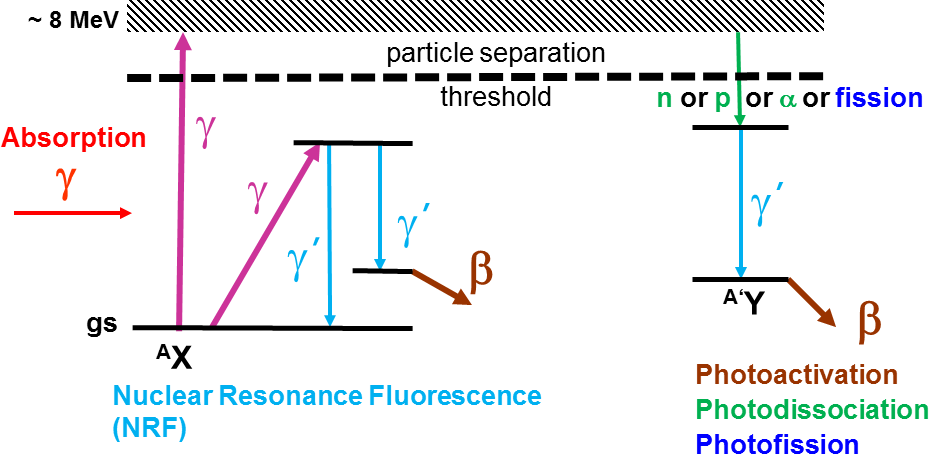}
    \caption{Possible photonuclear reactions following the absorption of a photon
             provided by the GF: excitation below the particle separation threshold will allow for nuclear
             spectroscopy via nuclear resonance fluorescence (NRF), potentially followed by $\beta$ decay
             (photoactivation). Excitation into the continuum above the particle-separation threshold will
             be followed by photodissociation (i.e. emission of neutrons, protons or $\alpha$ particles)
             or by photofission.}      
    \label{fig:photonuclear}
\end{figure}

Perhaps the most obvious application of the GF is nuclear spectroscopy, which will benefit from the narrow
spectral bandwidth achievable with collimated GF photons, their energy tunability and high intensities.
Figure\,\ref{fig:photonuclear} displays a schematic overview of possible photonuclear reaction pathways
following the absorption of an impinging photon. Nuclear spectroscopy experiments have been performed since the advent of betatron particle accelerators and Schiff's seminal
paper from 1946 \cite{Schiff1946}, proposing to use electron bremsstrahlung (converted from energetic
electron beams) to detect nuclear resonance fluorescence (NRF) off bound excited nuclear states. NRF is discussed in 
Sec.\,\ref{subsec:narrow-res} on gamma spectroscopy of narrow resonances at the GF. 

As indicated in Fig.\,\ref{fig:photonuclear}, in parallel to NRF, deexcitation of an intermediate level may proceed not directly
towards the ground state of the excited nucleus, but via $\beta$ decay to a daughter nucleus (photoactivation).
Alternatively, an excitation into the continuum above the particle-separation threshold will lead to subsequent
deexcitation via particle emission (neutrons, protons, $\alpha$ particles), hence to photodissociation, addressed in more detail in Sec.\,\ref{subsec:photonucl_res_above_trheshold} or
to photofission, discussed in Sec.\,\ref{Subsec:Photofission}.

\begin{figure}[!hpb]\centering
    \includegraphics[width=1.0\linewidth]{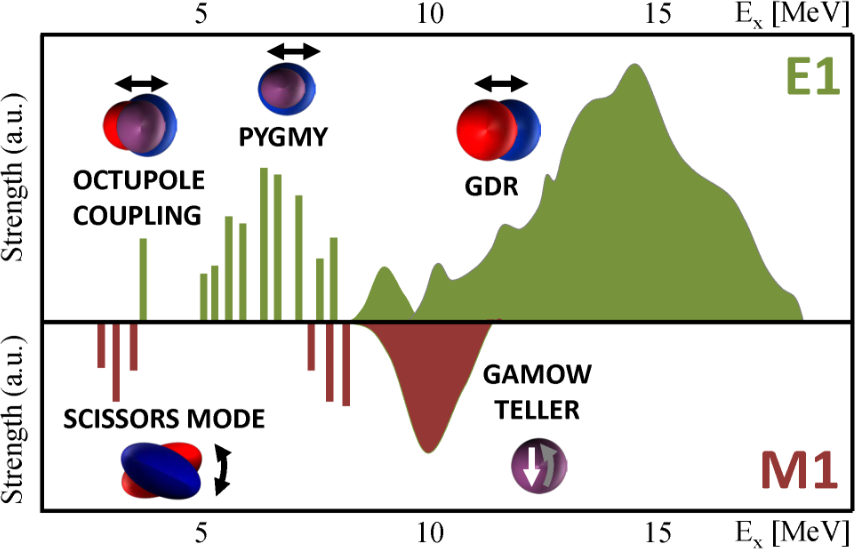}  
    \caption{Illustration of the photoresponse of deformed atomic nuclei as a function of the
             photon energy, divided into phenomena involving electric ($E1$) and magnetic ($M1$)
             excitations. Figure by A. Zilges.}      
    \label{fig:dipole-photoresponse}
  \end{figure}

In Fig.\,\ref{fig:dipole-photoresponse}, possible strong (electric and magnetic) dipole excitations
in heavy, deformed nuclei are schematically shown as a function of their excitation energy. We concentrate on dipole excitations since at the few MeV energies of interest, the gamma-ray wavelength is much larger than the nuclear size. On the magnetic dipole ($M1$) side, at lower energies,
the orbital magnetic dipole  excitation corresponds
in a simplistic geometric, macroscopic picture to a scissors-like vibration of the deformed proton and neutron
fluids against each other \cite{Kneissl1996}. Therefore this $M1$ mode is referred to as the `scissors mode', the magnetic analogue of the GDR. We further discuss the importance
of these transitions in the context of the GF in Sec.\,\ref{subsec:narrow-res}.
At higher energies of $\approx$10\,MeV, Gamow--Teller (GT) transitions \cite{Note_GT} appearing as $M1$ excitation strength  are related to the common weak-interaction processes of spin--isospin-type ($\sigma\tau$, i.e., a product of the spin and isospin operators) in atomic nuclei. These are of interest not only in
nuclear physics, but also in astrophysics; they play an important role in supernovae explosions and
nucleosynthesis \cite{Moreh1982,Fujita2011,Suzuki2011}.

On the electric dipole ($E1$) side of Fig.\,\ref{fig:dipole-photoresponse}, the GDR  represents a collective excitation mode where protons and neutrons vibrate against each other
along or perpendicular to the symmetry axis of the nucleus. The double-humped structure of the GDR in deformed
nuclei appears typically at energies of around 77$\cdot A^{-1/3}$\,MeV ($A$ is the mass number), which corresponds to
about 14\,MeV in rare earth nuclei. This isovector GDR was initially found as a spectral
feature corresponding to the excitation of the nuclear ground state, but soon Brink and Axel concluded that giant resonances could be associated with any nuclear state, independent of the microscopic structure of this state \cite{Brink1955_PhDThesis,Axel1962_GDR}. This led to intense studies of GDR properties of excited states, the
     so-called GDR in hot nuclei \cite{Santonocito2020_hotGDR}. Further prospects of GF studies of the GDR and potential multiphoton excitation 
     are presented in  Sec.\,\ref{Subsect:GDR-multi}.

In addition, enhanced electric dipole excitations are expected at lower
excitation energies. Several  experimental studies using electromagnetic probes have found an enhancement of electric dipole strength between about 5 and 10\,MeV. This phenomenon is denoted as pygmy dipole resonance (PDR)
and has been intensively studied in recent years \cite{Savran2013_pigmy,Schwengner2013,Zilges2015,Wilhelmy2020,Muscher2020,
Tsoneva2019,Ries2019,Tamkas2019}. In a simplified geometrical picture, the PDR is described as neutron skin
oscillating against a $N=Z$ core. The PDR phenomenon is discussed in more detail in
Sec.\,\ref{Subsect:pygmy}.

Moving towards even higher secondary-beam energies, and towards hadron physics, the second part of the Section goes beyond traditional gamma spectroscopy to address Compton and photo-induced processes on light nuclei in Sec.\,\ref{Subsec:Photopys_protons_light_nucl}, pion photoproduction in Sec.\,\ref{Subsec:Pion_photoproduction}, and the Delta-resonance region and continuum effects in Sec.\,\ref{subsec:Delta}. The Section concludes with prospects of more exotic processes at the GF involving parity-violation \ref{sec:P-violation}.


\subsection{Narrow resonances}
\label{subsec:narrow-res}
Narrow resonances in a wide energy range up to the particle-separation threshold are the subject of NRF. 
Resonance fluorescence in general refers to the resonant excitation of an upper state by absorption of electromagnetic
radiation and decay of the excited level by emission of radiation. Via NRF, nuclear states are excited in photon-scattering experiments and their deexcitation through electromagnetic transitions is studied via gamma-ray spectroscopy \cite{Metzger1959,Kneissl1996}.
From these experiments, photoabsorption cross sections and the related photon strength functions \cite{Goriely2019,
Isaak2019} are deduced, which are important for the description of photonuclear reactions and the inverse
radiative-capture reactions. 

In NRF experiments, the nucleus is excited by the absorption of a real photon. As these photons only carry a
small angular momentum, excitation favors transitions where the change of the nuclear angular momentum is small. The strongest transitions are typically of dipole character ($E1$ or $M1$). 
The excited states may subsequently decay either
directly or via intermediate lower-lying states back to the ground state. The latter transitions are typically
weak and difficult to resolve from the background when using energetically continuous (broadband) bremsstrahlung.
Therefore, for example, the determination of the total transition width $\Gamma$ is not always possible with
bremsstrahlung beams. An alternative way to perform NRF experiments uses quasi-monochromatic photon beams
as provided by existing or upcoming Compton-backscattering facilities \cite{HIGS,VEGA} and in the
future by the GF.

In general, for all nuclear shapes where the centers of mass and charge do not coincide,
one expects an electric dipole moment leading to enhanced electric dipole transitions. Examples are deformed
shapes due to octupole deformations or any kind of cluster configurations. Rotations on top
of the octupole vibrations coupled to the quadrupole deformed core lead to the so-called octupole vibrational
bands. In spherical nuclei close to magic proton or neutron shells the lowest-lying excitations are quadrupole
and octupole vibrations of the nuclear shape. A coupling of these two single-`phonon' excitations leads to
a two-phonon quintuplet with spins $J'=1^--5^-$. The 1$^-$ member of this multiplet can, in principle, be
excited via an electric dipole transition from the ground state in NRF experiments.
The energies of all these dipole excitations ($E1$ and $M1$) lie in the same range of roughly 2--4\,MeV, see also Fig.\,\ref{fig:dipole-photoresponse}. This
emphasizes the crucial necessity for parity assignments for the interpretation of observed excitations.
NRF is by far the most sensitive tool to detect such dipole excitations and to
(model independently) determine their characteristics: excitation energies, spins, parities, decay energies,
level widths, lifetimes, decay branchings, multipole mixing ratios and absolute transition strengths.

\begin{figure}[!hpb]\centering
    \includegraphics[width=1.0\linewidth]{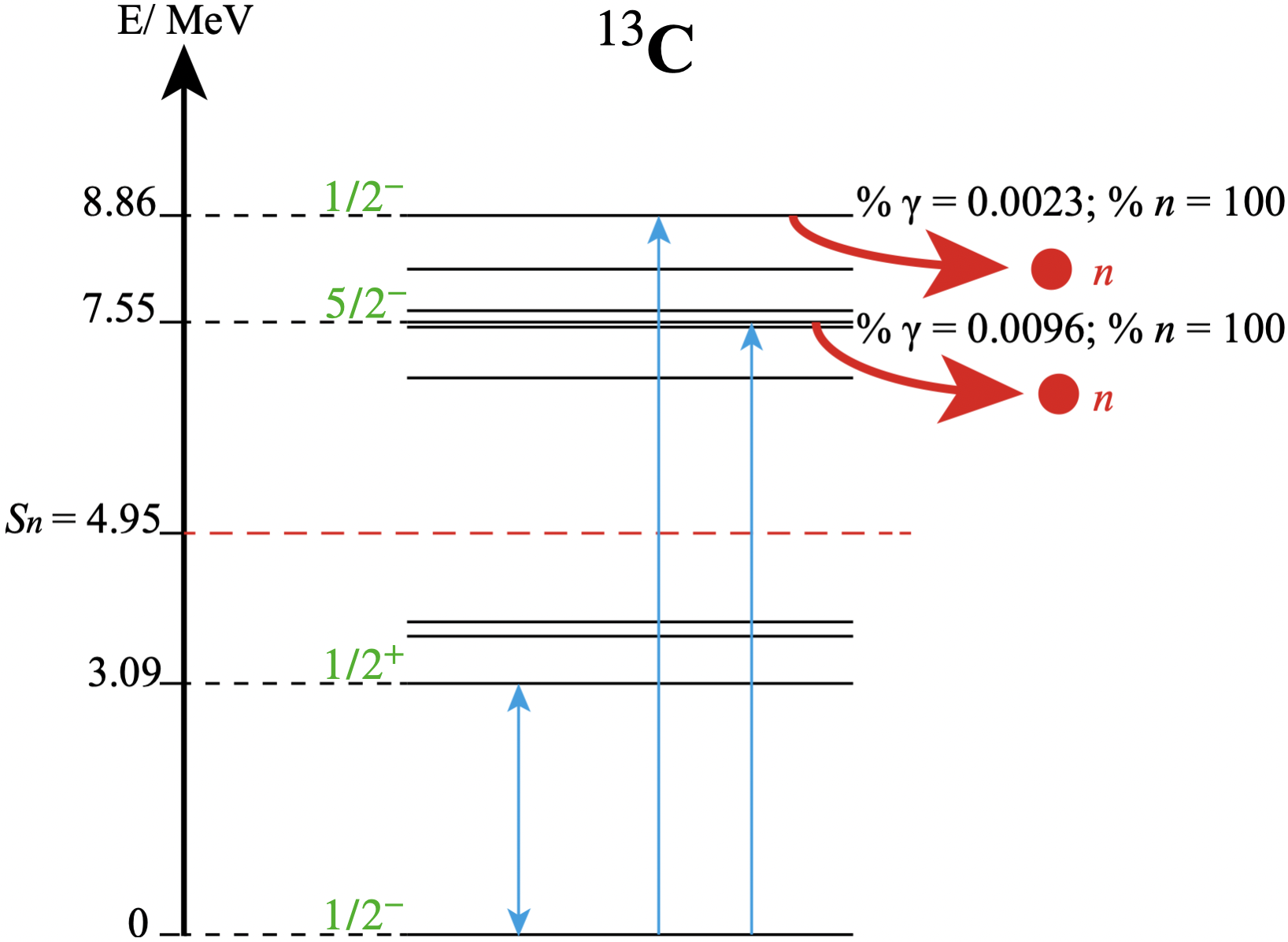}
    \caption{Nuclear energy levels below 8.86\,MeV in $^{13}$C. Energies, $I^{P}$ and decay modes \cite{NNDC}
      are given for the levels discussed in this work. The arrows show excitation to be induced by the
      secondary GF photons. For the lowest-energy gamma resonance, the double-sided arrow indicates that the
      only decay channel is via photon emission.}
    \label{fig:C13_En}
\end{figure}

There are many specific examples of transitions that can be studied at the GF taking advantage of its monochromaticity, tunability and unsurpassed photon flux. An interesting case is the one of $^{13}$C,
which exhibits many accessible resonances displayed in the partial level scheme in Fig.~\ref{fig:C13_En}. The case of $^{13}$C will be discussed in the context of polarimetry with narrow resonances in Sec.~\ref{subsec:Polarimetry_with_narrow_res}. The interesting neutron emission channels at 7.55 MeV and 8.86 MeV can be selectively excited by the GF and used in turn to generate tertiary monoenergetic fast neutron beams. This topic is the subject of Sec.~\ref{subsusbsec:Production_Monoenerg_neutrons}.

Another interesting case is the longest lived ($\tau_{1/2} \gtrsim 1.2\cdot10^{15}$\,y) isomer $^{180m}$Ta lying 77 keV above a short-lived ground state (see also Table\,\ref{tab:Isomer_dark_matter}). 
This isomer has been extensively studied in  order to understand its decay mechanisms via photoexcitation to a gateway level above the isomer
\cite{Auerbach2014_180Ta_m}, following the scheme presented in Sec.~\ref{sbsec:Low_En_Nucl_Trans}. 
Previous experiments \cite{Belic2002_180Ta_m} using broadband bremsstrahlung radiation could neither identify the gateway levels nor could they
provide information about the involved transition probabilities. The high flux and excellent monochromaticity of the GF secondary beam could finally shed light 
on this matter. 

Finally, we note that scattering of linearly polarized photons on unpolarized or spin-zero nuclei is an effective method to determine the parity of excited nuclear states, see, for example, Ref.~\cite{Krishichayan2015Parity}. The advantage of the GF for such studies is the ability to resolve relatively narrow resonances combined with high statistical sensitivity and tunability of the photon energy.

\subsection{Pygmy dipole resonances}
\label{Subsect:pygmy}


The dipole response for stable (medium to heavy) nuclei with a moderate neutron excess is almost
entirely concentrated in the giant dipole resonance (GDR) that exhausts almost 100$\%$ of the
Thomas-Reiche-Kuhn (TRK) sum rule (which is proportional to the energy-weighted sum rule) \cite{Kuhn1925,Reiche1925}. 
For this collective excitation, the nuclear ($N-Z$) asymmetry and the corresponding symmetry energy provide the restoring force for the oscillation between the excess neutrons against the ($N=Z$) nuclear core \cite{Piekarewicz2011}.
Thus only a minor part of the $E1$ strength is expected at lower
excitation energies. 
The electric dipole strength distribution at low energies (below $\approx 5$\,MeV) is dominated by multi-phonon excitations originating from the coupling between $J^{\pi} =2^+$
quadrupole and $J^{\pi} = 3^-$ octupole vibrations of the nuclear shape, being the lowest single-phonon excitations in even-even nuclei near closed neutron or proton shells. The two-phonon excitations $2^+ \otimes 3^-$, called quadrupole-octupole coupling, lead to a multiplet consisting of five states with spins and parities $J^{\pi} = 1^-,...,5^-$. In photon-scattering experiments, the 1$^-$ member of the $2^+ \otimes 3^-$
multiplet can be excited selectively by an $E1$ transition from the ground state.
Its subsequent decay is predominantly back to the ground state via an $E1$ transition~\cite{Kneissl2006}.

At energies between 5\,MeV and a few hundred keV
above the neutron-separation threshold, an enhancement of dipole strength was observed in many
semi-magical and magical nuclei \cite{Savran2013_pigmy}. It was found that the $E1$ dipole strength at energies
below the GDR range is of the order of 1$\%$ of the TRK sum for stable nuclei, and up to about 5$\%$ for the exotic nuclei studied so far \cite{Zilges2015}. Owing to its relatively low strength, this low-lying $E1$ transition is
often denoted as the pygmy dipole resonance (PDR). The first PDR observation was made by Bartholomew
in 1961 who found enhanced gamma-ray emission after neutron capture \cite{Bartholomew1961}. The PDR
name was coined in 1969, when its impact on calculations of neutron-capture cross sections was
reported \cite{Brzosko1969}. In 1971, Mohanet et al. proposed a description of the PDR in a three-fluid
hydrodynamical model \cite{Mohan1971}. One interpretation is an oscillation of excess neutrons (the neutron skin)
against the $N=Z$ core. The PDR is of great interest because it provides information on the neutron skin and on
the nuclear symmetry energy as a crucial ingredient to the nuclear matter equation-of-state \cite{Bertulani2019}.
On theoretical grounds, employing the relativistic quasiparticle random phase approximation (RQRPA) approach, a one-to-one correlation was found between the pygmy
dipole strength and parameters describing the density dependence of the nuclear symmetry energy, and in
turn with the thicknesses of the neutron skin \cite{Klimkiewicz2007}.

As discussed in Ref.~\cite{Piekarewicz2011}, the determination of the neutron radius of a heavy nucleus, i.e.
its neutron distribution, is a fundamental problem related to the equation-of-state of nuclear matter with
far-reaching consequences in areas as diverse as atomic PV \cite{Pollock1992,Sil2005},
nuclear structure \cite{Brown2000,Furnstahl2002,Danielewicz2003,Centelles2009,Centelles2010}, heavy-ion
collisions \cite{Tsang2004,Chen2005,Steiner2005,Shetty2007,Tsang2009}, and neutron-star
structure \cite{Horowitz2001,Horowitz2002,Carriere2003,Steiner2005_Iso,Li2006}.
Hence, the far-reaching impact of understanding the $E1$ strength distribution of the PDR motivated still ongoing
extensive experimental and theoretical studies in a wide mass range \cite{Savran2013_pigmy,Paar2007,Scheck2013,Paar2010,Ries2019_Cr_pigmy,Tamkas2019, Tsoneva2019,Zilges2015,Muscher2020,Wilhelmy2020,Romig2013,Spieker2020}. In a recent work at GSI \cite{Ries2019_Cr_pigmy}, PDR were studied in a chain of chromium isotopes, see Table\,\ref{tab:Isotope_chains}.

As photon scattering is a key experimental technique to study the $E1$ dipole strength in nuclei,
and in particular the PDR, an intense and high-quality (i.e. monoenergetic) photon beam
as envisaged for the GF would turn out highly beneficial for this active research field.

\subsection{GDR and multiphoton excitation}
\label{Subsect:GDR-multi}

The high photon flux in the GF secondary beam can be used at photon energies of $10-20$~MeV   to excite the GDR in medium-mass
and heavy nuclei.  In a shell-model picture the GDR is a superposition of particle-hole excitations out of the ground state and is not an eigenstate of the nuclear Hamiltonian.  These particle-hole excitations actually do
  not all have the same energy, leading to a spreading of the GDR
  often referred to as Landau damping \cite{Fiolhais1986,Speth1991}. The residual two-body interaction mixes the particle-hole excitations with each
  other and with other shell-model configurations and, thus, spreads
  the GDR over the eigenstates of the nuclear Hamiltonian. This leads to a Lorentzian
  distribution of the dipole strength with a so-called ``spreading'' width
  $\Gamma^\downarrow$ of around 5~MeV \cite{Feshbach1964, Herman1992}. Ground-state deformation in axially-symmetric deformed nuclei leads to a splitting of the GDR Lorentzian into two components. The GDR spectrum, with its strength, centroid, and Lorentzian parametrizations thus offer information on the nuclear shape \cite{Gaardhoeje1992}, and in addition, on the nuclear size, the nuclear symmetry energy (important for the study of neutron-star structure) and the viscosity of the neutron and proton fluids \cite{Pandit2013}. In a time-dependent picture, the spreading of the
GDR over nuclear eigenstates  can be viewed as statistical equilibration \cite{Agassi1975} with a characteristic time scale $\hbar / \Gamma^\downarrow$.

\begin{figure}[hb]
\vspace{-0.4cm}

\includegraphics[width=0.5\textwidth]{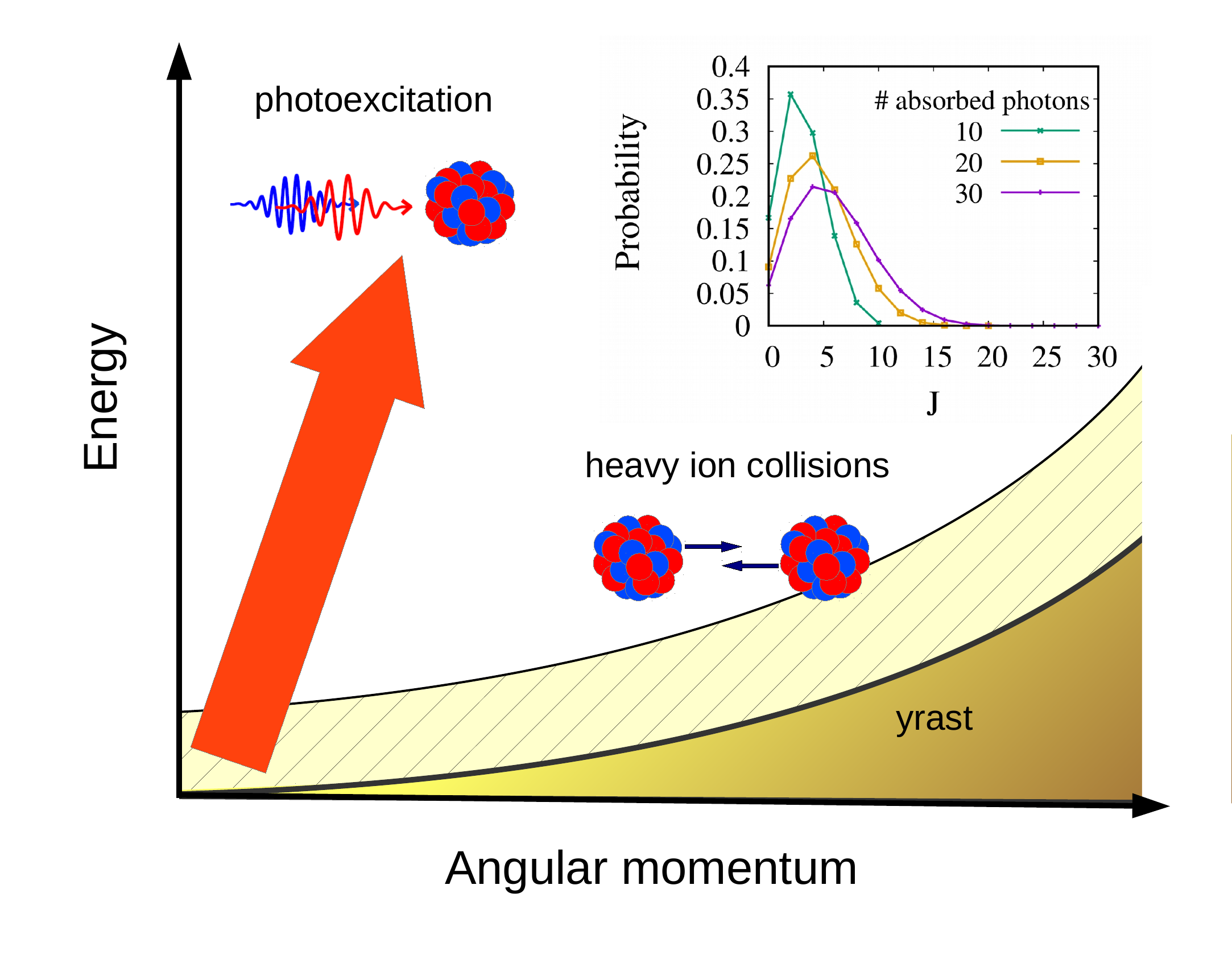}
  \caption{\label{fig:yrast}  Qualitative illustration of two
    regimes of nuclear excitation. The yrast line defines the minimum 
    energy of a nuclear state with a certain angular momentum. Heavy-ion 
    collisions preferentially excite states close to the yrast line (region depicted by
    hatched area).  Multiple absorption of multi-MeV dipole
    photons involves small transfer of angular
    momentum and could lead to compound states several hundred MeV above
    yrast (red arrow). The inset shows the angular-momentum distributions for
    $N_0=10, 20$ and $30$ absorbed dipole photons starting from an initial state with $J=0$. Adapted from Ref.~\cite{Pal14}.}
\end{figure}

According to the Brink-Axel hypothesis, the GDR is a mode of excitation not only of the nuclear ground state, but of any excited state as well \cite{Brink1955_PhDThesis,Axel1962_GDR}. GDRs built on excited states are termed ``hot GDRs'' and have so far been produced in heavy-ion collisions leading to fusion or incomplete fusion \cite{Santonocito2020_hotGDR}. As a result of the collision, a compound nucleus is formed, i.e., a highly excited nucleus in which the excitation energy is distributed over several or many particle-hole excitations \cite{Agassi1975}. Heavy-ion collisions involve large angular momenta, leading to high angular momentum states in the compound nucleus. Completely new prospects would be opened by the possibility to reach the same excitation energies by multi-photon absorption. Absorption of a dipole photon transfers one unit of angular momentum. The total angular momentum in a multiple dipole absorption process typically increases with the square root of the number of absorbed photons \cite{Pal14}. Multiple dipole absorption would therefore lead to excitation of the compound nucleus at high energy and at comparatively low angular momentum, far above the yrast line, as illustrated in Fig.\,\ref{fig:yrast}. The spectral region of high excitation energy and small angular momentum is so far mostly unexplored. Little is known experimentally in that region about state densities, the widths of the GDRs, decay properties of highly excited states, and, specifically, about the nuclear equilibration process.
  
The identification of $\hbar / \Gamma^\downarrow$ with the equilibration time becomes questionable at high excitation
  energy well above particle threshold. For example, hot GDRs
  \cite{Santonocito2020_hotGDR} are characterized by high neutron-evaporation rates and have correspondingly short lifetimes \cite{Bortignon1991}. Neutron evaporation may happen prior to complete equilibration. The increase of the GDR width due to neutron evaporation then does not correspond to a shorter equilibration time.   
  Such increase is considered to be one of the possible reasons for the damping of the hot GDR \cite{Bortignon1991,Gaardhoje1992nuclear,Santonocito2020_hotGDR}. Sokolov and Zelevinsky confirmed the interpretation of Ref.\,\cite{Bortignon1991} and 
  attributed the disappearance of the collective GDR strength observed experimentally to the complex interplay between internal and external dynamics with decay into the continuum \cite{Sokolov1997_collective}.  A different explanation postulating an increase of the spreading width with temperature was also put forward \cite{Yoshida1990}.
  \pagebreak
  
Due to its large width, the GDR can in many cases be investigated with broadband gamma sources. In the following we identify special cases for which the high photon flux and/or monochromaticity of the GF may be of advantage. 

First, the high photon flux would be beneficial for the study of GDR excitations in exotic nuclei far from the valley of stability. Due to their short lifetime, such targets are bound to produce small excitation yields and are hard to access experimentally. A notable exception is the PDR and GDR measurement of the neutron-rich unstable $^{68}$Ni nucleus \cite{Rossi2013}. Nuclear gamma spectroscopy on the GDR  at the GF could bring new insights into the shapes and properties of exotic nuclei. 

Second, the GF will enable photofission studies of fission isomers in actinides. A nuclear reaction such as (d,p) or fragmentation would produce highly deformed excited states in the second minimum of the fission barrier. Photoabsorption would excite the GDR built upon these states.  The high photon flux and in particular the monochromaticity of the GF would allow for photofission studies of these strongly deformed actinides with a much better sensitivity than so far available to the fission barrier landscape. This is the subject of Sec.\,\ref{Subsec:Photofission}.

Finally, one may consider the possibility of multiple absorption of dipole photons, where for instance a second absorbed photon excites a GDR built upon the GDR reached by the first photon. As shown  in Ref.\,\cite{Lewenkopf1994}, a comparison of single and multiple GDRs could probe statistical aspects  of the resonance width, exhibiting either a Lorentzian or Gaussian shape.  For the focused laser pulses envisaged at the Nuclear Pillar of the Extreme Light Infrastructure in Bucharest \cite{ELI-NP-WB}, with pulse durations of tens of zeptoseconds ($1\,\textrm{zs}=10^{-21}$\,s), theoretical studies based upon the Brink-Axel hypothesis predict multi-photon absorption of up to $\approx$hundred photons.  This would lead to excitation energies in the range of several $100$\,MeV, to multiple neutron evaporation and to the formation of neutron-deficient nuclei far from the valley of stability \cite{Wei09,Pal14,Pal15,Kobzak2020}. 
 In comparison, focusing of the secondary beam does not appear to be feasible at the GF and the envisaged pulse durations are many orders of magnitude longer. Thus,  even the probability of two-photon excitation remains  small at the GF.


\subsection{Photonuclear response above particle threshold}
\label{subsec:photonucl_res_above_trheshold}

The excitation-energy region around the particle-separation threshold is, on the
one hand, of interest from a theoretical perspective because the coupling of
bound quantum states to the continuum of unbound states requires an extended
formalism, which is particularly intricate for exotic nuclei near the
drip-lines where all structures are weakly bound~\cite{Jonson2004}.
On the other hand, this energy region covers the Gamow-window of thermally
driven reactions of nucleons with heavy nuclei. Its understanding is a
prerequisite for modelling nuclear-reaction cascades in hot
cosmic objects and thus for understanding nucleosynthesis. Below the threshold, all excited
resonances decay predominantly by gamma-ray emission, with rare cases of $E0$ transitions involving internal conversion or electron-positron pair production. 

In this regime, the knowledge of the nuclear level densities and the gamma-ray
strength function is of crucial importance as an ingredient, for example, to model nuclear-reaction-network calculations of
nucleosynthesis. From an experimental perspective,
the so-called ``Oslo method'' was established as a reliable tool to derive these properties from measured gamma-ray energy spectra.
The Oslo method comprises a set of analysis techniques designed to extract the nuclear level density and average gamma-decay strength
function from a set of excitation-energy tagged gamma-ray spectra.
The method was first introduced in Ref.~\cite{Schiller2000} and since then continuously
further developed~\cite{Tornyi2014}. Recently a new software implementation of the
method was provided~\cite{Midtbo2021}.

Above the threshold, the
particle-decay channel opens up. Either no gamma rays can be observed at all
or their intensity cannot be used as a measure of the total electromagnetic
excitation strength of the resonance due to the unknown particle-decay
branching ratio. An intense narrow-energy-band photon beam will open
up new horizons for investigation of nuclear photo-response at and above the separation threshold~\cite{Pietralla2010}.

\subsubsection{Neutron-capture cross sections of s-process branching nuclei}
\label{Subsubsec:s-process}

Cosmic nucleosynthesis of heavy elements above the so-called iron peak mainly proceeds via neutron-capture processes; the r process (r: rapid neutron capture) is connected to scenarios of high neutron densities well above
10$^{20}$\,cm$^{-3}$ and temperatures on the order of 2-3$\cdot$10$^9$\,K as
occurring in explosive scenarios such as, for example, supernovae or neutron-star
mergers~\cite{Wallerstein1997,Hotokezaka2018}. In contrast, average neutron
densities during s-process nucleosynthesis (s: slow neutron capture) are rather
small ($\approx$10$^8$\,cm$^{-3}$), i.e. the neutron capture rate $\lambda_n$ is
normally well below the $\beta$-decay rate $\lambda_{\beta}$ and the reaction
path is close to the valley of $\beta$
stability \cite{Kaeppeler1999,Kaeppeler2006,Dillmann2008}.
However, when the s-process reactions occur at peak neutron densities,
reaction path branchings occur at unstable isotopes with half-lives as low as
several days. The half-lives at these branching points are normally known
with high accuracy, at least under laboratory conditions. One relies on theory
for the extrapolation to stellar temperatures~\cite{Takahashi1983}; the neutron-capture cross sections are  accessible to direct experiments only in special cases. Besides the limited availability of sufficient amounts of target
material, intrinsic activity of the target hinders experimental
access especially in the case of branching points involving isotopes with short half-lives.
Moreover, the predictions in the Hauser-Feshbach statistical model yield
different results depending on the underlying parameter sets. Additionally,
studies of branching points involving long-lived isotopes (e.g. $^{147}$Pm, $^{151}$Sm, $^{155}$Eu)
showed that the recommended values of neutron-capture cross sections in the
Hauser-Feshbach statistical model~\cite{Bao2000} differ by up to 50$\%$ from
the experimentally determined values~\cite{ELI-NP-Whitebook2010b}.
Therefore, further and precise experimental constraints on the theoretical
predictions of these crucial values are needed. The inverse ($\gamma$,n)
reaction could be used at the GF to determine optimized model-input parameters
for improved predictions.

\subsubsection{(\texorpdfstring{$\gamma$}{g},p), (\texorpdfstring{$\gamma,\alpha$}{g,a}) cross sections for p-process nucleosynthesis}
\label{Subsubsec:p-process}

In the framework of cosmic nucleosynthesis, `p-nuclei' are certain proton-rich,
naturally occurring isotopes of some elements which cannot be produced in
either slow (s-process) or rapid (r-process) neutron-capture processes.
In the seminal paper by Burbidge, Burbidge, Fowler and Hoyle
(`B$^2$FH')~\cite{B2FH1957}, accompanied by~\cite{Cameron1957}, the s- and
r-processes for the production of intermediate and heavy nuclei beyond iron were
introduced. It was immediately realized by the authors of both papers that a number of proton-rich
isotopes can never be synthesized through sequences of only neutron captures
and $\beta^-$ decays. This required the postulation of a third process.
It was termed p-process, because it was thought to proceed via proton captures
at high temperature, however, later findings shed doubts on the feasibility to
use proton captures for producing all of the nuclides missing from the s- and
r-process mechanisms. In the literature, `p-process' is sometimes used as a general term denoting whatever production mechanism(s) is/are found to be responsible for the p-nuclides.
Historically, there were 35 p-nuclides identified, with $^{74}$Se being the
lightest and $^{196}$Hg the heaviest~\cite{Rauscher2010}.
Compared to the bulk of natural isotopes, p-nuclei generally show
abundances which are 1-2 orders of magnitude lower.
Photodesintegration rates, like ($\gamma$,n), ($\gamma$,p) and
($\gamma$,$\alpha$) play a crucial role in the nucleosynthesis of these nuclei.
As the p-reaction network calculations comprise several hundred isotopes and
the corresponding reaction rates, theoretical predictions of these rates,
normally obtained in the framework of the Hauser-Feshbach statistical model
theory~\cite{HauserFeshbach1952}, are necessary for the modelling.
The reliability of these calculations should be tested experimentally for selected isotopes.

While the ($\gamma$,n) cross sections in the energy regime of the Giant Dipole
Resonance around 15\,MeV were measured extensively already several decades
ago (see, for example, Ref.\,\cite{BermanFultz1975}), many efforts using continuous
bremsstrahlung spectra have been made~\cite{Vogt2001,Sonnabend2004,Erhard2006}
to determine the reaction rates without any assumptions on the shape of the dependence of the
cross section on energy  in the astrophysically relevant energy region just above the reaction threshold. Also laser Compton backscattering to produce
monoenergetic photon beams was used to determine reaction rates by an
absolute cross section measurement \cite{Utsunomiya2003}.
In contrast, experimental knowledge about the ($\gamma$,p) and
($\gamma$,$\alpha$)
reactions in the corresponding astrophysical Gamow energy window is much more scarce.
In fact, the experimental data are based on the observation of the time
reversal (p,$\gamma$) and ($\alpha$,$\gamma$) cross sections,
respectively \cite{Sauter1997,Oezkan2002,Rapp2002,Gyurky2005} for the
neutron-deficient
nuclei with mass numbers around 100. Therefore, being able to measure these rates directly would represent a significant progress.
It should be noted that deeper insight into the p-process nucleosynthesis
would not emerge from the measurement of only a few selected reactions, but via
the development of a comprehensive database. 
This necessitates rather short
individual measurement times for studying the stable p nuclei despite their typically
low natural abundances~\cite{ELI-NP-Whitebook2010c}. 
This will only
become possible using intense
$\gamma$ beams as will be provided by the VEGA facility at ELI-NP and, ultimately, by the Gamma Factory.

However, the majority of nuclides in the p-process network calculations are radioactive. Here, the access to stored secondary beams at the GF, see Sec.\,\ref{Sec:Radioisotope_Storage_Ring}, will enable a unique possibility to investigate photo-induced reactions on short-lived species. Furthermore, direct $(p,\gamma)$ and $(\alpha,\gamma)$ reactions can be as well studied by employing in-ring, pure, thin, gaseous H$_2$ and He targets \cite{MeiBo-2015,Glorius-2019}.
In this way, direct and reverse reactions can be measured for the same pairs of nuclides, thus enabling immediate testing of Hauser-Feshbach calculations.

\subsubsection{Direct measurement of astrophysical \texorpdfstring{$S$}{S}-factors}
\label{subsec:AstroSfactors}

Helium burning in stars, as has been realized at the dawn of nuclear astrophysics \cite{B2FH1957}, is the key process to understanding the abundances of chemical elements in the universe. 
This process proceeds via the  triple-$\alpha\to ^{12}$C reaction, 
enhanced by the Hoyle resonance \cite{Hoyle:1954zz}, and  followed by the $\alpha$ radiative capture on carbon, $^{12}$C$ (\alpha, \gamma) ^{16}$O. The rate of the former process at stellar temperatures is known to $\sim10\%$. In turn, a $\approx100\%$ uncertainty in the rate of the latter reaction represents the main source of uncertainty in stellar evolution models. A direct measurement of the rate of the $^{12}$C$ (\alpha, \gamma) ^{16}$O reaction at astrophysical energies is considered `the holy grail' of nuclear astrophysics, but is extremely challenging. The radiative process is rare due to the weakness of the electromagnetic force, and may be suppressed by 3-6 orders of magnitude with respect to other processes (e.g., elastic scattering). 

Attempts to measure the inverse process generalized to a virtual-photon-induced disintegration $^{16}$O$ (e,e'\alpha) ^{12}$C showed the possibility to access this process experimentally \cite{DeMeyer:2001zgp}, albeit no satisfactory agreement of theoretical predictions with the data was achieved. There is a revived interest in studying this process at electron-scattering facilities \cite{Friscic:2019eow}.  While one wins in the rates due to high intensity of electron sources, the price to pay are electromagnetic background processes and the necessity to extrapolate to the real photon point. 
At GSI, Coulomb dissociation of a 500\,MeV/nucleon $^{16}$O beam colliding  with a Pb target is used to address this reaction.
This is an extremely challenging experiment. Apart from tiny reaction rates at low center-of-mass energies, the dissociation products have nearly identical magnetic rigidities \cite{Goebel-2020}, making it hard to separate them. 
The Gamma Factory will allow to directly study the $^{16}$O$ (\gamma,\alpha) ^{12}$C photodisintegration process  avoiding the aforementioned complications.

To be of direct use for understanding stellar nucleosynthesis, measurements
  at the GF should be done close to the energy
  relevant to the helium-burning conditions in stars, where the ignition plasma
  temperature is T$\approx 2\cdot 10^8$\,K. The kinetic energy of
  the $\alpha$ particle corresponding to the astrophysics-relevant Gamow window is
  $E_{\alpha}\approx$300\,keV.
  The respective threshold for the (inverse) photodisintegration process
  $^{16}$O($\gamma,\alpha$)$^{12}$C is $E^{th}_{\gamma}\approx$7.6\,MeV.
  For comparison, the competing proton-knockout reaction threshold lies at $\approx$11.6\,MeV.


\subsubsection{Alpha clustering in heavy nuclei}
\label{Subsec:Alpha_Clustering}

Although alpha clustering in light nuclei is well known and the famous Hoyle state in $^{12}$C plays an important role in the evolution of the Universe, there are many open questions concerning the physics of clusterization in nuclei \cite{Karamian2014}.
Even the geometrical structure of the most clusterized nuclei, such as ${^{6}}$He in the ground state and ${^{12}}$C in the Hoyle state (and higher similar states), is still under debate. In the case of ${^{6}}$He, one critical feature is the correct prediction of the $\beta$-decay to $^6$Li that is sensitive to the details of the wavefunction of ${^{6}}$He. The case of ${^{8}}$He is also important as its size is somewhat smaller than that of $^6$He due to the pressure of outer neutrons.

Much less is known concerning the alpha clustering in heavier nuclei. At the Budker Institute of Nuclear Physics (Novosibirsk), more than 30 years ago, an experiment with a super-thin nuclear jet target crossing the electron beam in a storage ring \cite{Dmitriev1987} 
at electron energy $E_e = 130$\,MeV demonstrated a significant cluster component in the double magic $^{16}$O nucleus. There were observed various cluster-decay channels, with the final states in ${^{12}}$C, ${^{8}}$Be and four alpha particles. 
These promising experiments were discontinued and this storage ring does not exist anymore.
Also the electron-ion scattering experiment ELISe, aiming at reaction studies on stored short-lived nuclides, is significantly delayed at FAIR \cite{Antonov-2011}.

A series of experiments
on alpha clustering in heavier nuclei were performed at Jyv\"askyl\"a with alpha scattering from heavy nuclei \cite{Norrby2011_alpha_clustering}.
This reference is the latest detailed publication on the experiment with ${^{36}}$Ar$+\alpha$ at 150\,MeV energy. There is a systematic accumulation of collective states, like cluster of cluster states with the same spin going from $^{16}$O all the way to $^{40}$Ca. In the shell-model language, such states are regular combinations of many selected states from higher shells. The only currently available picture is that of a quasicrystal structure where the analogs of crystal bands, probably overlapping, are formed built on alpha constituents. The bands are split by alpha tunneling as electron bands in normal crystals. 

The study with gamma rays of high energy and therefore appropriate wavelength, a distant relative of X-ray diffraction from crystals, can shed light on this structure that has no analogs in other nuclear phenomena. 
The $(\gamma,\alpha)$ and possibly also $(\gamma, 2\alpha)$ reactions \cite{Smith2020_gamma_alpha}, studied as a function of photon energy and polarization can provide unique information concerning specific features of alpha-clusterization, especially in even-even $N=Z$ nuclei \cite{Sen'kov2011}.

The development of nuclear physics away from the stability band brought to the experimental light such exotic nuclei as those in vicinity of $^{100}$Sn, the heaviest double magic nucleus with $N = Z$. Here, there are indications of alpha structure continuing up to such heavy systems. This should be a regular part of the equilibrium nuclear structure in distinction to alpha decay of superheavy elements. There are various arguments in favor of the existence of such a substructure including the quadrupole collective states in heavier tin isotopes, in spite of the absence of valence protons in this shell according to the primitive shell-model scheme. Especially characteristic are extremely strong alpha decays of nuclei near ${^{100}}$Sn \cite{Auranen2018_100Sn}. The decay of ${^{105}}$Te to ${^{101}}$Sn is one of the strongest alpha decays in the whole periodic chart, with a half-life of only 620\,ns \cite{Liddick2006}. This is comparable to the 229\,ns half-life of ${^{212}}$Po which is just a bound state of the alpha particle and the core of ${^{208}}$Pb. The strongest known $\alpha$ decay of ${^{104}}$Te to ${^{100}}$Sn has an upper limit half-life of only 18\,ns \cite{Auranen2018_100Sn}. A detailed study of alpha structures in heavy nuclei would also be a driving force for the development of corresponding theory (some directions are outlined in Ref.\,\cite{Sen'kov2011}
as the nuclear analog of the transition between the Bardeen–-Cooper-–Schrieffer (BCS) picture of superconductivity and boson condensation).

\subsubsection{Fano effect in nuclear gamma spectroscopy}
\label{susubsec:Fano}

The Fano effect \cite{Fano1961}, well known in atomic and molecular physics, is a characteristic asymmetric spectral lineshape that arises due to the interference of the resonant and nonresonant transition amplitudes in the vicinity of a resonance.
In nuclear physics, the Fano effect was studied, for example, in $^{15}$N($^{7}$Li,$^{7}$Be)$^{15}$C \cite{Orrigo2006} and d($^{9}$Li,$^{10}$Li)p \cite{Orrigo2009} reactions.  However,  to the best of our knowledge, the Fano effect has not as yet been observed  in nuclear gamma transitions. A loosely related experiment \cite{Heeg2015}  on X-ray interference with $^{57}$Fe M\"{o}ssbauer nuclei observed  Fano interference in the X-ray reflectivity of thin-film cavities around the nuclear resonance energy. However,   in that case the nonresonant channel was given by the electronic scattering of the cavity, while only the resonant channel was stemming from the driving of the nuclear M\"ossbauer transition. 

A study of nuclear Fano resonances via photoexcitation was proposed in the context of three-body Efimov states \cite{Efimov1970}. Photoabsorption on the ground state of $^{20}$C ($\tau_{1/2}\approx16$\,ms) should be sensitive to the presence of the $n+^{19}$C and $n+n+^{18}$C states embedded in the $^{20}$C continuum and display Fano resonances around the respective energies  \cite{Mazumdar2006}. 

With tunable narrow-band gamma rays from the GF, one can envision a systematic study of Fano interference in the vicinity of nuclear gamma transitions, for example the ones in $^{13}$C discussed in Appendix\,\ref{Subsec:Appendix:Gamma_Res_13C}.

\subsection{Photofission with monochromatic gamma beams}
\label{Subsec:Photofission}

\begin{figure}[!hpb]\centering
     \includegraphics[width=1\linewidth]{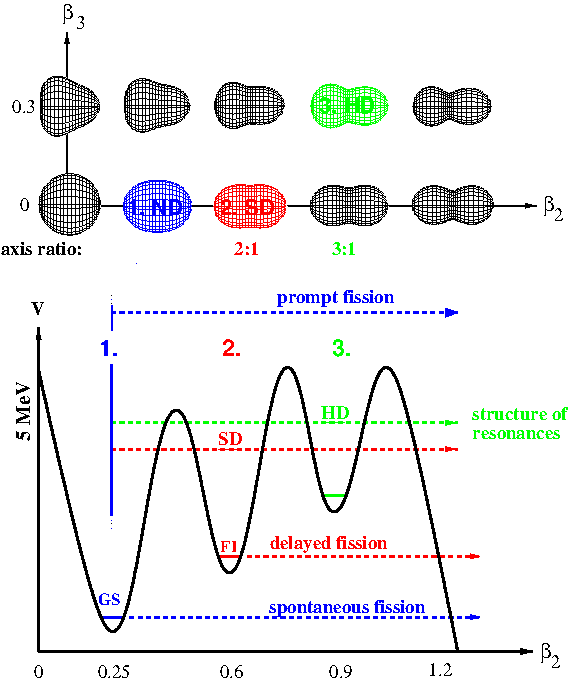}
     \caption{Schematic overview of the multiple-humped fission barrier in light actinide isotopes
              together with the corresponding nuclear shapes. Lower part: cut through the potential
              energy surface along the fission path, revealing - besides the normal deformed (ND)
              first minimum with the nuclear ground state (GS) - a superdeformed (SD) second minimum
              at an axis ratio of 2:1 (with a fission isomer (FI) as its ground state) and a
              hyperdeformed (HD) third minimum at an axis ratio of 3:1. In the upper part the
              corresponding nuclear shapes are displayed as a function of the quadrupole and
              octupole degrees of freedom. Figure adapted from \cite{Thirolf2002}.}
     \label{fig:multiple-humped-PES}
\end{figure}

Photofission measurements enable selective studies of extremely deformed nuclear states in light
actinides and can be utilized to better understand the landscape of the multiple-humped potential
energy surface  in these nuclei. The selectivity of these measurements originates from the
low amount of angular momentum transferred during the photoabsorption process.
Studies of fission isomerism in the actinide mass region provided evidence already in the 1960s
for the existence of a second, minimum of the potential energy
surface \cite{Polikanov1962} corresponding to superdeformation (SD), with a ratio of 2:1 between the long and short principal
axes of the strongly deformed nuclear shape, parameterized as a rotational ellipsoid. This corresponds to a value of $\beta_2\sim 0.7$ of the quadrupole-deformation parameter in the nuclear surface
parameterization in terms of a spherical harmonic multipole expansion. Three decades later also the
existence of a hyperdeformed (HD) third minimum of the fission barrier (axis ratio 3:1,
$\beta_2\approx$ 0.9, octupole deformation $\beta_3\approx$ 0.3) could be established (see
review articles \cite{Bjornholm1980,Thirolf2002}). The lower part of
Fig.\,\ref{fig:multiple-humped-PES} illustrates such a triple-humped fission barrier (expressed
by the potential energy as a function of the nuclear quadrupole deformation $\beta_2$).

\begin{figure*}[!hpb]\centering
     \includegraphics[width=1.0\textwidth]{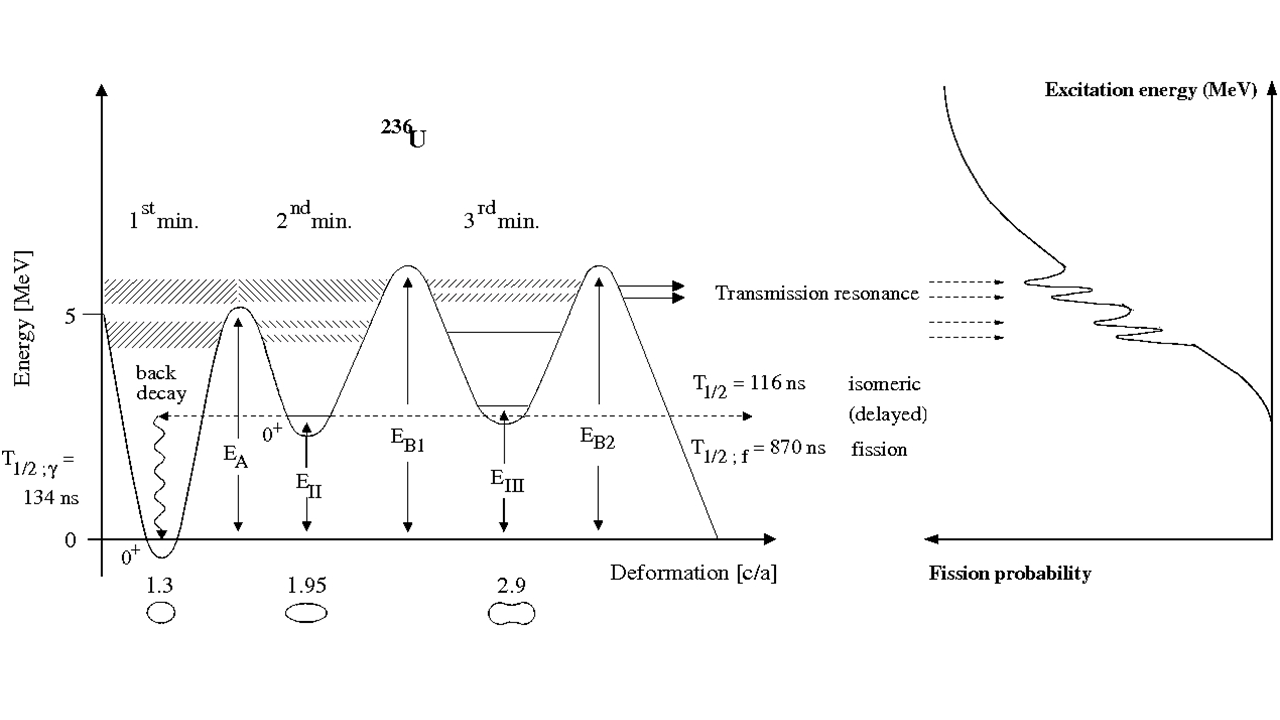}
     \caption{Left part (adapted from \cite{Csatlos2005}): The triple-humped potential-energy surface of $^{236}$U. Also damped compound nuclear states in the normal deformed (ND) first, superdeformed (SD) second and hyperdeformed (HD) third potential minima are shown as hatched areas.
              For strongly mixed ND and SD states, transmission resonances of HD states can occur as visible in the prompt fission probability shown in the right part.}
     \label{fig:U236-3humped-fissbar}
   \end{figure*}

Spectroscopic information on the properties of these extremely deformed nuclear states was obtained
for the superdeformed second minimum via direct decay studies (conversion electron and $\gamma$
spectroscopy) and isomeric (delayed) fission, while for the third minimum, transmission-resonance
spectroscopy was performed, analyzing resonances in the prompt fission probability.
Figure\,\ref{fig:U236-3humped-fissbar} displays the fission barrier of $^{236}$U (left)
and, in the right part, the related prompt fission cross section with transmission
resonances \cite{Csatlos2005}.

\begin{figure*}[!hpb]\centering
    \includegraphics[width={\textwidth}]{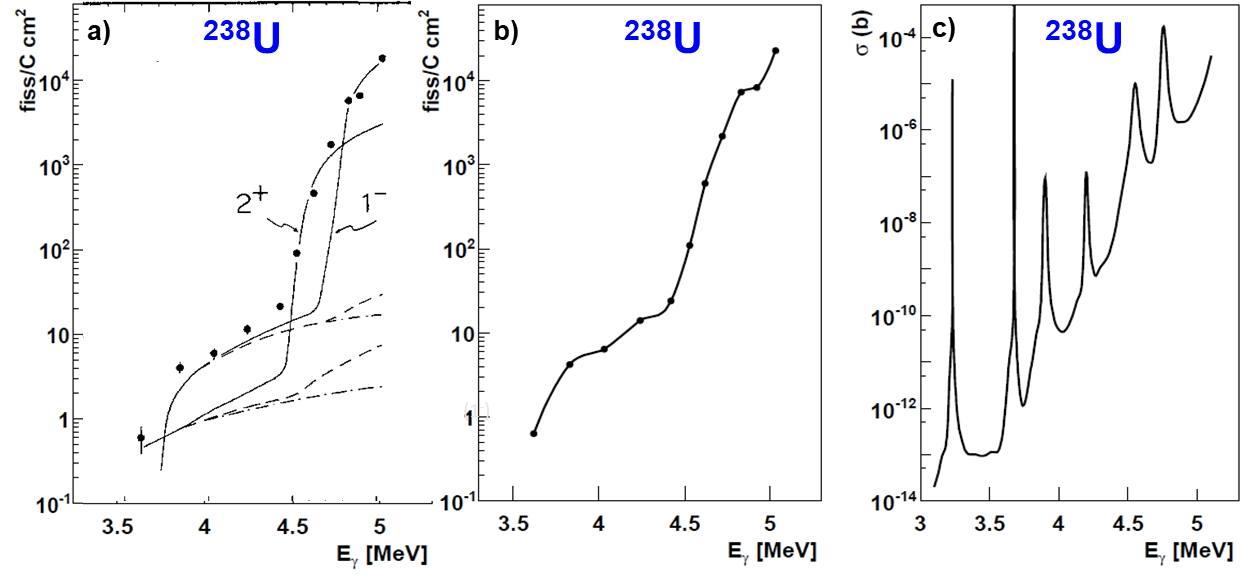}
     \caption{a) Photofission yield for $^{238}$U as a function of the excitation energy:
            experimental data (full symbols) and 2$^+$ and 1$^-$ contributions from model calculations
            (labelled solid lines) \cite{Bellia1983}. b) Photofission yield data from a)
            (solid line to guide the eye) as accessible with bremsstrahlung photons of an
            effective bandwidth $\Delta E\sim$\,300\,keV. c) Expected photofission yield of $^{238}$U
            when using a $\gamma$ beam of $\Delta E/E\sim$ 10$^{-6}$, based on resonances tentatively
            reported in an early photofission experiment with limited resolution \cite{Zhuchko1977}. Figure adapted from \cite{Bellia1983}.}
     \label{fig:isomeric-shelf}
\end{figure*}

While direct decay spectroscopy is limited to the second minimum (due to the thin outer
barrier of the third well and thus too short decay lifetimes compared to fission) and to
an excitation-energy range up to about 1.5\,MeV above the SD ground state, transmission
resonance spectroscopy in the third minimum probes the energy region of about 1-1.5\,MeV below
the barrier top. Thus the intermediate range of excitation energies between 3.5\,MeV and 4.5\,MeV
(representing the region of the second and third vibrational phonon) remained so far inaccessible
to high-resolution studies that could have provided deeper insight, for example, into the harmonicity
of the nuclear potential at these extreme deformations. Instead, early photofission studies
in actinides performed with bremsstrahlung photons with an effective energy bandwidth of about
$\Delta E =$200-300\,keV revealed the existence of a so-called `isomeric shelf' in the photofission
cross section of $^{238}$U in this energy region, as visible below the sharp bend at ca. 4.5\,MeV in
Fig.\,\ref{fig:isomeric-shelf}a) \cite{Bellia1983}.
This isomeric shelf is interpreted as the result of a competition of prompt and delayed (isomeric)
fission, following the $\gamma$ decay to the isomeric ground state. Due to the high selectivity
of the ($\gamma$,f) reaction in terms of spin and parity, only 1$^-$ and 2$^+$ states are formed
following the absorption of $E1$ and $E2$ multipolarity, respectively. Therefore\YAL{,}
Fig.\,\ref{fig:isomeric-shelf}a shows an analysis of the experimental data in terms of quadrupole
and dipole contributions, exhibiting a dominant 1$^-$ component for the isomeric shelf.

While the photon bandwidth achievable with bremsstrahlung limits the spectral resolution
as shown for the existing data in Fig.\,\ref{fig:isomeric-shelf}b, using narrow-bandwidth photons
of the monochromatized beam from the GF could allow resolving a
resonance structure underlying the isomeric shelf, as illustrated in Fig.\,\ref{fig:isomeric-shelf}c.
The resonance positions and strengths are based on tentative findings of an early photofission
experiment with limited resolution \cite{Zhuchko1977}, that could conclusively be performed with
the envisaged high-resolution photon beam of the GF.

\begin{figure}[!hpb]\centering
     \includegraphics[width=1\linewidth]{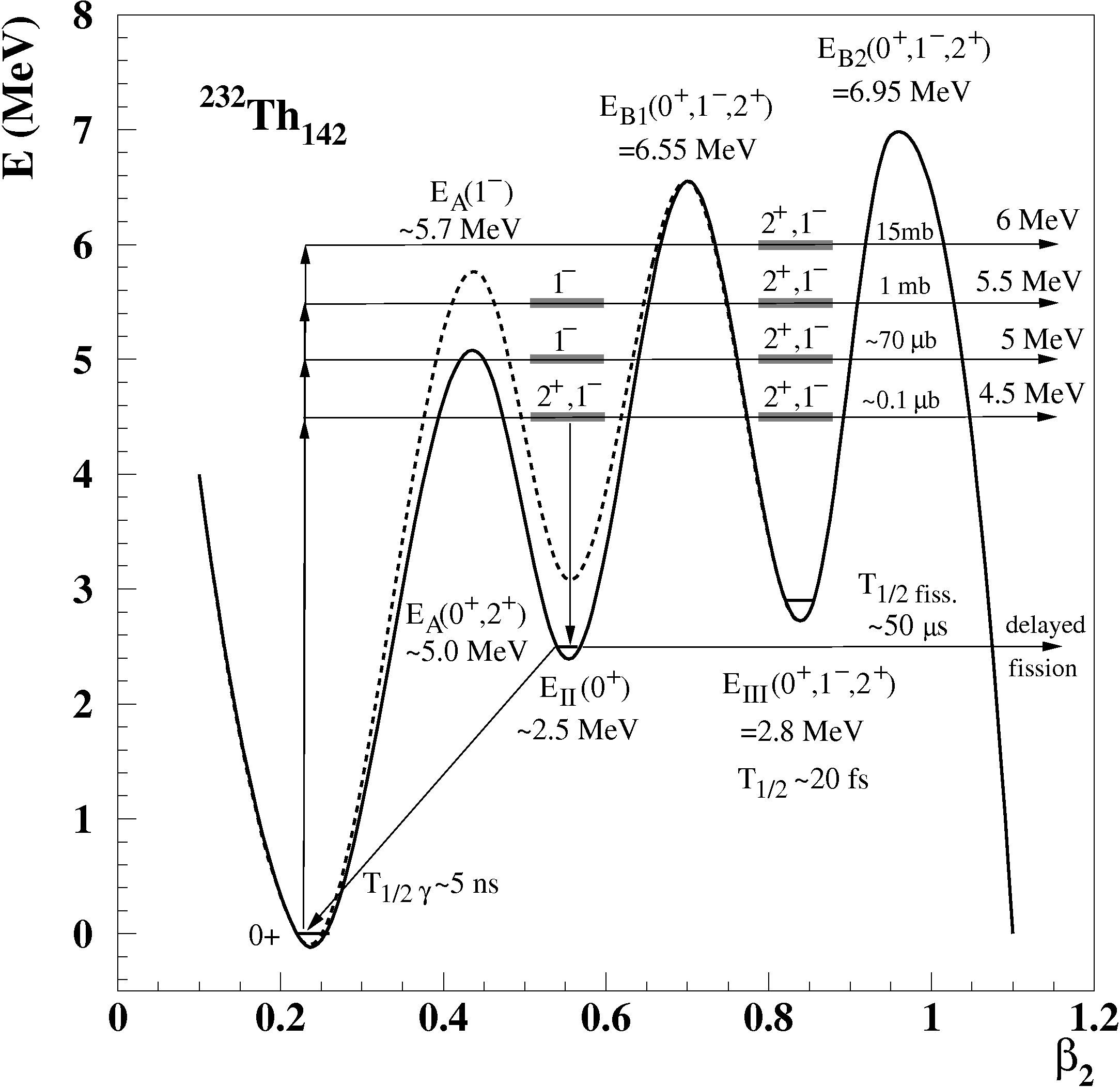}
     \includegraphics[width=1\linewidth]{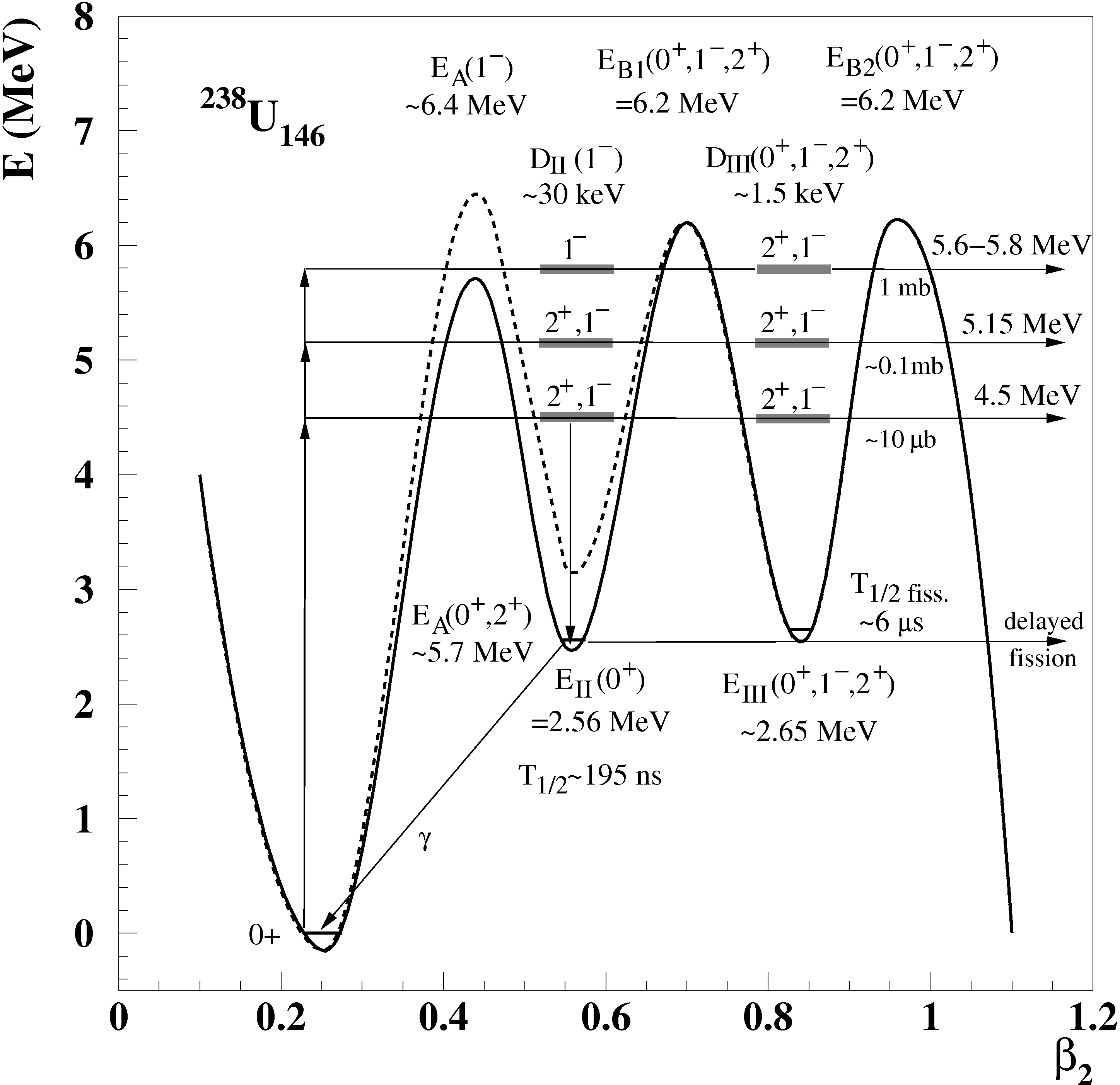}
     \caption{Triple-humped potential energy landscape for $^{238}$U (top) and $^{232}$Th (bottom) as
              a function of the quadrupole deformation $\beta_2$. The solid lines show a parameterization
              in the harmonic model, where potential barriers and minima are described by joint parabolas.
              The dashed curves represent the spin-dependent barriers for dipole (1$^-$) and quadrupole
              (2$^+$) excitations, differing at the inner but degenerate at the middle
              barrier \cite{Vandenbosch1973}. Expected transmission resonances in various potential
              wells are indicated with their energies, spins and cross sections. For $^{232}$Th
              extrapolated partial lifetimes for delayed fission and ($\gamma$) back decay are also
              indicated.}
     \label{fig:232Th-236U-barrier-expectation}
\end{figure}

Candidates for photofission studies at the GF are illustrated in
Fig.\,\ref{fig:232Th-236U-barrier-expectation}.
For $^{232}$Th and $^{238}$U these plots show an expected triple-humped fission barrier
structure as a function of the quadrupole deformation $\beta_2$.
For a long time only a shallow third potential minimum was assumed to exist, until it could be
demonstrated for $^{234}$U~\cite{Krasznahorkay1999}, $^{236}$U~\cite{Csatlos2005} and
$^{232}$U~\cite{Csige2009} that the outer third minimum is in fact as deep as the second minimum.
This also leads to a new interpretation of early photofission data, where a fine structure in the
6.0\,MeV resonance of $^{232}$Th was measured by Zhang et al. \cite{Zhang1984,Zhang1986}.
In view of our improved knowledge on the triple-humped barrier we have to conclude that in these
measurements in fact the depth of the third well instead of the second minimum was determined.
For $^{238}$U also the potential landscape based on a triple-humped fission barrier is drawn.
For a long time, photofission in $^{238}$U was only studied via measurements using bremsstrahlung,
where resonances could not be resolved, and with a weak tagged-photon beam of 100\,keV bandwidth
resulting in low-statistics resonances reported by Dickey and Axel \cite{Dickey1975}.
The early $^{236}$U(t,pf) (pf here indicates that a proton is emitted before the residual nucleus undergoes fission) measurements by Back et al. \cite{Back1974} show several pronounced transmission
resonances between 5 and 6\,MeV, while a whole sequence of further (yet unresolved) transmission
resonances at lower energies is expected to explain the isomeric shelf \cite{Bellia1983}.

Using a brilliant photon beam for photofission studies on strongly deformed actinides will provide
many advantages compared to the experimental approach pursued over decades and lead to a renaissance
of photonuclear physics in general and photofission in particular. So far the potential
energy surface of actinides was mapped via particle induced reactions like (d,pf), (d,tf)
or ($^3$He, df), resulting in a statistical population of states \cite{HauserFeshbach1952} in the
second and third potential well. In these cases, the typical population probability amounts to
$10^{-4} - 10^{-5}$, equivalent in typical experimental scenarios to about one isomeric fission event
per second. Therefore only strong resonances with resonance strengths $\sigma\Gamma\sim$\,10\,eVb
could be studied, in the vicinity of a strong background from prompt fission.
When using a monochromatic $\gamma$ beam with a spectral density of about $10^4 - 10^6 \gamma$/eV/s,
the spin selectivity of the photonuclear reaction together with the narrow energy bandwidth of
$< 10^{-3}$ would result in a strongly increased isomeric fission rate of $10^2 - 10^6$ isomeric fission
events per second. Moreover, due to the almost complete absence of background from prompt
fission, clean spectra would emerge, granting access also to weak resonances with
$\sigma\Gamma\sim$~0.1\,eVb and widths from about a few 100\,eV to 100\,keV.

Using a monochromatized $\gamma$ beam with $\Delta E/E\approx10^{-6}$ would provide a $\gamma$-beam
linewidth approaching the thermal Doppler broadening limit of about 5\,eV at E$_{\gamma}=$5\,MeV for
heavy nuclei of $A\approx$240 at 300\,K, thus enabling a new era of photofission studies with
the GF: exploring the fission barrier landscape of actinides would enable tests of
macroscopic-microscopic nuclear models with unprecedented sensitivity, allowing, for example, to improve
the input to astrophysical nucleosynthesis reaction-network calculations. High-resolution $\gamma$ spectroscopy of nuclear
configurations at large deformations could be used to study Nilsson orbitals
with energies decreasing as a function of
nuclear deformation.
In the Nilsson-orbital framework, such studies will provide information on the structure
of (less deformed) superheavy elements. Dynamical aspects of the fission process could be revisited
with much improved sensitivity to angular, mass and charge distributions. Finally, also applications
of the fission process would benefit from such high-quality photofission studies, for example, from the understanding of
strong $E1$ resonances as doorway states to fission in order to better understand and optimize the
mechanisms of minor actinide transmutation in nuclear waste treatment (see also Sec.\ref{Subec:Nuclear_Waste}).

Another application of nuclear photofission, considered also for the ELI-NP facility \cite{Kaur2019_Photofission}, is measurement of magnetic moments of isomeric states of neutron-rich nuclei. 

\subsection{Odd harmonics in angular distribution of fission fragments} 

Odd harmonics in angular distribution of fission fragments appear due to interference of the fission amplitudes from nuclear states of opposite parity. They may appear due to mixing of opposite-parity nuclear states by the weak interaction or, in the case of neutron-induced fission, due to population of opposite-parity states by capture of neutrons in s- and p-waves. In the case of photofission, opposite parity states may be populated by the capture of $E1$ and $M1$ or $E2$ photons. Observation of odd harmonics requires separation of light and heavy fission fragments ($A_{fragment}< A/2$ and $A_{fragment}> A/2$).

For many years it was believed that observation of odd harmonics is impossible because of the large number of final states of the fragments (fragments are formed in excited internal states) so that any interference effect should average out, see for example, the classic nuclear physics book by  A.\,Bohr and B.\,Mottelson \cite{BohrMottelson}. However, pioneering work by the Danilyan group and others \cite{Danilyan1977,Andreev1978,Vesna1980,Petukhov1980,Sushkov_1982} discovered that the parity violating correlation ${\vec{\sigma}\cdot \vec{p}_l}$ is not suppressed in the neutron-induced fission (here $\vec{\sigma}$ is the neutron spin and $\vec{p}_l$ is the light-fragment momentum), it is actually about $10^{-4}$, i.e. it is enhanced by three orders of magnitude relative to the ratio $10^{-7}$ of the weak to strong interactions in nuclei.

The theory of this phenomenon was developed in Refs.\,\cite{Sushkov_1982,Sushkov1980,Sushkov1981}. Interference between opposite-parity amplitudes is not suppressed since the orientation of the strongly deformed  nucleus is produced before the separation of the fragments due to mixing of the doublet of the opposite-parity rotational states.  The enhancement of PV happens at the initial stage due to mixing of very close opposite-parity states in the spectrum of compound resonances which are formed after neutron capture. Earlier it had been assumed that this mixing is ``forgotten'' during the  complicated fission process and does not appear at the final fission stage. Indeed, it is the case where temporal description of the fission process is applicable, with different stages of fission separated in time. However, there is an uncertainty relation between the energy resolution and time resolution, $\delta E \delta t \gtrsim \hbar$.  To temporally separate different stages of fission, we need a large energy spread $\delta E$. For the case of nearly monochromatic thermal neutrons ($\delta E \sim T \ll D$ where $T$ is neutron temperature and $D$ is the distance between the compound resonances), we have all components of the nuclear wavefunction present simultaneously, the time dependence  is given by a common factor $\exp( -i E t/\hbar)$.   As a result, the mixed-parity compound state contains also the mixed rotation doublet in the strongly deformed nucleus before the separation of the fragments. The details of the corresponding calculations may be found in Refs.\,\cite{Sushkov_1982,Sushkov1981}.    

Odd harmonics in fission may also appear due to the interference of $s$- and $p$-wave neutron capture resulting in the correlations  ${\vec{p}_n \cdot \vec{p}_l}$ and ${\vec{\sigma} \cdot (\vec{p}_n \times  \vec{p}_l)}$, where $\vec{p}_n$ is the neutron momentum \cite{Sushkov_1982}. The correlation  ${\vec{\sigma} \cdot (\vec{p}_n \times  \vec{p}_l)}$ has been observed with the predicted magnitude.

The GF allows us to investigate both parity violating and parity conserving odd harmonics in photofission. The corresponding theory is presented in Ref.\,\cite{Flambaum1985odd}. The variable spread of the photon energies will allow investigation of the transition from small $\delta E$ regime, where odd harmonics are not suppressed, to the ``lost-memory'' regime where there is a suppression by a factor of $\sqrt{\delta E/D}$.  The explanation of this suppression factor is simple: the number of compound states captured  in the interval $\delta E$ is $N \approx \delta E/D$, the  amplitudes of the fission from different  compound states have random signs, so the effect is suppressed by $\sqrt{N}$.

\subsection{Photoinduced processes on the proton and light nuclei}
\label{Subsec:Photopys_protons_light_nucl}

\begin{figure}[ht] 
  \centering 
\begin{minipage}[t]{0.49\textwidth}
    \centering 
       \raisebox{0.00cm}{\includegraphics[width=0.9\textwidth]{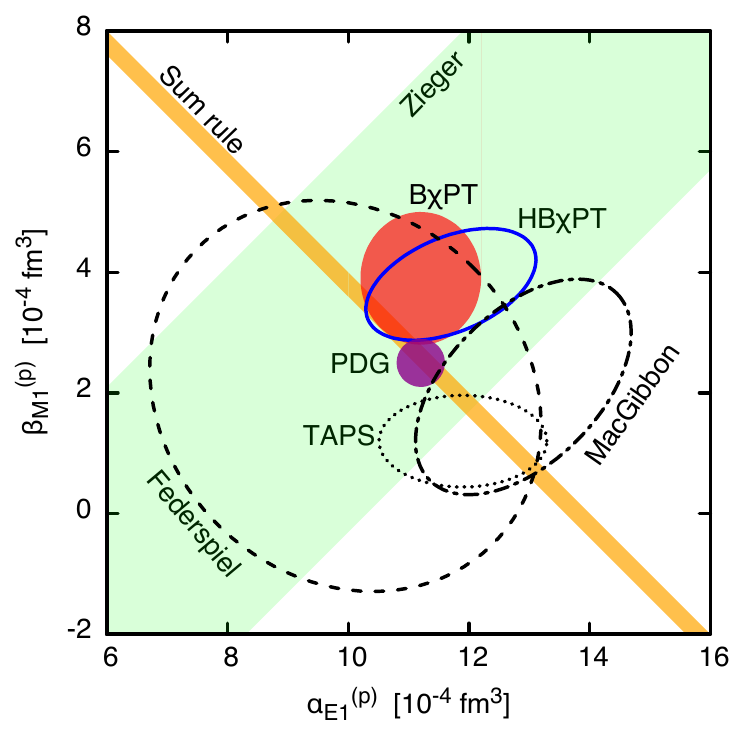}}
\end{minipage}
\begin{minipage}[t]{0.49\textwidth}
    \centering 
       \raisebox{0.1cm}{\includegraphics[width=0.9\textwidth]{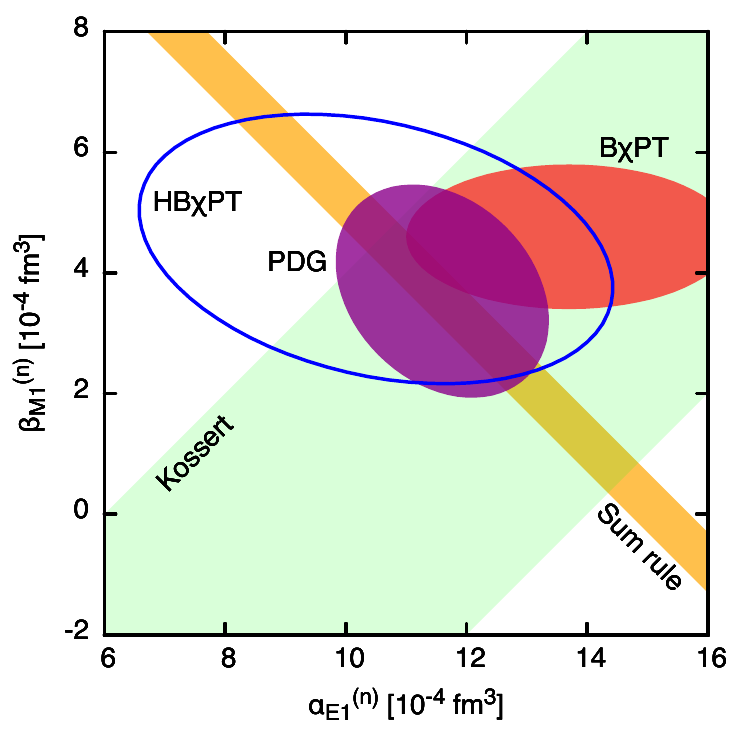}}
       \end{minipage}
 \caption{Plot of $\alpha_{E1}$ versus $\beta_{M1}$ for the proton (upper panel) and neutron (lower panel).   The orange band represents the latest Baldin sum-rule evaluation \cite{Gryniuk:2015aa}. `PDG' denotes the latest Particle Data Group summary. Other references can be found in Ref.\,\cite{Hagelstein201629}.\label{fig:alphaVSbeta}}
\end{figure}

Low-energy Compton scattering off the proton p$(\gamma,\gamma')$p and light nuclei (viz. deuteron or helium), is traditionally used to 
measure the nucleon polarizabilities, see the recent reviews \cite{Schumacher:2005an, Griesshammer:2012we, Hagelstein201629, Hagelstein2020}. The present status of the dipole electric ($\alpha_{E1}$) and magnetic ($\beta_{M1}$) polarizabilities of the proton and neutron is shown in Fig.\,\ref{fig:alphaVSbeta}. 
The GF could improve upon the latest measurements at the operating photon-beam facilities (HI$\gamma$S and MAMI) given its broader energy range, high energy resolution and photon flux. Nucleon polarizabilities enter as input to precision atomic physics, most notably, spectroscopy of muonic atoms
\cite{Pohl:2010zza}. Better precision of the polarizability measurements would reduce the current
uncertainties in atomic calculations \cite{Birse:2012eb,Alarcon:2013cba,Carlson:2015jba}.

Compton scattering above the pion-production threshold  provides better sensitivity to  polarizabilities compared to the case of lower energies. In addition, a strong correlation between the values of nucleon polarizabilities and the properties of nucleon excited states, in particular, the ratio of the electric quadrupole to magnetic dipole $E2/M1$ strengths at the $\Delta (1232)$ resonance has been observed~\cite{Schumacher:2005an}. The delta resonance \YAL{topic} is discussed further in Sec.\,\ref{subsec:Delta}.

The optical theorem, in conjunction with other general principles such as micro-causality, fully determines the amplitude of forward Compton scattering in terms of integrals of photoabsorption cross sections \cite{GellMann:1954db}. 
For example, the spin-independent  forward Compton scattering amplitude, $f(\nu)$, as function of the photon energy $\nu$, is given  in terms of total photoabsorption cross section, $\sigma(\nu)$, as:
\begin{eqnarray}
f(\nu) &=& - \frac{(Ze)^2}{M} + \frac{2\nu^2 }{\pi} \int_{\nu_\pi}^\infty d \nu' \frac{\sigma(\nu')}{\nu^{\prime \, 2}-\nu^2}, 
\end{eqnarray}
where the first term is the classical Thomson amplitude characterized by the ratio of the target charge and mass, and the lower integration limit reads $\nu_\pi=m_\pi(1+m_\pi/2M)$, with $m_\pi$ the pion mass. Expanding this expression in powers of $\nu/m_\pi$ yields a number of useful sum rules; first of all, the Baldin sum rule for the sum of dipole polarizabilities \cite{Baldin:1960}:
\begin{eqnarray}
\alpha_{E1} + \beta_{M1} &=&  \frac{1}{2\pi^2} \int_0^\infty d \nu \frac{\sigma(\nu)}{\nu^2}.
\end{eqnarray}
Our current knowledge of the empirical cross sections, by means of this sum rule, provides stringent constraints
on the nucleon polarizabilities, as shown by the yellow band in Fig.\,\ref{fig:alphaVSbeta}. 

Similar considerations for spin-dependent Compton scattering lead to  the Gerasimov-Drell-Hearn (GDH) sum rule \cite{Gerasimov:1965et,Drell:1966jv} which relates the anomalous magnetic moment $\kappa$ of a particle of spin $S$ and mass $M$ to an energy-weighted integral over the spin-dependent photoabsorption cross section via,
\begin{equation}
    I^{\rm GDH}\equiv\int_0^\infty\frac{d\nu}{\nu}\left[\sigma^P(\nu)-\sigma^A(\nu)\right]=4\pi^2\kappa^2\frac{\alpha}{M^2}S,
\end{equation}
where $\sigma^P$ ($\sigma^A$) stands for the photoabsorption cross section with circularly polarized photons and the target polarization parallel (antiparallel) to the photon momentum, respectively. The GDH sum rule was shown to hold exactly for the electron, order by order in QED \cite{Altarelli:1972nc, Dicus:2000cd}. For the nucleon, a perturbative QCD calculation is not helpful, but the comparison has been done based on the measured helicity-dependent cross  section $\sigma^P-\sigma^A$~\cite{Helbing:2006zp,Gryniuk:2015aa,Gryniuk:2016gnm}. For the proton, the result for the r.h.s is $I^{\rm GDH}_p\simeq 204\,\mu$b (1\,b=10$^{-24}$\,cm$^2$), to be compared to the l.h.s. value $205(21)\,\mu$b based on the empirical cross section. Likewise, for the neutron $I^{\rm GDH}_p=233\,\mu$b to be compared to the l.h.s. value $225(\dots)(\dots)\,\mu$b.

For the case of the deuteron the empirical verification of the GDH sum rule is much more delicate. 
The deuteron anomalous magnetic moment is small, $\kappa_d\simeq -0.143$ 
(compared to proton $\kappa_p\approx1.79$ and neutron $\kappa_n\approx-1.91$), leading to a tiny sum-rule value, $I^{\rm GDH}_d=0.65\,\mu$b. This small sum rule value implies an almost complete cancellation of the contributions from the hadronic range $\sim I^{\rm GDH}_p+I^{\rm GDH}_n$ vs. the near-threshold photodisintegration $\gamma d\to p n$. At present, an account for all channels leads to $I^{\rm r.h.s.}_d=27.3\,\mu$b, deviating from the sum-rule expectations.
Recent experimental data at photon energies between the breakup threshold and 10\,MeV from HI$\gamma$S \cite{Ahmed:2008zza}, while compatible with the sum rule, feature significant a uncertainty. 

The spin asymmetry of deuteron photodisintegration \cite{Wojtsekhowski2001} as well as
pion photoproduction on the deuteron is of great interest not only in view of the GDH sum rule but also in view of a sensitive test of present day theoretical models, although we are not aware of any measurements of the beam-target spin asymmetry in the d($\gamma$,p)n reaction.

A polarized deuteron (spin-1) contains polarized proton and neutron. Absorption of a circularly-polarized photon is only possible in an antiparallel helicity configuration, whereas only scattering (a Compton process suppressed by one power of $\alpha$) is possible in the parallel configuration. The threshold for photodesintegration of the deuteron is 2.22\,MeV, the maximum is reached around 5\,MeV. A 100\% asymmetry is expected. Nuclear effects are not expected to play a significant role; in addition, they are calculable in chiral effective field theories (EFT) \cite{Schwamb2001}. 

In general, polarization observables offer sensitive tests of the theoretical understanding \cite{Nikolenko2017}. For example, the polarization of the outgoing neutron P${}_y$(n) in d($\gamma$,n)p
at low energies shows a discrepancy between the theory and existing data.

\subsection{Pion photoproduction}
\label{Subsec:Pion_photoproduction}
Several main directions in contemporary nuclear science are addressing problems related to astrophysics and investigating weakly bound nuclei at the limit of nuclear stability. Such nuclei usually have a neutron halo or other weakly bound external nucleons \cite{Thoennessen2016discovery_isotopes,Tanihata_1996_neutron_halo,Mukha2018_proton}. Loosely bound nuclei have been studied by electroinduced two-nucleon knockout \cite{Ryckebusch1997_two-nucleon}. Several new approaches  can be envisaged for the future GF. One approach is to use photoexcitation with variable gamma energy at wavelength comparable to the size of the exotic orbitals. Ideally, one would perform a series of ($\gamma$,NN) experiments on a sequence of isotopes from stable to exotic that could provide a ``photo image'' of the evolution of valence orbitals as a function of the mass number. Another related image can be derived from pion photoproduction with the registration of the pion and emitted nucleons on a series of such isotopes.
One can expect that pion production on such nuclei will have a component similar to this process on a free nucleon.

Investigation of ($\gamma,\pi$) reactions is one of the rare techniques to probe the wavefunction of loosely bound nucleons, for example by the difference of pion production from the halo and from normal strongly bound isotopes. The idea is to check how the $\pi$ production depends on the isospin asymmetry of the nucleus. The $(\gamma,\pi)$ cross section at several hundred MeV scales with the atomic weight $A$ approximately as $A^{\alpha}$  dependence with $\alpha \approx 2/3$ \cite{Arends1986pion0}, close to what one might expect for an incoherent surface effect.  
$\pi$ production on the proton and deuteron is well studied, and this reaction on a free neutron is known with a minimal model dependence, providing reference for measurements on other systems. The  targets of interest are the long chains of stable isotopes in Sn, Yb, Sm, and Dy, Table\,\ref{tab:Isotope_chains}. We note that the isotopic dependence of atomic PV was recently measured in the Yb isotopic chain \cite{Antypas2018NP}.
\begin{table*}[t]
    \centering
    \begin{tabular*}{\textwidth}{@{\extracolsep{\fill}} p{0.1cm} p{16cm}}
    \hline
    \hline
    $_Z$X  &Isotopes  A: abundance and/or $T_{1/2}$\\
    \hline \\[-0.2cm]
        $_{24}$Cr  &48: 21.6\,h, 50: 4.34\% $\geq1.3\times10^{18}$\,y, 51: 27.7\,d, 52: 83.8\%, 53: 9.50\%, 54: 2.36\%\\
        \hline
        $_{50}$Sn  &110: 4.11\,h, 112: 0.97\% $\leq1.3\times10^{21}$\,y, 113: 115\,d, 114: 0.66\%, 115: 0.34\%, 116: 14.5\%, 117: 7.68\%, 118: 24.2\%, 119: 8.59\%, 120: 32.6\%, 121: 27.0\,h, 122: 4.63\%, 123: 129\,d, 124: 5.79\% $\geq1.2\times10^{21}$\,y, 125: 9.64\,d, 126: $2.30\times10^5$\,y, 127: 2.10\,h\\
        \hline
        $_{62}$Sm  &142: 72.5\,m, 144: 3.07\%, 145: 340\,d, 146: 10.3$\times10^7$\,y, 147: 15.0\% $1.06\times10^{11}$\,y, 148: 11.2\% $7\times10^{15}$\,y, 149: 13.8\%, 150: 7.38\%, 151: 90\,y, 152: 26.8\%, 153: 46.3\,h, 154: 22.8\%, 156: 9.4\,h\\
        \hline
        $_{66}$Dy  &152:2.38\,h, 153: 6.4\,h, 154: $3.0\times10^6$\,y, 155: 9.9\,h, 156: 0.056\%, 157: 8.14\,h, 158: 0.095\%, 159: 144\,d, 160: 2.33\%, 161: 18.9\%, 162: 25.5\%, 163: 24.9\%, 164: 28.3\%, 165: 2.33\,h, 166: 81.6\,h\\
        \hline
        $_{70}$Yb  &164: 75.8\,m, 166: 56.7\,h, 168: 0.123\%, 169: 32.0\,d, 170: 2.98\%, 171: 14.1\%, 172: 21.7\%, 173: 16.1\%, 174: 32.0\%, 175: 4.18\,d, 176: 13.0\%, 177: 1.91\,h, 178: 74\,m\\

    \hline
    \hline
    \end{tabular*}
    \caption{Examples of isotope chains. Isotopes in the table have half-lives longer than 1\,hr \cite{NNDC}.}
    \label{tab:Isotope_chains}
\end{table*}

\subsubsection{Photoproduction of bound \texorpdfstring{$\pi^-$}{pi-}}
 



Secondary photons at the GF which have highly tunable energy and narrow width obtained by proper collimation provide a unique opportunity to realize resonance photoproduction of pionic atoms.
It is realized through the reaction $\gamma+ n \rightarrow p + \pi^-$ inside a nucleus, i.e.,
\begin{equation}
    \gamma + {}^A_{Z_i}\textrm{X} \rightarrow (^A_Z\textrm{X}' +\pi^-)_{nl},
\end{equation}
where ${}^A_{Z_i}$X and $^A_Z\textrm{X}'$ are the initial and final nucleus, respectively (both in their nuclear ground state \footnote{In principle, pionic atoms with final nuclei in excited states can also be resonantly produced with higher photon energies.}), $Z_i$ and $Z= Z_i+1$ are the corresponding atomic numbers,  $A$ is the number of nucleons which is the same for the initial and final nucleus, and $l$ is the orbital angular momentum of the bound pion. 

Estimates of total widths of bound $ns$ states in pionic atoms, 
the effective photoproduction cross section of the initial nucleus, dominant contributions to photon-attenuation background,
maximal photoproduction rate $p_{max}$ of pionic atoms with atomic numbers $Z$ up to 92, and suggested tuning of the photon-energy spread are presented in Ref.\,\cite{Flambaum2020_resonance}. These estimates indicate that $p_{max}$ can reach up to $\sim 10^{10}$ atoms per second, which offers orders-of-magnitude improvement compared with other production methods, e.g.,  producing pionic atoms by capturing free pions as in Ref.\,\cite{Hori2020_spectra_helium}, where $\approx10^5$ pionic helium atoms per second are produced at a 590\,MeV proton facility.

For the estimates presented in Ref.\,\cite{Flambaum2020_resonance}, cross sections for producing free pions with final nuclei in the ground state are needed. While there exist such experimental data for light nuclei \cite{Singham1981_pion_threshold_light_nuclei,Bernstein1976_pion_threshold_12C_12N,Bosted1979_pion_threshold_cross-section,DeCarlo1980_pion_threshold_14N_14O,Min1976_pion_threshold_11B_11C}, we are not aware of such data for heavy nuclei. There are studies of charged-pion photoproduction from heavy nuclei \cite{Sakamoto1989_51Cr_from_51V,Sakamoto1990_133Ba_from_133Cs,Blomqvist1978_197Hg_from_197Au} using bremsstrahlung-photon beams and radiochemical measurements, where the measured cross section corresponding to the final nucleus in the ground state actually include contributions of decays from excited states. To measure the cross section for producing free pions with final nuclei in the ground state, monochromaticity of the GF photons would be useful, especially for heavy nuclei where the energy difference between the nuclear ground state and the first excited state is typically smaller than in light nuclei. We can tune the energy of the narrow-band gamma rays at the GF precisely so that excited nuclear states cannot be produced. These cross section data would be invaluable for estimating the rates of bound-pion photoproduction discussed in this section.

We note that gamma rays at the GF can be utilized to study coherent photoproduction of $\pi^0$ \cite{Boffi1991_coherent_pion0,Krusche2002_pion0} near threshold including the cross section dependence on $A$, the mass number of the nucleus. An advantage is that, with narrow-band gamma rays tuned close to the threshold, contributions other than coherent photoproduction (i.e., incoherent production and breakup reactions \cite{Krusche2002_pion0}) are energetically suppressed. Coherent pion photoproduction is further discussed in the following sections.
 

\subsubsection{Neutron-skin measurements in coherent (\texorpdfstring{$\gamma,\pi^0$}{g,pi0}) reactions}
\label{sec:NskinPi0Prod}
The all-important nuclear equation of state (EOS) relates objects with orders-of-magnitude different sizes, describing the structure and stability of nuclear matter as well as the properties of neutron stars \cite{Thiel:2019tkm}. In particular, the authors of Ref.~\cite{Fattoyev:2020cws} used the EOS to likely rule out the neutron-star nature of an object with 2.6 solar masses observed via detecting gravitational waves \cite{Abbott:2020khf}. The recent LIGO observation of a neutron-star merger constrains the tidal neutron-star polarizability  \cite{Chatziioannou:2020pqz}. 

\begin{figure}[!hpt]\centering
    \includegraphics[width=0.45\textwidth]{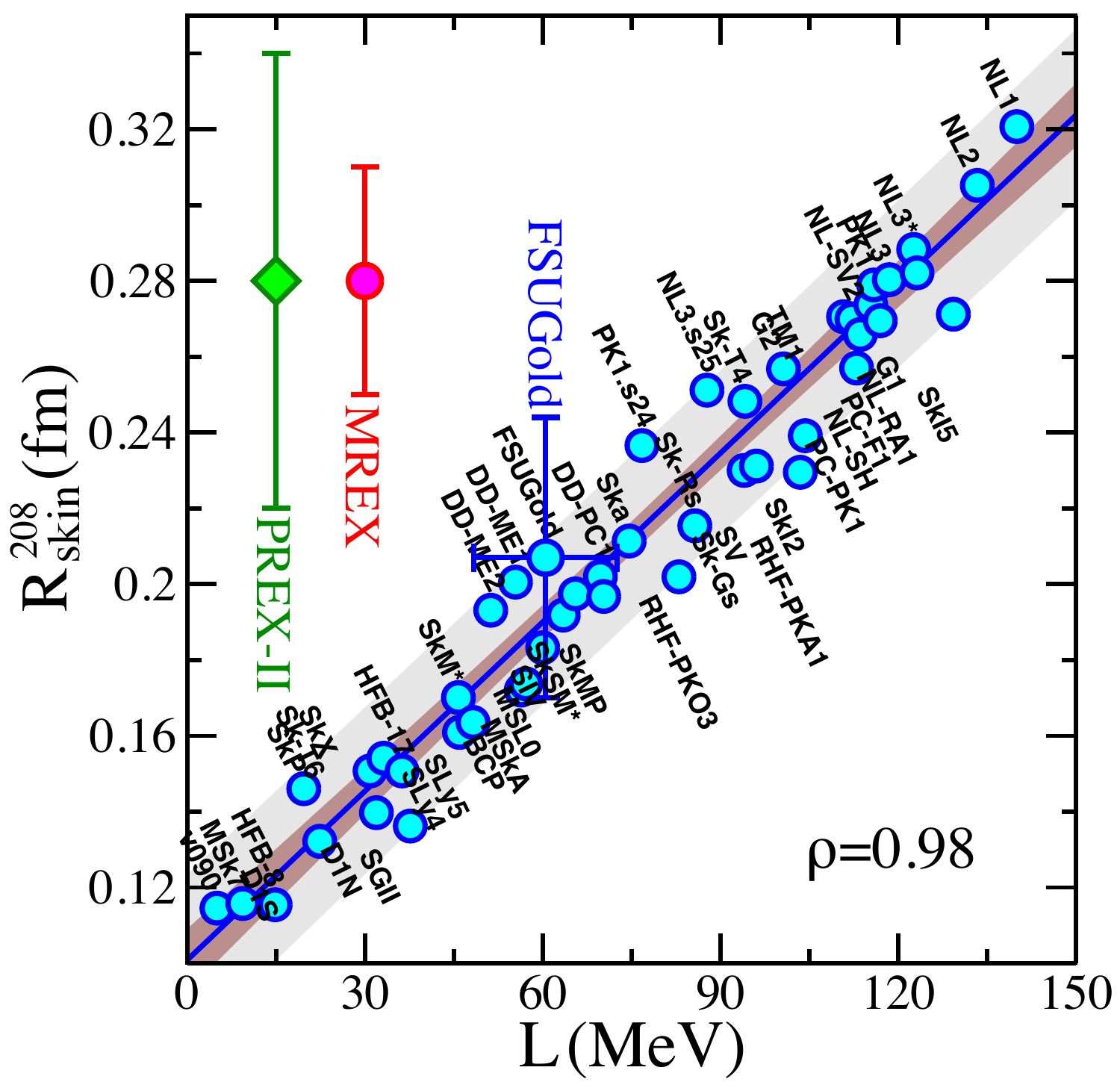}
    \caption{Correlation of the neutron skin of $^{208}$Pb with the slope $L$ of the density dependence of symmetry energy. Blue circles indicate predictions of various models. The 
    precision 
    of the experimental measurements of the neutron skin of $^{208}$Pb (but not the central values) is indicated by the error bars at the green diamond~\cite{Abrahamyan:2012gp,Adhikari:2021phr} and the red/magenta circle~\cite{Thiel:2019tkm,Becker:2018ggl}. The plot is adopted from Ref.~\cite{RocaMaza:2011pm}, and is copyrighted by the American Physical Society.}
    \label{Fig:SkinvsL}
\end{figure}

In turn, tidal polarizability crucially depends on the linear slope $L$ of the density dependence of the symmetry energy. The neutron skin $R_n-R_p$, the difference of the radii of the neutron and proton distributions within a nucleus, is also a sensitive probe of isovector interactions within nuclei and is strongly correlated with $L$ \cite{RocaMaza:2011pm}, as shown in Fig.~\ref{Fig:SkinvsL}.
While proton distributions are known with high precision across the nuclear chart~\cite{DeJager:1987qc}, information on neutron distributions is scarce. The idea of accessing neutron skins with parity-violating electron scattering on nuclei relies on the fact that the neutral weak boson $Z^0$ couples predominantly to the neutrons. 
Recent measurements at the Jefferson Lab~\cite{Abrahamyan:2012gp,Adhikari:2021phr} furnished precise information on the neutron skin of lead, leading to tighter constraints on the parameters of the EOS~\cite{Reed:2021nqk}. Further experiments to extract neutron skins from electron scattering are under analysis \cite{CREX,Kumar:2020ejz} or planned~\cite{Thiel:2019tkm,Becker:2018ggl}.

Additional ways to access neutron skins include measuring PV in atoms \cite{Viatkina2019NeutronSkins}, 
and coherent $\pi^0$ photoproduction that was used by the A2 collaboration at Mainz to investigate neutron skin of $^{208}$Pb using photons in the energy range of 
180 to 240\,MeV \cite{Tarbert:2013jze}. 

\begin{figure}[!hpt]\centering
    \includegraphics[width=0.35\textwidth]{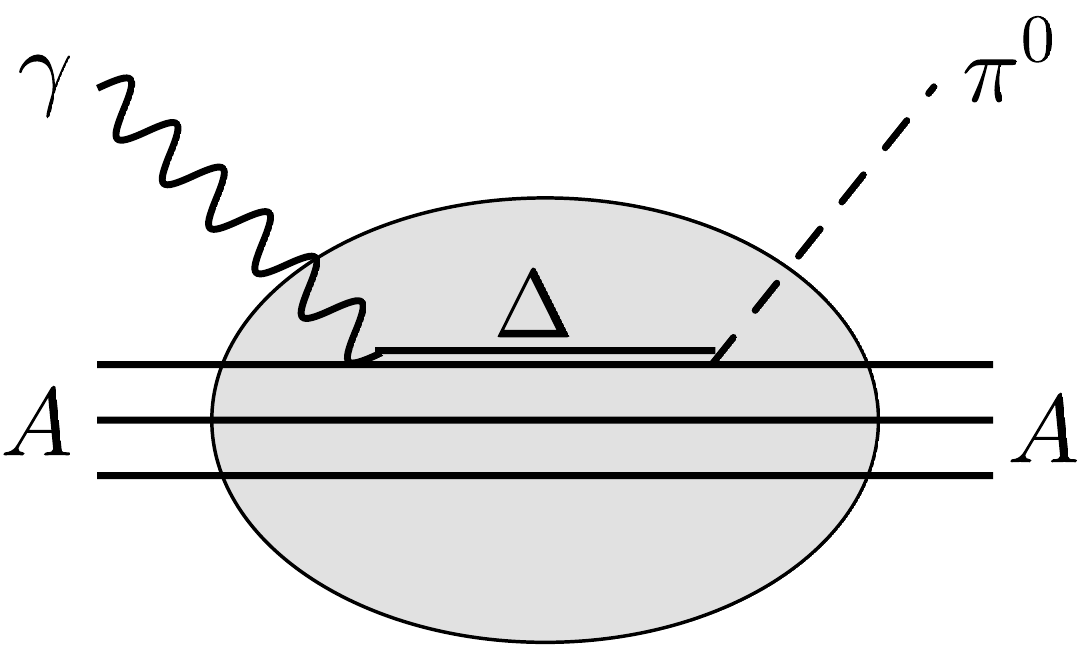}
    \caption{Coherent neutral pion photoproduction in the $\Delta(1232)$ region.}
    \label{Fig:CoherentPI0}
\end{figure}
The reaction mechanism is mainly the $\Delta(1232)$ resonance excitation, as shown in Fig.\,\ref{Fig:CoherentPI0}, by isospin symmetry the same on protons and neutrons, so that the cross section can be brought in correspondence with the baryon density $Z\rho_p(r)+N\rho_n(r)$. Using precise knowledge of nuclear charge (i.e. proton) densities~\cite{DeJager:1987qc}, one can then deduce the neutron density and the neutron skin $R_n-R_p$, using the definition of the proton and neutron radii,
\begin{equation}
    R_{n,p}^2=\int d^3 \vec r\, r^2\rho_{n,p}(r).
\end{equation}
The GF with its unique characteristics can allow one to measure coherent $\pi^0$ photoproduction on various nuclear targets with an unprecedented precision and over a wide range of photon energies and momentum transfers. In order to reconstruct a $\pi^0$ from the two decay photons one will need a $4\pi$-detector similar to the Crystal Ball used in the Mainz experiment~\cite{Tarbert:2013jze}. Importantly, the 100\% polarization of the photon beam in GF will allow to suppress $\pi^0$'s coming from incoherent processes which do not give access to the neutron skin. These incoherent contributions constitute a slowly-varying background that extends over a wide range of momentum transfer, limiting the applicability of the form factor fit beyond the second diffraction minimum. 
As a result, they contribute a significant part of the systematic uncertainty (see, for example, Ref.~\cite{Miller:2019btv} which discusses additional theory inputs that may affect the interpretation of the experimental data). A measurement of coherent $\pi^0$ photoproduction at GF thus offers a promising avenue for extracting accurate information about neutron skins alternative to parity-violating electron scattering and atomic parity violation. 



\subsubsection{Pion decay constant from the Primakoff effect}
\begin{figure}[!hpt]\centering
    \includegraphics[width=0.25\textwidth]{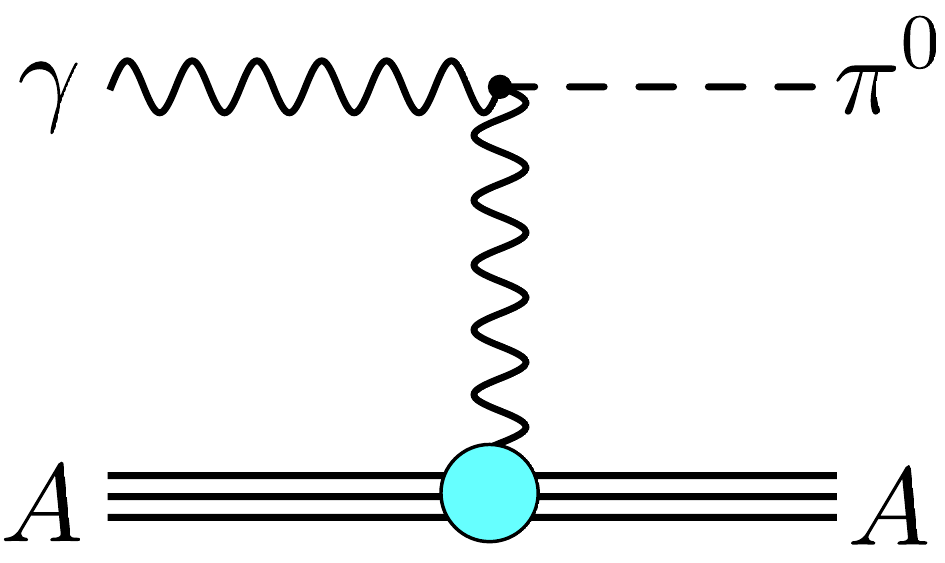}
    \caption{Neutral pion photoproduction via the Primakoff effect.}
    \label{Fig:Primakoff}
\end{figure}

The same process, 
coherent $\pi^0$ photoproduction, but measured at very forward angles,  provides a direct way to measure the radiative $\pi^0$ width $\Gamma_{\pi^0\gamma\gamma}$ via the Primakoff effect, see Fig.~\ref{Fig:Primakoff}.

\begin{figure}[!hpt]\centering
    \includegraphics[width=0.45\textwidth]{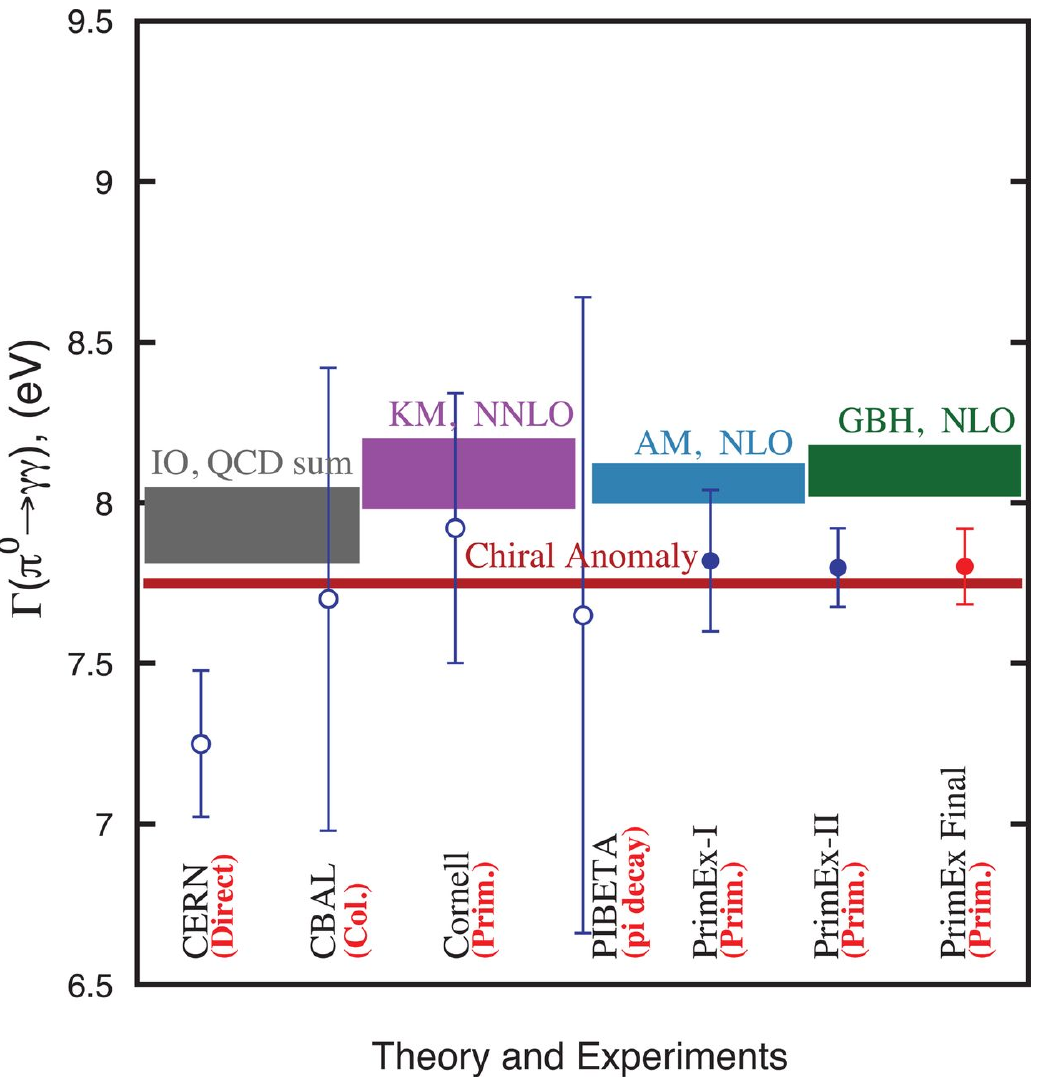}
    \caption{Theoretical predictions for the $\pi^0$ decay constant: Ref.~\cite{Bell:1969ts} (Chiral Anomaly), Ref.~\cite{Ioffe:2007eg} (IO QCD) Ref.~\cite{Kampf:2009tk} (KM NNLO) 
    Ref.~\cite{Ananthanarayan:2002kj} (AM NLO) and Ref.~\cite{Goity:2002nn} (GBH NLO)
    in comparison with results from earlier experiments \cite{Atherton:1985av,Williams:1988sg,Browman:1974cu,Bychkov:2008ws} along with the recent ones by the PrimEx collaboration \cite{Larin:2010kq,Larin:2020bhc}. The figure is adopted from Ref.~\cite{Larin:2020bhc}, and is copyrighted by Science Magazine.}
    \label{Fig:ChiralAnomalyExpTh}
\end{figure}

The differential cross section for this process is given by
\begin{equation}
    \frac{d\sigma^{\rm Prim}}{d\Omega}=\frac{8\alpha\Gamma_{\pi^0\gamma\gamma}}{m_\pi^3}Z^2F_{\rm Ch}^2(t)\frac{\nu^4\beta_\pi^2}{t^2}\sin^2\theta_\pi,\label{eq:Primakoff}
\end{equation}
with $m_\pi$ being the pion mass and $Z,\,F_{\rm Ch}$ the charge and charge form factor of the nucleus, respectively. Furthermore, $\nu$ is the photon energy, $\beta_\pi=\sqrt{1-m_\pi^2/\nu^2}$ the pion velocity, $t$ the momentum transfer to the nucleus and $\theta_\pi$ the angle the pion makes with respect to the incoming photon momentum. The most recent
PrimEx-II experiment at JLab  \cite{Larin:2020bhc} combined with older experimental determinations leads to $\Gamma(\pi^0\to\gamma\gamma)=7.806(052)(105)$\,eV, with the first and the second uncertainty referring to the  statistical and systematic one, respectively. This 1.5\% determination improved over its predecessor PrimEx-I \cite{Larin:2010kq} by almost a factor of 2. 
Historically, the chiral anomaly $\pi^0\to\gamma\gamma$ has served as an important milestone for establishing the chiral perturbation theory (ChPT), the low-energy effective theory of quantum chromodynamics (QCD), and has been considered the perfect place for comparing ChPT predictions to experimental results ever since. Currently, the most precise next-leading order (NLO) ChPT prediction for the $\pi^0$ radiative width \cite{Goity:2002nn,Ananthanarayan:2002kj} reads $\Gamma(\pi^0\to\gamma\gamma)=8.10(8)$\,eV with a 1\% uncertainty, showing a mild $2\sigma$ tension with the experimental result, see Fig.\,\ref{Fig:ChiralAnomalyExpTh}.

The main challenge in extracting the $\pi^0$ radiative width from a pion-production experiment on a nucleus is due to the presence of coherent and incoherent strong-interaction background channels,
\begin{equation}
    \frac{d\sigma}{d\Omega}= \frac{d\sigma^{\rm Prim+Coh}}{d\Omega}+ \frac{d\sigma^{\rm Incoh}}{d\Omega} ,
\end{equation}
 which need to be separated from the Primakoff signal.
 The latter is suppressed with respect to those background processes by $\alpha^2$, but has a characteristic $1/t^2$ behavior which leads to a strong peak at forward angles. This peak resides close to the minimal possible momentum transfer $t_{min}\approx m_\pi^4/4\nu^2$, while strong-interaction contributions reside at larger values of $t$. This kinematic separation is the lever arm that is used to reliably extract the Primakoff signal from the pion-production data. As an example for this separation, in the PrimEx kinematics ($\nu\approx5$\,GeV) for the $^{208}$Pb target the maximum of the Primakoff peak of $d\sigma/d\theta_\pi$ resides at $\theta_\pi\approx0.03^\circ$, while the coherent nuclear contribution is peaked at $\sim1^\circ$. For comparison, for $\nu\approx400$\,MeV accessible for secondary photons in the Gamma Factory the peak positions are $\theta_\pi\approx4.5^\circ$ and $\sim11^\circ$, respectively. The shift of the Primakoff peak to larger values of $t$ automatically leads to its amplitude being reduced according to the $1/t^2$ dependence in Eq.\,\eqref{eq:Primakoff} with respect to the background signal that has a less steep $t$-dependence. To remedy this situation, the GF may offer a significantly higher statistics as compared to PrimEx-II that uses a tagged Bremsstrahlung photon source. With a higher statistics and assuming that the systematic effects are well under control (which is likely the case due to the lower energy), one can envision the possibility to independently determine all the ingredients by a fit over a wide angular range. This nicely connects the here envisioned Primakoff effect 
 measurements with the measurements of neutron skins, the subject of the previous subsection, within the same GF experiment. 
 
If a significant flux of tertiary photons can be produced with the GF, one can go to higher energies to conduct a Primakoff experiment in a more conventional setting.

\subsection{Precision measurement of nuclear \texorpdfstring{$E1$}{E1} polarizabilities}
\label{Subsec:E1_nucl_polariz}

Precise knowledge of electric dipole nuclear polarizabilities (discussed in Sec.\,\ref{Subsec:Photopys_protons_light_nucl} for light nuclei) is also important for eliminating uncertainties in determination of the neutron skins \cite{Piekarewicz2012_E1_pol_n_skin,Piekarewicz2021electric}. Polarizability determinations via measuring gamma-ray transmission are part of the scientific program of the ELI-NP facility \cite{ELI-NP-WB}. The $^{208}$Pb$(\gamma,\gamma'$) reactions were studied also at the electron-bremsstrahlung source ELBE \cite{Tamii2011_E1}.  
The GF will offer a possibility of carrying out such studies on a broad  range of targets with much higher statistical sensitivity and spectral resolution. A measurement program of nuclear electric polarizabilities will complement coherent $\pi^0$ photoproduction method described in Sec.\,\ref{sec:NskinPi0Prod}.

\subsection{Delta-resonance region and continuum effects}
\label{subsec:Delta}

The Delta resonance that can be excited in all nuclei is located \YAL{at} around 300\,MeV excitation energy and has a large width close to 100\,MeV. Of special interest is the pionic decay of this resonance. First of all, pion production on loosely bound nucleons can provide unique information on the wavefunction of exotic nuclei, especially if it would be possible to follow the evolution of this process along a chain of nuclei, from stable to unstable isotopes.

\begin{figure}[!hpt]\centering
    \includegraphics[width=0.45\textwidth]{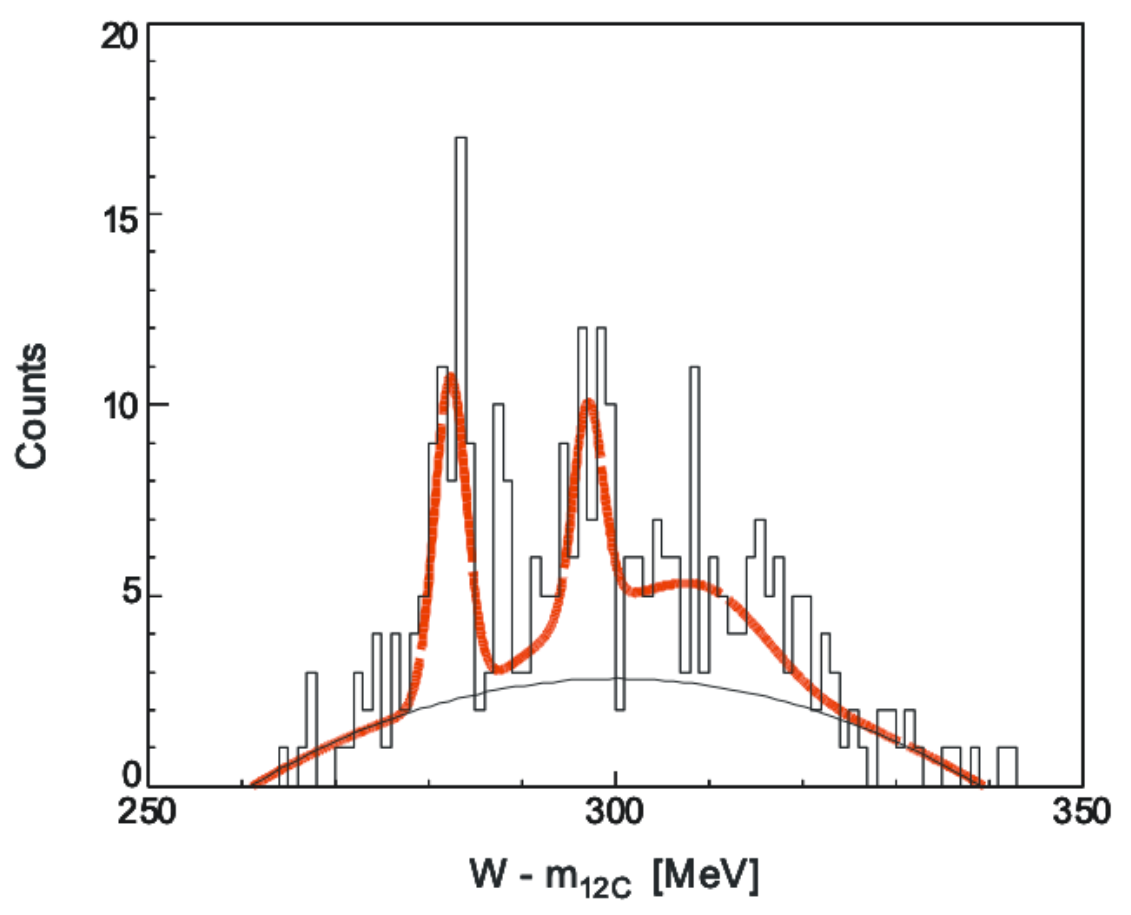}
    \caption{Narrow resonances on top of the broad Delta resonance as claimed in Ref.\,\cite{Bartsch1999}. Figure adapted from Ref.\,\cite{Auerbach2004_Super}.}
    \label{Fig:Mainz_Narrow_Delt}
\end{figure}

There is another non-trivial interest in the detailed study of the Delta-resonance region. A Mainz experiment in the late 1990s \cite{Bartsch1999}
claimed the existence of narrow peaks on the background of the broad Delta resonance (see Fig.\,\ref{Fig:Mainz_Narrow_Delt}). Later, this result was put into doubt, and the experimental situation remains unclear. Simple theoretical arguments \cite{Auerbach2002_Super,Auerbach2004_Super}
show that in the energy region of the Delta resonance there are several possible nuclear Delta + nucleon hole states with the same quantum numbers which are strongly coupled to the pion channel.

This is a typical situation of several intrinsic states interacting both through internal nuclear forces and through the virtual decay to the continuum channel. In such cases, a strong continuum coupling leads to the phenomenon of superradiance when a special collective combination (state) acquires a broad width (short superradiant lifetime), while other states with the same quantum numbers become long-lived narrow resonances. The theory of such many-body superradiance, an analog of the theoretically predicted and later discovered electromagnetic Dicke superradiance in optics, was developed earlier \cite{Sokolov1988,Sokolov1989,Sokolov1992},
supported by numerical simulations for various systems  with quantum signal propagation and interaction through the continuum, as
reviewed in Ref.\,\cite{Auerbach_2011_RPP}.
The simplest explanation, according to P. von Brentano \cite{vonBrentano1996}, is in the fact that interaction of quantum states through a common decay mode, in contrast to the usual Hermitian perturbation, leads to the width (imaginary part of the complex energy) repulsion and vastly different lifetimes. Therefore, the presence of narrow pionic resonances on the background of a ``superradiating'' Delta peak is not excluded theoretically and should be further explored experimentally. Narrow resonances of similar nature are possible also at higher energy as a result of specific quark structures. 

Detailed studies of the fine structure of the Delta resonance possible with an intense gamma beam of appropriate energy with different nuclear targets would open a new branch of nuclear physics of relatively high energy where particle physics is combined with details of the nuclear many-body problem.

\subsection{Parity-violation in photo-nuclear processes}
\label{sec:P-violation}

    PV signatures in the interaction of real photons with nucleons and nuclei arise due to the SM weak interaction among quarks. This underlying mechanism, however, needs to be matched onto hadronic observables. As a result of this matching, a number of effective PV hadron-hadron couplings are generated.  Those involving the lightest mesons $\pi,\rho,\omega$ are generally expected to govern low-energy processes, and a minimal set was introduced by Desplanques, Donoghue and Holstein (DDH) \cite{Desplanques:1979hn}. Among these couplings the PV $\pi N$ coupling $h^1_\pi$ is expected to play a major role since it is the only non-derivative pion-nucleon coupling.  Naive dimensional analysis leads to an expectation $h^1_\pi\sim G_F\Lambda_\chi F_{\pi}\sim10^{-6}$ with $\Lambda_\chi\approx1$\,GeV the chiral symmetry breaking scale, and $F
    _\pi=92.4$\,MeV the pion decay constant. The ``DDH preferred value'' reads $h^1_\pi\sim5\times10^{-7}$~\cite{Desplanques:1979hn}.
    
    \begin{table*}[ht]
    \centering
\begin{tabular*}{\textwidth} {@{\extracolsep{\fill}} cccccccc c cc}
    \hline 
    \hline
      Isotope & Transition & $T_{1/2}$ & Type & Admixture   & $T_{1/2}$ & Type & $\Delta E$\,(keV) &  10$^4$ $A^{\rm PVTC}$& Ref. \\
    \hline \\[-0.2cm]
     ${}^{18}$F\,($1^+$) & $0^-(1081)\to{g.s.}$ & 19\,ps & $E1$  &0$^+(1042)\to g.s.$  & 1.8\,fs & $M1$ & 39  & $-10\pm18$ & \cite{Ahrens:1982vfn} \\
     ${}^{19}$F\,(${1}/{2}^+$) & ${1}/{2}^-(110)\to g.s.$ & 0.6\,ns  & $E1$  & ${1}/{2}^+-{1}/{2}^-$ &  &  & 110 & $-0.68\pm0.18$ & \cite{Elsener:1984vp,Elsener:1987sx}\\
     ${}^{21}$Ne\,(${3}/{2}^+$) & ${1}/{2}^-(2789)\to g.s.$ & 81\,ps  & $E1/M2$  & ${1}/{2}^+(2795)\to g.s.$ & 5.5\,fs & M1 & 5.7 & $24\pm29$ & \cite{Haxton:1980iu}\\
     &&&&&&&& $8\pm14$ & \cite{Earle:1983ji}\\
     ${^{180}}\textrm{Hf}\,(0^+)$ & $8^-(1142)\to6^+(641)$ & 5.53\,h & $M2/E3$ & $8^+(1085)\to6^+(641)$ & 2\,ps & $E2$ & 57 & $-148\pm26$& \cite{Stone:2007xt} \\
    \hline
    \end{tabular*}
    \caption{Gamma transitions with an enhanced PV component in isotopes of fluorine, neon and hafnium. Energies are given in keV. Parameters are from the
     ENSDF database \cite{ENSDF}.}
    \label{tab:NuclAsymmmetries}
\end{table*}

The hierarchy of DDH couplings has recently been questioned from a new perspective that combines chiral effective theory with the large-$N_c$ approach \cite{Gardner:2017xyl}, where $N_c$ is the number of colors. 
Within this new paradigm, couplings associated with isovector transitions involving spin (or isoscalar transitions without spin) are large, while the opposite spin-isospin correlation is suppressed. In meson-nucleon interactions without PV, this pattern is supported by a large $\pi N$ (spin-isovector), vector $\omega N$ (no spin-isoscalar) and magnetic $\rho N$ (spin-isovector) couplings. From this perspective, $h^1_\pi$ (no spin-isovector) should be suppressed.  
The first non-zero determination of $h^1_\pi$ in the $\vec n+p\to d+\gamma$ reaction \cite{Blyth:2018aon} obtained $h^1_\pi=(2.7\pm1.8)\times10^{-7}$, about half the DDH best value, leaving the question, whether or not the large-$N_c$ hierarchy is realized in nature, open. 

In Sec.\,\ref{sec:PCP_Compton} low-energy, nonresonant Compton processes with a PV signature enhanced by low-lying parity doublets is considered, with primary laser photons backscattered off the ion beam, with the limitation $\omega\lesssim60$\,keV in the rest frame of the ion. When using secondary photons on a fixed target, higher energies will be achieved with a possibility to tune the photon energy to resonance transitions of interest (see also Sec.\,\ref{Sec:Nuclear_spectrosc_stored_ions}). 
Parameters of relevant resonance transitions with a significant enhancement due to parity doublets are listed for several nuclei in Table\,\ref{tab:NuclAsymmmetries}. The resonance will de-excite by emitting photons which will need to be counted in a $\approx4\pi$ integrating photon detector, and the dependence of the photon flux in the detector on the circular polarization of the incident photon beam will give access to the PV asymmetry. Due to the resonance kinematics, the enhancement is much stronger than in the nonresonant case, compare to Table\,\ref{tab:Parity_Doublets}. On the other hand, lower rates will be achieved: the necessity to excite a narrow resonance will reduce the rate accordingly, due to the energy-angle resolution correlation of the secondary-photon GF beam. Still, much larger asymmetries will dominate over lower rates in FOM$=\mathrm{Rate}\times A^2$. While there is a substantial overlap with Table\,\ref{tab:Parity_Doublets}, the PV asymmetries listed here are significantly enhanced due to the resonant nature of the process. In the past, PV in $^{19}$F and $^{21}$Ne (see the respective entries in  Table\,\ref{tab:NuclAsymmmetries}) was studied with nuclear, rather than photon, polarization. As shown in Ref.\,\cite{Titov:2006bg}, the two observables are fully analogous and, when integrated over the full solid angle of the emitted photon, are equal.

In general, a global analysis of PV observables, typically performed with nuclear systems, is complicated by the need to embed these PV meson-nucleon couplings into PV nuclear potentials and solve nonrelativistic equations which may bear significant uncertainties. 
As a result, different couplings are intertwined with complicated nuclear effects and may be difficult to disentangle. An alternative way is to move to higher energies where one can directly produce light mesons. Tuning the energy of the photons to the relevant domain one may hope to  enhance--if not isolate--the dominant contribution.
    
Threshold $\pi^+$  photoproduction on the proton with circularly polarized photons was proposed in Ref.\,\cite{Chen:2000hb} as a way to access $h^1_\pi$. The PV asymmetry is  practically isotropic and is given by
\begin{equation}
    A_\gamma\approx\frac{\sqrt{2}F_\pi (\mu_p-\mu_n)}{|g_A| M_N}h^1_\pi\approx 0.52h^1_\pi,
\end{equation}
with $g_A\approx-1.27$ being the nucleon axial coupling and $\mu_{p,n}$ being the proton and neutron magnetic moment, respectively.
The sensitivity to the value of $h^1_\pi$ compares well to that in the nuclear $\vec n+p\to  d+\gamma$ process for which the asymmetry is $A_\gamma\approx-0.11 h^1_\pi$. To estimate the rate and FOM, we note that the total cross section for $\pi^+$ photoproduction off the proton at 180\,MeV photon energy is $\approx 100\,\mu$b. Since the asymmetry is roughly energy-independent, no particularly high energy resolution is necessary. Let us assume $10^{-3}$ out of $10^{17}$ photons arrive on target; hydrogen target is probably not the best option, so let us assume it is carbon -- then we will be looking at $\pi^\pm$ in the detector. The asymmetry is roughly the same but the cross section is $Z^2=36$ times that on the proton, for each pion species. Assuming a 1\,cm$^3$ target, its cross section will have $\approx10^{23}$ $^{12}$C nuclei to interact with. So, we get $10^{14}\times(3.6\times{10^{-27}}$\,cm$^2)/1\,$cm$^2\times10^{23}=3.6\times10^{10}$ $\pi^+$ per second. For asymmetries $\approx 1.4\times10^{-7}$ we get FOM$\sim7\times10^{-4}\,$s$^{-1}$.

A similar process was proposed with a semi-inclusive $\pi^+$ electroproduction at threshold $p(\vec e,\pi^+)e'n$. Photoproduction has an advantage over electroproduction in that there is no background from $Z$-exchange, and all the signal is due to $h^1_\pi$. 

As one moves to the resonance region, a direct PV  photoexcitation of a nucleon and $\Delta$ resonances become possible. Most notably the PV  $\gamma N\Delta$ coupling $d_\Delta^+\,(d_\Delta^-)$ \cite{Zhu:2001br} responsible for a $E1$ $\gamma p\to\Delta^+$  ($\gamma n\to\Delta^0$) transition, respectively, gives rise to the asymmetry 
    \begin{equation}
     A_\gamma^\pm\approx-\frac{2d_\Delta^\pm}{C_3^V}\frac{M_N}{\Lambda_\chi},
    \end{equation}
with $C_3^V=1.6$ being the usual parity-conserving coupling inducing the M1 $\gamma N\Delta$ transition.
A measurement of $\pi^-$ electroproduction on the $\Delta$ resonance on a deuterium target by the G0 collaboration \cite{G0:2011aa} 
obtained 
\begin{equation}
    A_\gamma^-=-(0.36\pm1.06\pm0.37\pm0.03)\,{\rm ppm},
\end{equation}
with the three uncertainties being the experimental statistical, theoretical, and experimental systematic, respectively. 
Accordingly, a loose bound $d_\Delta^-=(3\pm10)\times10^{-7}$ was extracted. It can be expected that the GF will be able to measure these 
small asymmetries $\sim10^{-8}$ due to
 high intensity and $100\%$ photon polarization, together with relatively high resonant cross sections $\approx300\,\mu$b. 

Another promising task for the GF is the resonant PV $\Sigma$ hyperon production  $\vec\gamma+p\to\Sigma^+$ with a subsequent decay to the $\pi N$ final state. 
Until now, the reverse $\Sigma^+\to\vec\gamma+p$ decay with a branching ratio $1.23(5)\times10^{-3}$
    has been accessible, with the scope of studying the asymmetry $a_\gamma=0.76\pm0.08$ \cite{Zyla2020_Particle_Data}. 
    The competing process is the  nonresonant $\gamma p\to\pi N$ process which, at the $\approx$200\,MeV  incident beam energy, is well understood, and the asymmetry for the inclusive reaction is due to the PV coupling $h^1_\pi$ leading to $10^{-8}$ asymmetries. While the contribution of $\Sigma$ production is suppressed with respect to nonresonant pion photoproduction, the figure of merit introduced in Sec.\,\ref{sec:PCP_Compton} favors $\Sigma$ photoproduction: it depends on the cross section linearly, and quadratically on the asymmetry. With a precise tune of the photon energy to the $\Sigma$ hyperon position, the large asymmetry more than compensates for low rate.


    

   

\section{Experimental considerations with fixed targets}
\label{Sec:ExptlConsid}
\subsection{Thermal-load issues}
\label{Subsec:ExptlConsid_Thermal_load}
In designing fixed-target experiments at the GF, it is important to realize that the secondary photon beam carries significant power. With the maximum gamma-ray energy and a total photon flux (at all energies), the power carried by the photons is 3\,MW. To make such energy load manageable for the target, it may be beneficial to position the target as far away as possible from the interaction region. With $\gamma\approx 10^3$ and neglecting the initial PSI angular spread, positioning a 2\,m diameter target at a distance of 1\,km would match the target size to the diameter of the ``spot'' produced by the gamma rays emitted in the $\approx1/\gamma$ cone. 

\subsection{Parallel spectroscopy with spatially resolved detection}
\label{Subsec:ExptlConsid_Parallel_Spectrscpy}
\begin{figure*}[!htpb]\centering
    \includegraphics[width=1.0\textwidth]{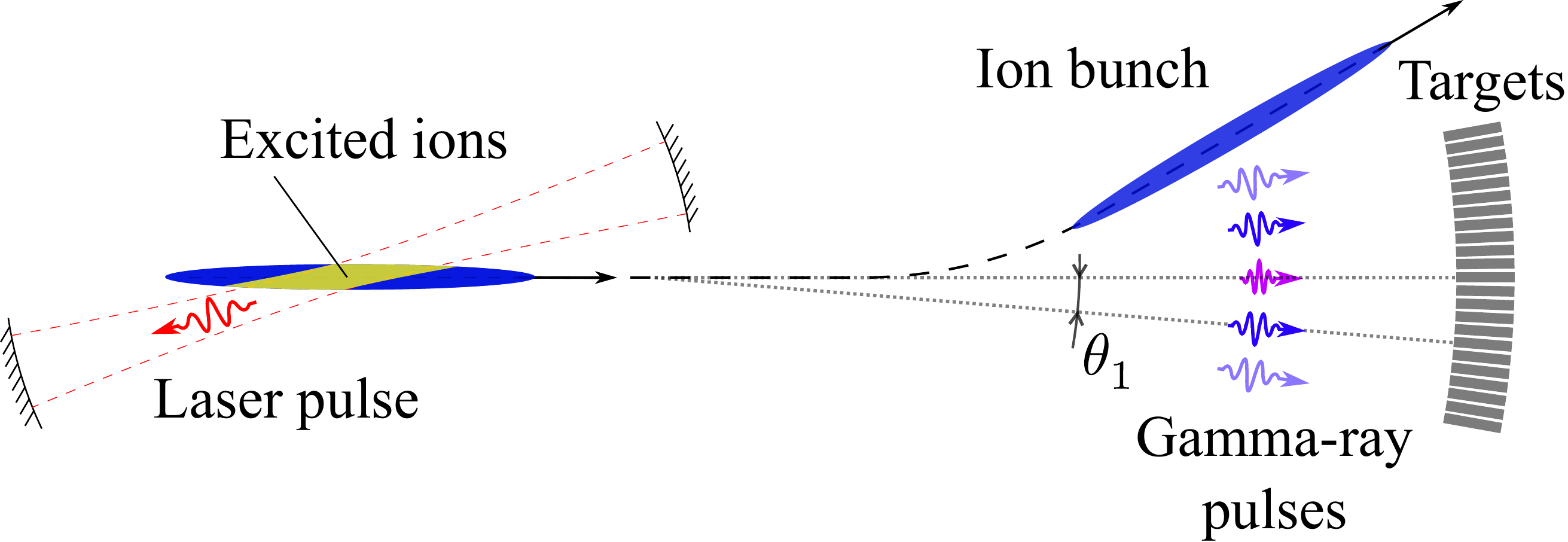}
    \caption{The parallel spectroscopy configuration.}
    \label{Fig:parallel_spectroscopy}
\end{figure*}

The strong correlation of the photon energy with its angle of propagation (see Sec.\,\ref{subsec:The Gamma Factory}) 
suggests a possibility of conducting experiments utilizing the entire flux of the GF secondary photons. This requires a detector providing spatial resolution (Fig.\,\ref{Fig:parallel_spectroscopy}). With this arrangements, a photon with a given energy will hit a ring-shaped area on the target.
Parallel spectroscopy offers the advantage of efficient utilization of the GF running time as opposed to scanning arrangements, where a narrow slice of the gamma-ray energies is selected via collimation and the energy is tuned, for instance, by simultaneously adjusting the PSI $\gamma$ factor and tuning the primary-photon energy.

\subsection{Gamma-ray/X-ray and gamma-ray/gamma-ray pump-probe spectroscopy}
\label{Subsec:ExptlConsid_Pump_Probe}

Pump-probe techniques are ubiquitous in atomic, molecular, and condensed matter physics, and are used in many variants, both in time and spectral domain. Here the system under study is first subject to the ``pump" photons that cause the system to undergo a transition, for example, an excitation of an atomic or molecular state or melting a crystalline lattice. The changes in the system are then monitored via its interactions with the ``probe" photons. 

\begin{figure*}[!htpb]\centering
    \includegraphics[width=1.0\textwidth]{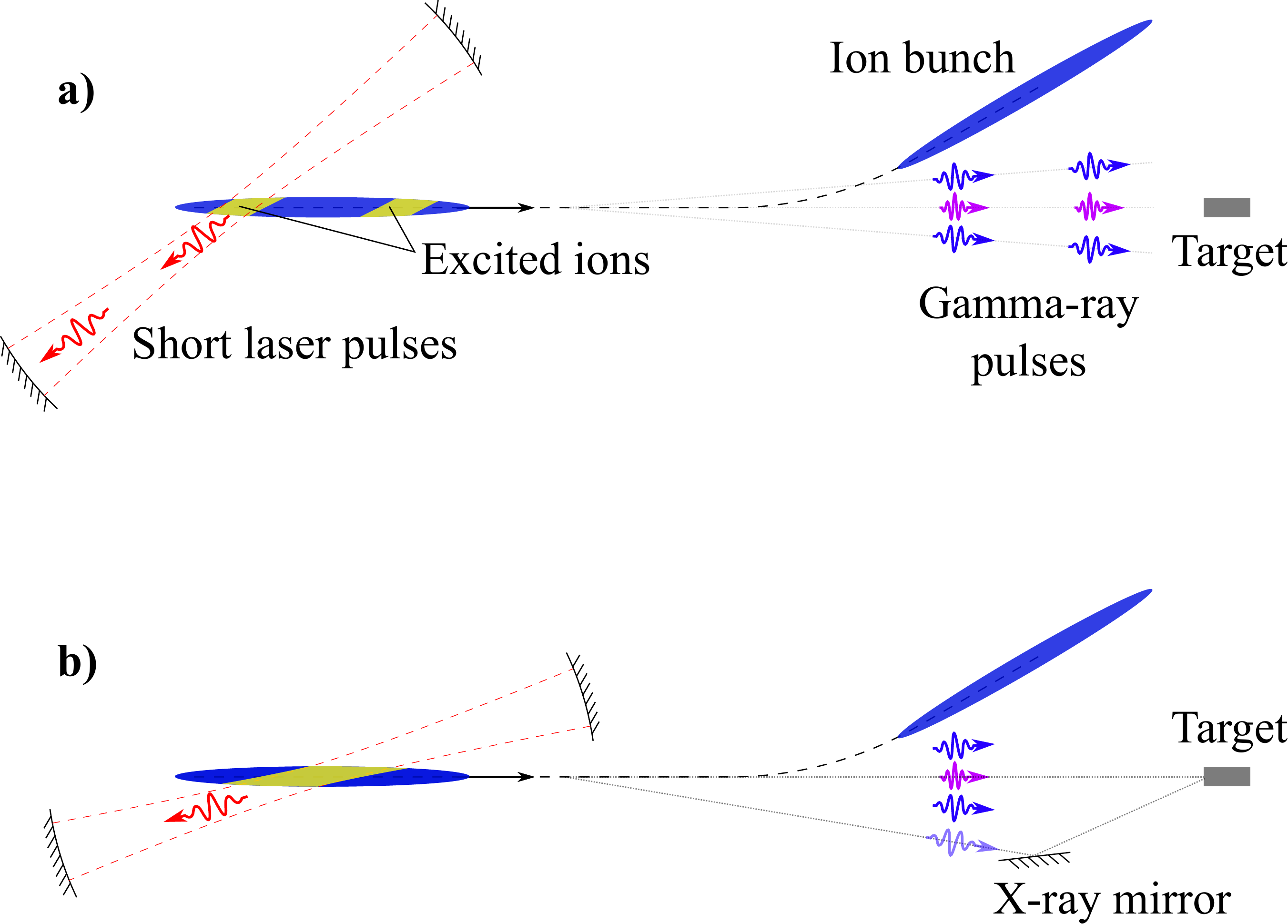}
    \caption{The gamma factory configurations suitable for the pump-probe experiments: a) sufficiently large angle of collision between the laser and the ion beam allows for the laser temporal profile to be imprinted into the gamma-ray pulses; b) X-ray probe pulse can be selected and energy-tuned using a small-angle X-ray mirror.}
    \label{Fig:laser_imprinting}
\end{figure*}

Pump-probe spectroscopy, in novel regimes, is also possible with the GF (Fig.\,\ref{Fig:laser_imprinting}). The pump and probe pulses of secondary GF photons can be produced  
using two laser pulses. This gives an option for both the pump and the probe being gamma rays. We can, for example, have two primary laser pulses interacting with the same ion bunch. The energy of the pump and probe can be tuned within $\approx10^{-3}$ (the energy spread of the PSI that can be maintained in the ring) by selecting different parts of the ion energy distribution. One can also do ``gross'' tuning by choosing different atomic transitions in the PSI. It could be possible to imprint a temporal structure onto the gamma-ray pulse in order to perform pump-probe experiments with sub-ps resolution. The duration and timing of the gamma-ray pulses will be determined by geometric overlap between the sufficiently short laser pulse and the ion beam. With a $\approx10\,\mu m$-wide ion beam in a typical LHC interaction point, the minimum duration of the gamma-ray pulse could be $\approx30\,$fs, 
which comes from the 10\,$\mu$m divided by the speed of light, assuming that the laser beam propagates perpendicularly to the ion beam. This requires a correspondingly short laser pulse, but generation of such pulses is routine with modern lasers. Different-energy photons could come in overlapping pulses (down to 30\,fs) or be separated by a fraction of a ns if they are derived from the same ion bunch, or can come from different ion bunches and have correspondingly long separation. 

Some of the options for GF based pump-probe spectroscopy include:
\begin{itemize}
    \item The light pulses could be tuned to two different transitions of the PSI;
    \item Pump and probe can  be  of the same energy (depending on the experiment);
    \item The pump and probe can be produced from different PSI bunches;
    \item Pump and probe pulses do not necessarily need to be temporally separated;
    \item Pump-probe spectroscopy with nuclear isomers. If the produced nuclear states are sufficiently long-lived, one could move the production target [Fig.\,\ref{Fig:laser_imprinting}(a)] to probe produced nuclear states using gamma rays with different energy at different spatial locations;
    \item Because the photons emitted at the periphery of the $\approx 1/\gamma$ cone are of low energies, one can design a pump-probe experiment with a gamma-ray pump and an X-ray probe [Fig.\,\ref{Fig:laser_imprinting}(b)] using the well-developed technology of grazing-incidence X-ray mirrors \cite{attwood2017Xbook}. An attractive feature of this scheme is that the X-ray energy can be tuned by tilting the mirror; a potential downside for some experiments would be the inherent time delay (due to the difference in the optical path length) between the pump and probe pulses.
\end{itemize} 
\begin{figure*}[!htpb]\centering
    \includegraphics[width=1.0\linewidth]{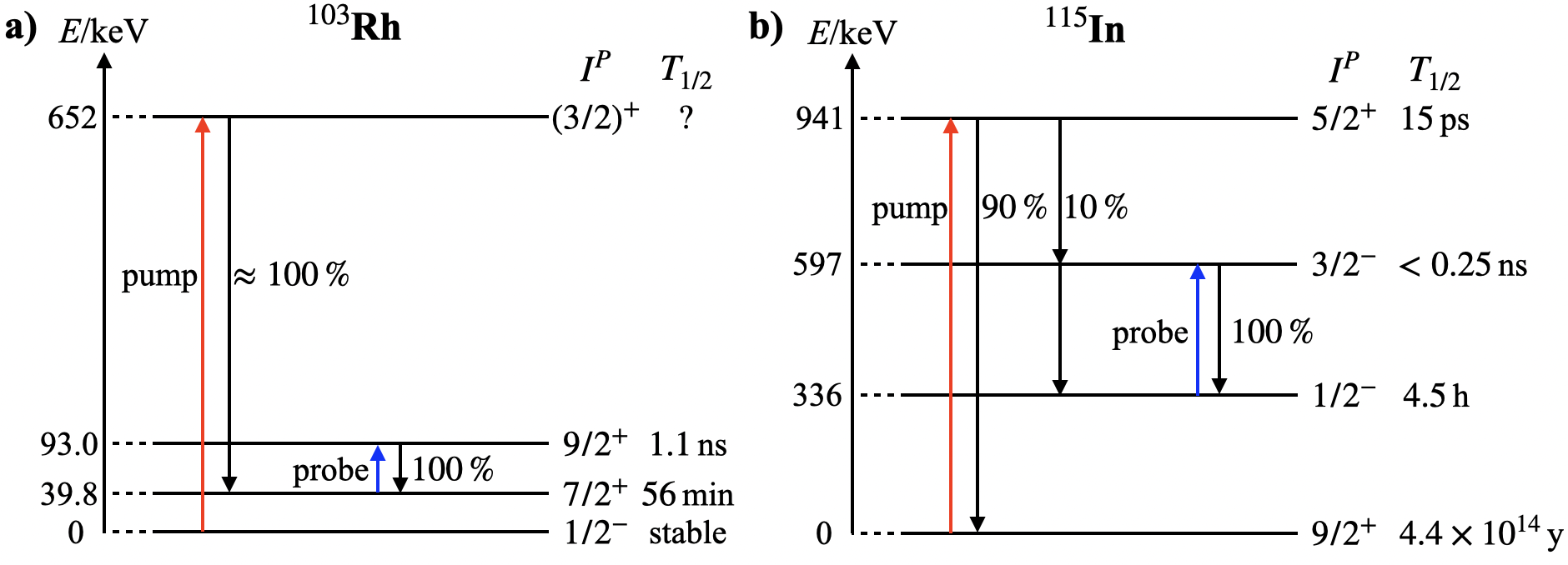}
    \caption{Pump-probe spectroscopy with a) $^{103}$Rh and b) $^{115}$In isomers. 
    The branching ratio after exciting the nucleus to a higher energy level, $I^P$ and $T_{1/2}$ of energy levels are shown (according to the NuDat 2 database \cite{NNDC}).}
    \label{Fig:103Rh_pump_probe}
\end{figure*} 

Two specific examples of a possible application of pump-probe spectroscopy are shown in Fig.\,\ref{Fig:103Rh_pump_probe}.
These examples involve isomers from Table\,\ref{tab:Isomer_(g,g')}. Pump-probe spectroscopy as discussed here could be used to prove the feasibility of producing the isomers at the GF and to study reaction cross sections and production rates. 

\subsection{Highly monochromatic gamma beams}
\label{Subsec:Highly_monochr_gammas}

Energy bandwidth is a key parameter when aiming for exciting narrow nuclear resonances
with typical intrinsic widths of a few eV. Exciting such a resonance with a $\gamma$
beam of a few tens to hundreds of keV, as currently available from existing facilities,
will predominantly result in background generated by the large amount of off-resonant
incident photons. So aiming for clean $\gamma$-spectroscopic conditions requires
reducing the energy bandwidth of the $\gamma$ beam ideally to the physical limit
set by (thermal) Doppler broadening.

The highly brilliant $\gamma$-ray source of the Gamma Factory will deliver a
$\gamma$ beam with a divergence in the range of milliradians, with a
photon flux of up to $10^{17}$ $\gamma$ rays per second, and a possibility of monochromatization on the
order of $\Delta E_{\gamma}/ E_{\gamma} \approx 10^{-3}$ over a wide energy range. Due to pile-up problems, this
intense beam would put serious constraints on detection systems if it were used without
further preparation. A substantial fraction of foreseen experiments at this facility will be
based on the concept of nuclear resonance fluorescence (NRF; see Sec.\,\ref{subsec:narrow-res}).
For NRF experiments the optimal monochromatization should be equal to the width of the
excited nuclear resonances, which is expected to be of the order of
$\Delta E_{\gamma}/ E_{\gamma} \approx 10^{-6}$, due to thermal motion of the target atoms.
Therefore, further monochromatization of the $\gamma$ beam would on the one hand provide
an improved signal-to-noise ratio for these experiments as well as solve eventual detection problems.
Therefore, when aiming for an ultimate quality of nuclear $\gamma$ spectroscopy, the unprecedented
intensity of the GF photon beam can serve as an asset to trade beam intensity for spectral
resolution. $\gamma$-ray optics allowing to further monochromatize the $\gamma$ beam is thus
of high importance for a future high-intensity $\gamma$-beam facility like the GF.

The most accurate method for the absolute determination of $\gamma$-ray wavelengths relies on crystal
diffraction from highly perfect flat crystals of silicon or germanium, allowing to obtain a
resolving power which is unequalled by any other $\gamma$-ray spectroscopic methods.
Over many years, the concept of
crystal-based photon diffraction has been pioneered and optimized at the GAMS facility at
the Institut Laue-Langevin (ILL) in Grenoble \cite{Deslattes1980,Dewey1989,Kessler2001} as a
technique to monochromatize $\gamma$ beams \cite{Jentschel2012}.

\begin{figure*}[!htpb]\centering
     \includegraphics[width=1.0\textwidth]{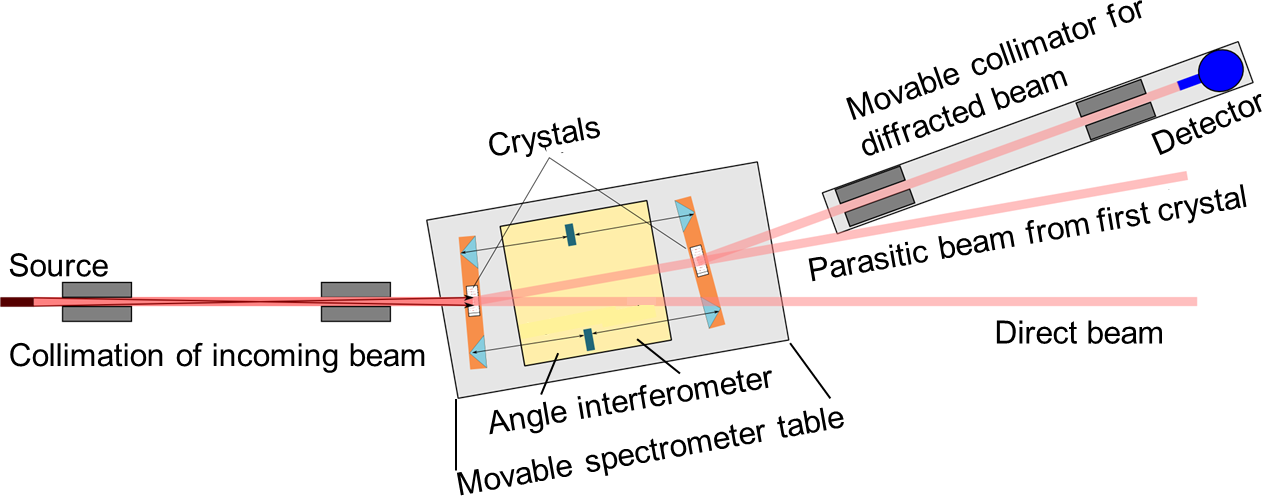}
     \caption{Layout of a double-crystal monochromator as realized at the GAMS facility of
              the ILL (Grenoble) \cite{Jentschel2012}. The consecutive action of the two crystals is
              seen. Both crystals are mounted on
              interferometer arms. The positions of these arms with respect to a movable spectrometer
              table are controlled with optical interferometers. The diffracted beam is separated with
              a movable collimation system away from the intense direct beam. Figure adapted from \cite{Jentschel2012}.}
     \label{fig:GAMS-sketch-2}
   \end{figure*}

Diffraction of photons is governed by Bragg's equation
\begin{equation}
  \label{eq:bragg}
    n\cdot\lambda = n \frac{hc}{E_{\gamma}} = 2d~sin\theta_B. 
  \end{equation}
Here $h$ represents Planck's constant, $c$ is the speed of light, $n$ is the diffraction order, $d$
the lattice spacing of the diffracting crystals and $\theta_B$ the diffraction angle. While this
equation provides information on the energy dependence of $\theta_B$, information on
resolution power is obtained from dynamical diffraction theory \cite{Zachariasen1945}, predicting
the dependence of diffracted intensity $I$ on the diffraction angle $\theta$. The result is
summarized by the following simplified expression
\begin{equation}
   \label{eq:diffract-intensity}
   I(\theta) \propto \frac{sin^2\left( A\sqrt{1+y^2}\right)}{1+y^2},~~
    y \propto \frac{hc}{E_\gamma} \left( \theta - \theta_B \right), ~~A \propto \frac{hc}{E_\gamma}.
\end{equation}
Hence the diffracted intensity consists of an oscillating term surrounded by a Lorenztian envelope
and its width, corresponding to the `acceptance width' of the monochromator can be estimated
to scale with $hc/E_{\gamma}$. Comparing the scaling of the acceptance width to Eq.\,\eqref{eq:bragg}
results in
\begin{equation}
   \label{eq:resolution}
   \frac{\Delta E_{\gamma}}{E_{\gamma}} = \frac{\Delta\theta}{\theta_B} \approx \frac{C}{n}  ,
\end{equation}
where $C$ is a constant. From Eqs.\,\eqref{eq:bragg} and \eqref{eq:diffract-intensity}, it
follows that the achievable energy resolution is independent of the incident photon
energy $E_{\gamma}$. The acceptance width of a perfect crystal can be as small as a few
nanoradians. Thus the number of $\gamma$ rays accepted from the incident beam for diffraction
is small. In fact, a perfect crystal acts for $\gamma$ rays as an excellent collimator,
accepting almost exclusively the non-divergent part of the beam and diffracting it in a different
direction with respect to the incoming beam. Since the initial beam is typically several
orders of magnitude more divergent, this selection results in a drastically attenuated
diffracted beam. For the divergence expected for the Gamma Factory,
one expects a loss factor of about nine orders of magnitude.
In order to fully exploit the properties of a perfect crystal for the production of monochromatic
$\gamma$ rays it would be required that the divergence of the incident photon beam were
comparable to the acceptance width of the crystal. This can be achieved by combining two crystals
to form a double-crystal monochromator. The first crystal will produce a multitude of monochromatic
beams, where each energy is diffracted in a low-divergence beam directed at its particular
Bragg angle. A particular energy may then be selected with the second crystal.
This scheme is realized in the GAMS facility at the ILL (Grenoble), as shown in
Fig.\,\ref{fig:GAMS-sketch-2} \cite{Kessler2001}.

A double-crystal spectrometer can be operated in two geometries. In the so-called
non-dispersive alignment mode, the two crystals are positioned in parallel, which means
that all $\gamma$ rays diffracted by the first crystal are accepted by the second crystal.
In this case, no energy selection is made and there is no monochromatization.
However, this geometry delivers a measurement of the intrinsic instrument resolution as it
measures the convolution of two single-crystal intensity profiles $I(\theta)$ of
Eq.\,\eqref{eq:diffract-intensity}. In the dispersive geometry, the spectrometer
realizes a dedicated Bragg angle $\theta_B$ between the crystals and therefore selects
specific energies. A detailed review on the two geometries can be found in Ref.\,\cite{Borner1993}.

\begin{figure}[!hpb]\centering
     \includegraphics[width=1.0\linewidth]{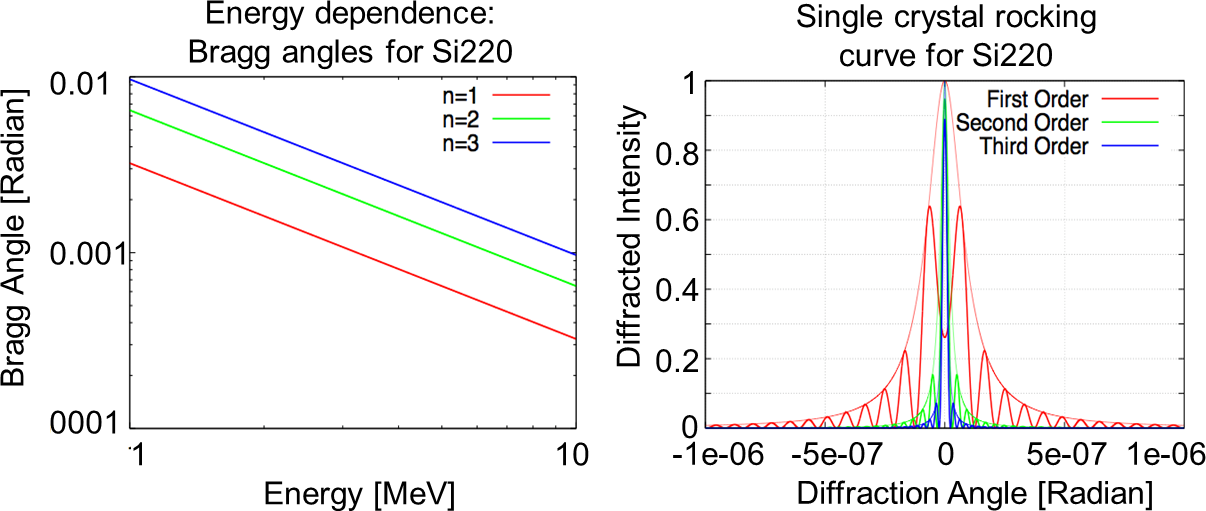}
     \caption{Left: Energy dependence of the Bragg diffraction angle $\theta_B$ for a (220)
              silicon crystal. Right: Rocking curve of a (220) Si crystal, i.e. distribution of the
              diffracted intensity as a function of the diffraction angle. }
     \label{fig:bragg-rocking}
\end{figure}

A typical measurement procedure involves rocking of the second crystal
(and detector) such that its orientation is $+\theta_B$ and $-\theta_B$ relative to the first
diffracted beam. Except for small ($\approx 10^{-7}$) corrections due to a finite vertical
divergence, the angular separation of these two diffracted beams is 2$\theta_B$. The diffraction
angles are measured by polarization sensitive Michelson interferometers which have a sensitivity
of $\approx 10^{-9}$\,rad \cite{Dewey1989} and which are calibrated using an optical
polygon \cite{Kessler2001}.
The left panel of Fig.\,\ref{fig:bragg-rocking} shows the $\gamma$-ray energy
dependence of the Bragg diffraction angle $\theta_B$ for a Si crystal cut along the (220) direction,
typically (depending on diffraction order and photon energy) ranging between 1\,mrad (at 1\,MeV)
to about 10\,mrad for $E_{\gamma}=$10\,MeV. The single-crystal rocking curve, i.e. the distribution
of the diffracted $\gamma$-beam intensity for the same type of (220) Si crystal is shown in the
right panel of Fig.\,\ref{fig:bragg-rocking}, exhibiting diffraction angles of the order of 10\,nrad.

Such a double-crystal monochromator can reach high resolution and instrumental diffraction
widths nearly equal to those predicted by dynamical diffraction theory \cite{Zachariasen1945}
have been obtained for energies up to about 6\,MeV.

From a practical perspective, to realize an ultimate $\gamma$-ray energy resolution
a three-stage measurement chain is needed to link the $\gamma$-ray wavelengths of interest to visible
wavelengths \cite{Deslattes1980}. In a first step, the lattice spacing of a Si crystal is measured
in terms of the wavelength of an iodine-stabilized HeNe laser operating near 633\,nm. This step
employs simultaneous X-ray and optical interferometry and yields a calibrated Si-crystal sample.
In the second step, the lattice spacings of various other crystals are compared to the calibrated
Si crystal to yield a family of crystals whose lattice spacings are known relative to the optical
wavelength. This method employs an X-ray crystal comparator. In the final step, which comprises the
double-crystal monochromator setup, $\gamma$ rays are diffracted by the crystals calibrated in the
second step and the diffraction angles are accurately measured. By combining the measured lattice
spacing and diffraction angles, $\gamma$-ray wavelengths are determined with high accuracy.

The difficulty in making sub-ppm wavelength measurements via this prescription may be seen by
considering the accuracy required in the measurement of the lattice-spacing parameter $d$ and
the diffraction angle $\theta$. The crystal-lattice constant $d$ must be determined with an accuracy
of $\sim 10^{-17}$\,m. The Bragg angle ($\sim 10^{-3}$\,rad at 5\,MeV) must be determined with an angular
precision of $\sim 10^{-10}$\,rad. In addition, the angular scale upon which the Bragg angle is
measured must have an accuracy of $\sim$ 1 in 10$^7$ \cite{Dewey1989}. Such properties have been
demonstrated at the GAMS facility and can thus serve as a `role model' for a potential monochromator
to be integrated into the GF experimental device suite with an envisaged relative energy resolution
$\Delta E_{\gamma}/E_{\gamma}\sim 10^{-6}$.


\section{Photophysics with a storage ring for radioisotopes}
\label{Sec:Radioisotope_Storage_Ring}

\begin{figure*}[!htpb]\centering
     \includegraphics[width=\textwidth]{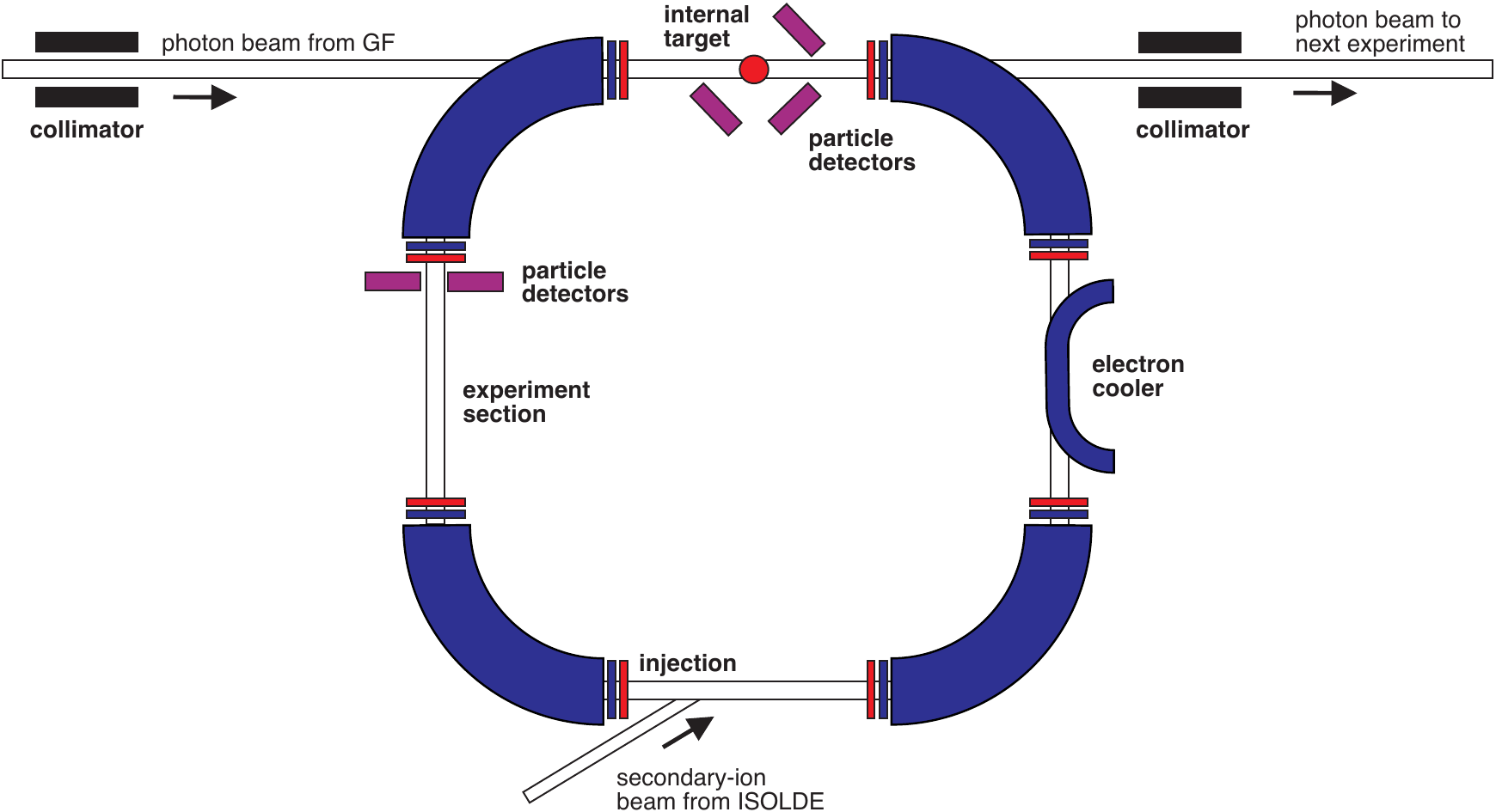}
     \caption{A sketch of a possible arrangement of the low-energy storage ring at the GF.
     The photon beam from the GF comes from the left and interacts with the stored secondary ion beam. The photon beam leaves the ring on the right and can be employed in a next experimental station. An internal gas target can be implemented as an option.}
     \label{fig:storage_ring}
\end{figure*}

The secondary beam of the GF can be used to irradiate solid-state targets of stable or long-lived nuclides.
In general, such experiments have been performed in the past and are planned at future photon facilities, though with photons at different energies and/or lower intensities as compared to the GF.
However, the ultimate interest and the highest discovery potential lie in investigations of exotic nuclei with large proton-to-neutron asymmetries. Nuclei far from the line of stability are currently at the center of interest in nuclear structure physics. They offer insight into the physics of loosely bound finite drops of Fermi liquid, and are intricately related to astrophysical processes of formation of stable nuclei with their observed abundances on Earth and in the Universe. 
Estimates predict the existence of more than 7000 nuclei stable with respect to particle decay \cite{Erler-2012}; the lifetime of those outside of the valley of stability is determined by weak decays in the direction of this valley.  Experimental access to these species is rendered possible by several new-generation facilities under construction including FAIR in Germany, SPIRAL2 in France, HIAF in China and FRIB in the USA.

Exotic nuclei are inevitably short-lived and their production rates in nuclear reactions are small. For instance, halo phenomena, where one or a few valence nucleons are weakly bound, occur in short-lived nuclei lying close to the limits of nuclear existence, the so-called driplines, see Sec.\,\ref{Subsec:Pion_photoproduction}. 
The ``classic'' halo nucleus $^{11}$Li has a half-life of merely 8.75(14)\,ms \cite{NNDC}, and no target can be produced out of it.  As known from  first detailed studies \cite{Tanihata-2013}, the size of external weakly bound neutron orbitals in $^{11}$Li is essentially the same as the size of the strongly bound doubly magic isotope $^{208}$Pb. There is still no general agreement on the mechanism leading to the binding at such large distances. The wavefunction of halo nucleons looks like that of one or a few Cooper pairs bound to a normal core by some special forces \cite{Barranco2001,Potel2010}. 
There are also competing mechanisms: 
the collective modes of the core coupled through the continuum may create loosely bound states, as for example in the heaviest oxygen isotopes \cite{Volya2006}.

The location of the GF at CERN has a unique advantage of having in its close proximity a state-of-the-art radioactive-ion-beam facility, the isotope separator on-line device (ISOLDE) \cite{Borge_2017}. 
The ISOLDE facility offers a large variety of secondary beams. As of today, an intense 1.4\,GeV  proton beam impinges on a thick, several 10\,g/cm$^2$, production target, such that the reaction products stop in the target material. Target spallation and fission are the major nuclear reactions giving access to exotic nuclei of interest. A variety of schemes have been developed to achieve clean secondary beams delivered to various experimental stations. In the present context, the major option is HIE-ISOLDE~\cite{Lindroos-2006} (HIE stands for high intensity and energy), which provides post-accelerated beams at energies of up to 10\,MeV/nucleon in high atomic charge states after charge breeding in a dedicated REXTRAP/REXEBIS system \cite{Ames-2004,Wenander-2011}. 

To bring exotic nuclei in collisions with photons, their storage is indispensable. About a decade ago, a proposal was put forward at ISOLDE to install a dedicated storage ring \cite{Grieser-2012}. Such storage ring is a part of the upgrade program of the ISOLDE, the EPIC project (Exploiting the Potential of ISOLDE at CERN) \cite{Catherall-2019}.
On the one hand, its installation will enable a broad range of unique physics experiments at HIE-ISOLDE, see \cite{Grieser-2012}; on the other hand, the research scope can be dramatically extended if merging with the photons from the GF is achieved. Since the photons cannot be transported to ISOLDE, the storage ring needs to be constructed next to the GF. 
A beamline from the ISOLDE to the GF site is required to transport secondary ion beams at energies of a few MeV/nucleon from the ISOLDE hall to the GF. Dependent on the exact location of the GF, such beamline can be 100-300 m long.

The interactions of the GF photons with antiprotons can also be envisioned. 
Various technical solutions can be considered.
A straightforward approach is to construct a beamline connecting the Antiproton Decelerator (AD) facility and the proposed storage ring at the GF. The PUMA project (antiProton Unstable Matter Annihilation \cite{Aumann-2019}), aims at transporting trapped antiprotons from AD to ISOLDE to study antiproton collisions with exotic nuclei. 
A similar methodology may as well be applied.

The design of the storage ring will be based on the experience gained in operation of the low-energy storage rings TSR in Heidelberg \cite{Grieser-2012} and CRYRING@ESR in Darmstadt \cite{Lestinsky-2016}. It will have a circumference of about 40\,m. Secondary ions stored at energies of a few MeV/nucleon will have typical revolution frequencies of a few hundred kHz. The ring will have four straight sections, where two of them will be used for injection/extraction and an electron cooler. 
The experimental straight section will be aligned with the photon beam, see Fig.\,\ref{fig:storage_ring}. 
In order to reduce heat load and stress on vacuum windows, an evacuated pipe can be used to connect SPS/LHC and the storage ring. 

Dependent on the specific physics case, coasting as well as bunched beams will be employed. 
If nuclear and storage lifetimes allow, accumulation of beam currents up to about 1\,mA is possible \cite{Grieser-2012}. 
The electron cooled ion beam has a momentum spread on the order of $10^{-4}$ or better and the transverse size is about 1\,mm. These parameters constrain the interaction region and thus offer excellent conditions for high energy and angular resolution in experiments. The beam is stored at a few MeV/nucleon energy which facilitates the detection of the beam-like recoils and reaction products.
Any product of a charge-changing (atomic or nuclear) reaction will be deflected differently by the dipole magnets as the primary beam and can thus be intercepted by a particle detector.
It should be emphasized that all tools developed at various storage rings \cite{Litvinov-2013, Steck-2020} can be available here as well, which allow, for example, to prepare the beam in a specific, well-defined atomic or nuclear state by employing internal targets or dedicated laser beams.

The photon beam from the GF will pass through the interaction region of the storage ring and can be used in a different experiment downstream the storage ring.

\section{Colliding-beam opportunities}
\label{Sec:Colliding-beam_opportunities}

\begin{figure}[!htpb]\centering
    \includegraphics[width=\columnwidth]{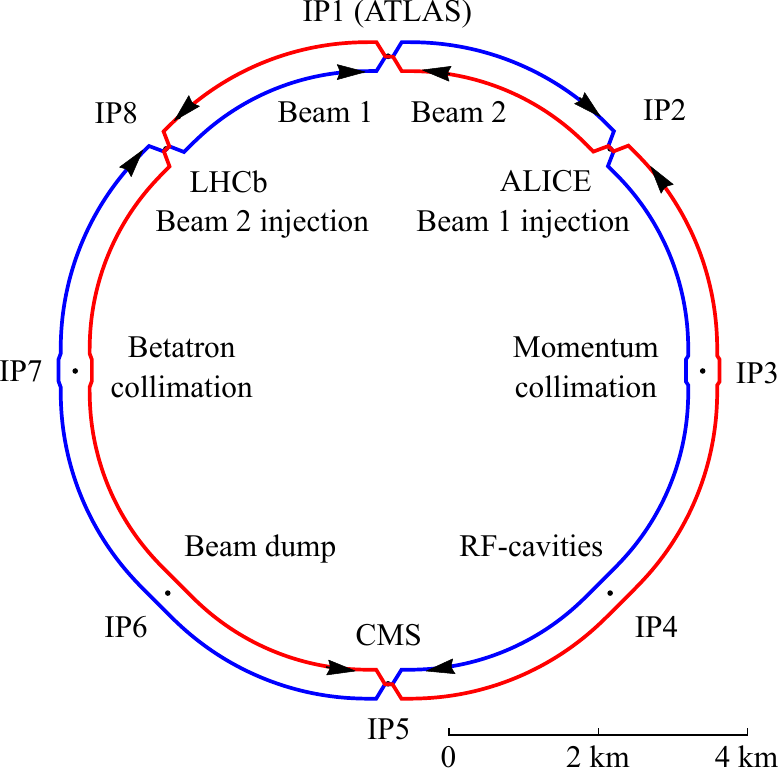}
    \caption{The LHC layout. Adapted from \cite{Hermes-2016}. The separation between the beams is exaggerated here (in reality it is 20\,cm). IP: interaction points.}
    \label{Fig:LHC_laytout}
\end{figure}

The LHC collider (see Fig.\,\ref{Fig:LHC_laytout}) has two counter-rotating beams which are normally separated by 20\,cm horizontally and brought to collision at several interaction points around the ring. Superconducting magnets guiding both beams share the same cryostat. Due to the magnet design, the strength of the magnetic field guiding each beam is the same (rigidity of both beams is the same). The LHC can operate in proton-proton, ion-ion, or ion-proton collision mode.

\begin{figure}[!htpb]\centering
    \includegraphics[width=\columnwidth]{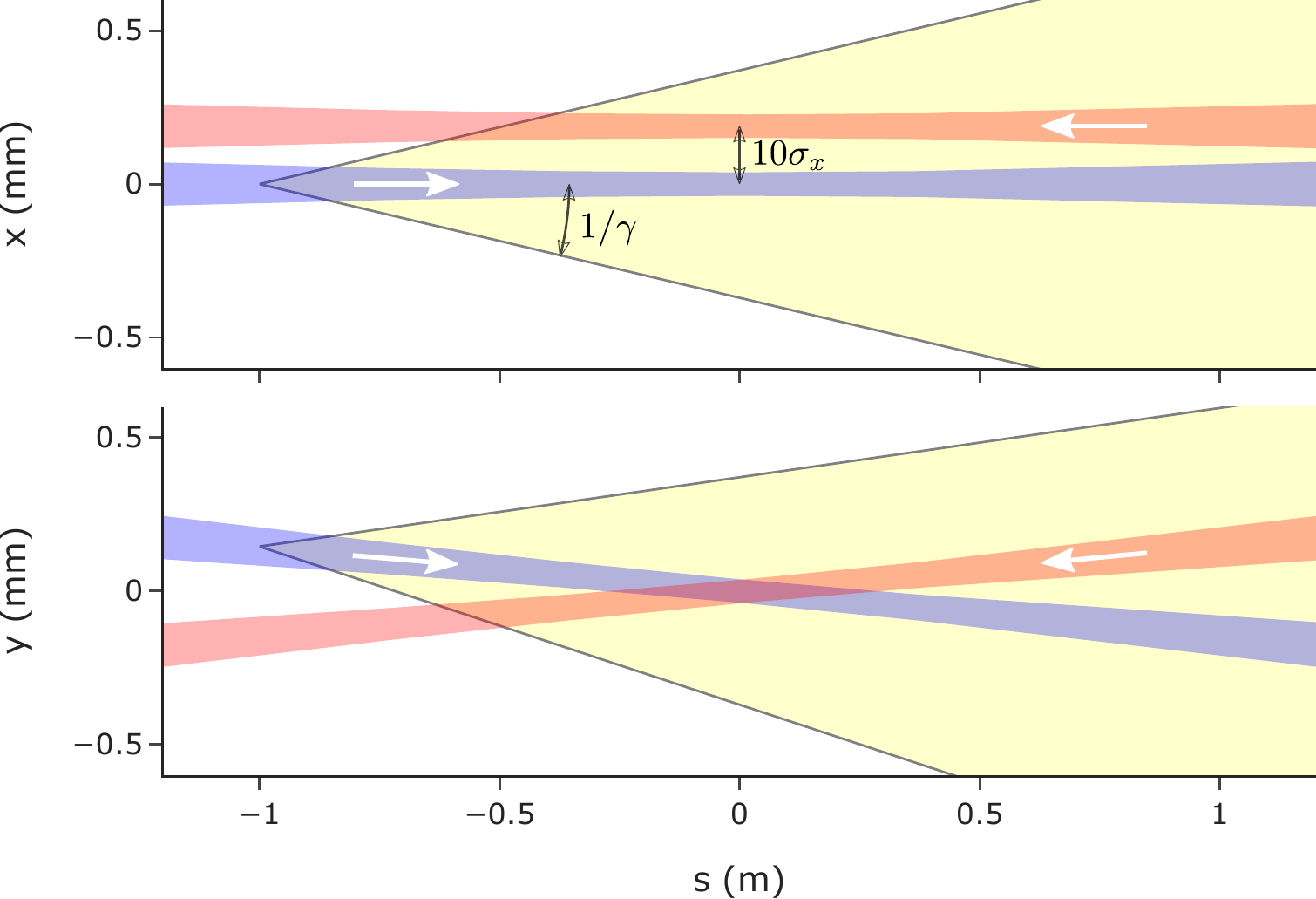}
    \caption{The typical LHC interaction point (IP) configuration compared to a $1/\gamma$ cone of gamma-radiation. The diameter of the beams in this picture is $4\sigma_{x,y}$ (at the IP $\sigma_{x,y} \approx 19~\mu\mathrm{m}$). The beams are separated by $10\sigma_x\approx0.2$~mm horizontally in order to avoid stripping of the PSI due to collisions with counter-propagating ion or proton beam. }
    \label{Fig:GF_collider}
\end{figure}

The secondary photons produced at the GF can themselves be directed onto a relativistic beam, for instance the counter-circulating beam of the LHC (see Fig.\,\ref{Fig:GF_collider}), thus benefiting from another Lorentz boost of photon energy (for secondary photons in the frame of the counter-circulating ions/protons) or two Lorentz boosts (for tertiary photons in the lab frame). This possibility, that may be realized at a later stage of the GF program, may open additional physics opportunities. We briefly discuss some of them here.

\subsection{Photoabsorption structure functions} 
\label{subsec:Photoabs_structure_finctions}

    Via scattering secondary photons from the GF head-to-head off a proton or ion beam, one can study polarized and unpolarized inclusive structure functions at high energies.
    
For secondary photons with energy $\omega$ in the laboratory frame and a proton beam with a relativistic factor $\gamma_p$ up to $\approx7000$ from the LHC,
the highest invariant mass amounts to
    \begin{equation}
    \sqrt s=\sqrt{4M_p\gamma_p\omega}\leq108\,{\rm GeV}\sqrt{\frac{\gamma_p}{7000}\cdot \frac{\omega}{400\,{\rm MeV}}}\,.
    \end{equation}
The total photoabsorption cross section can be written in terms of inclusive structure functions familiar in the context of inelastic electron scattering \cite{Zyla2020_Particle_Data}. Those surviving for real photons are $F_{1,3}$ and $g_{1,5}$, and denoting the photon circular polarization $\xi=\pm1$  and the proton helicity $h=\pm1/2$, we write 
    \begin{align}
        \sigma(\xi,h)&=\frac{8\pi^2\alpha}{s-M^2}
        \left[
        F_1-2h\xi g_1+\frac{\xi}{2}F_3+2hg_5
        \right].\label{eq:structurefunctions}
    \end{align}
 The structure functions $F_i,\,g_i$ are functions of $s$. The first two terms in Eq.\,\eqref{eq:structurefunctions} conserve parity, while the last two terms are parity-violating.

High-energy behavior of scattering amplitudes is governed by Regge theory and is economically described by $t$-channel exchanges. The leading contribution to the spin-averaged structure functions $F_1$ at asymptotically high energies is identified in QCD with exchanges of gluons which combine into colorless compounds, while quark-antiquark (meson) exchanges are suppressed \cite{Levin:1990gg}. The well-known pomeron that has quantum numbers of the vacuum and couples equally to particles and antiparticles ($C$-parity even), is identified with the colorless two-gluon exchange. An exchange of a colorless three-gluon state leads to a $C$-odd odderon which couples to particles and antiparticles with an opposite sign. Predicted nearly five decades ago \cite{Lukaszuk:1973nt}, this elusive kind of interaction of hadrons has just recently been observed in Tevatron/LHC data \cite{Abazov:2020rus}. 
While not accessible with the inclusive photoabsorption cross section by Furry's theorem (structure functions are the imaginary part of the forward Compton amplitude), the odderon can contribute to exclusive channels, e.g. photoproduction of axial vector mesons, or in parity-violating asymmetries, as outlined below. This would be the first observation of the odderon in electromagnetically induced scattering processes. 

Data on $F_1$ with real photons exist up to $\approx200$\,GeV~\cite{Zyla2020_Particle_Data}, nonetheless the new data at GF are expected to significantly improve the precision. Using ion beams one can study nuclear shadowing: it is observed that a high energy probe does not see all nucleons within a nucleus, but only part of them \cite{Bauer:1977iq}. GF can allow to study nuclear shadowing at highest attainable energies where only the leading pomeron exchange gives a sizable contribution (asymptotic regime), or below where other contributions are non-negligible  (sub-asymptotic regime).

Measuring the structure function $F_3$ entails shining a beam of circularly polarized photons onto an unpolarized proton/ion beam. No data on the purely electromagnetic $F_3$ exist. Data on this structure function in electron scattering are almost exclusively sensitive to interference of the $Z^0/\gamma$ exchange between the electron and the proton. A superconvergence relation $\int_0^\infty \frac{d\omega}{\omega^2}F_3(\omega)=0$ was derived in Ref.\,\cite{Kurek:2004ud}, but has never been verified experimentally. GF can provide the input to this sum rule from the inelastic threshold up to highest achievable energies. The structure function $F_3$ violates parity but conserves $CP$, therefore it is also $C$-parity odd. This feature removes the restriction of Furry's theorem and allows for the odderon direct contribution to $F_3$ at asymptotic energies: the odderon leads to slowly (e.g., logarithmically~\cite{Lukaszuk:1973nt}) growing cross section, while Regge meson exchanges lead to a $\sim1/\sqrt{s}$ behavior. 

The spin structure function $g_1$ with polarized proton/ion beams is relevant for the Gerasimov-Drell-Hearn  sum rule \cite{Gerasimov:1965et,Drell:1966jv} that equates the squared anomalous magnetic moment of the proton to an energy-weighted integral over $g_1$. Data on $g_1$ with real photons exist up to $\sqrt s\approx2$\,GeV, allowing to experimentally verify this important sum rule \cite{Dutz:2004zz}. However, the higher-energy part of the integral is estimated using a model-based parametrization of existing deep-inelastic data at low $Q^2$ \cite{Abe:1998wq,Airapetian:2007mh,Fersch:2017qrq,Aghasyan:2017vck} and their extrapolation to the real photon point \cite{Bianchi:1999qs,Bass:2018uon}. A direct measurement at the GF would serve as an explicit check of that model.

The parity-violating spin structure function $g_5$ requires a polarized proton/ion beam, and can be obtained from a single-spin asymmetry upon averaging over photon polarization. This structure function has generally been elusive even in deep-inelastic electron scattering. No data with real photons exist. Together with $F_3$, a measurement of $g_5$ at asymptotic and sub-asymptotic energies will be completely new in terms of addressing PV in the Regge domain: Regge theory operates with $t$-channel exchanges which have well-defined parity. In this context, it is worth mentioning the current $3\sigma$ tension in the unitarity of the Cabibbo-Kobayashi-Maskawa (CKM) matrix in its top row \cite{Zyla2020_Particle_Data}. This tension came about due to a recent re-evaluation of the electroweak radiative $\gamma W$-box correction to the neutron and nuclear $\beta$ decay rate in Refs.~\cite{Seng:2018yzq,Seng:2018qru}. The mechanism found in those Refs. is the Regge exchange contribution to the parity-violating structure function $F_3$.
        
\subsection{Production of ultrahigh-energy gamma rays} 
\label{subsec:Production_ultrahigh_en_gammas_colliding}

%

\subsubsection{Scattering of secondary photons off stored relativistic ions}
Consider a secondary photon beam generated by the GF that is  back-scattered (if possible, resonantly) from relativistic ions.
Nuclear levels have energies of several MeV. We take 15\,MeV as reference, as it is done in Sec.\,\ref{subsubsec:higher_en_gammas}. This places the energy of the secondary photons to $\approx15$\,MeV$/2\gamma$, between about 2.6\,keV and 40\,keV using the $\gamma$-factor range for the LHC and assuming head-on collisions.

An example of a configuration that yields parameters not too far from these is the following. 
Assume the LHC at injection energy with Li-like Xe beam ($\gamma\approx190$) using the 120\,eV $2s-2p_{1/2}$ transition (similar to the transition proposed for the GF Proof-of-Principle experiment at the SPS \cite{Krasny2019PoP}). The maximum energy of emitted secondary X-rays in this case will be 46\,keV. In the frame of reference of the counter-rotating ion beam these photons will appear to be of 17\,MeV energy suitable for excitation of some nuclear levels. The re-emitted gamma radiation from the nuclear level will be Lorentz boosted to 7\,GeV in the lab frame. The lifetime of the upper state of the $2s-2p_{1/2}$ transition in Li-like Xe is 186\,ps \cite{Theodosiou-1991}. This extends the 46\,keV X-ray emission region to 10\,m which is about 10 times longer than the focused ion beam region in the IP (see Fig.\,\ref{Fig:GF_collider}). Longer $2s$ lifetimes of lighter ions limits their efficiency for such a scheme.

A more efficient although much more expensive approach would be to use a dedicated X-ray FEL facility generating the required X-ray pulses.


If a 15\,MeV nuclear resonance decays via photons, with highest accessible relativistic factors, one would produce gamma rays with energies of up to $\approx 90$\,GeV in the lab frame. For most efficient gamma production, one would choose a resonance that decays in the photon channel with a high branching ratio.
How many such photons can be produced this way strongly depends on the specific system and the details of experimental arrangements; preliminary estimates show that on the order of 10$^6$ high-energy photons per second could be achievable in favorable scenarios.



\subsubsection{Scattering of secondary photons off a proton beam}
To achieve the highest possible photon energies, one may shine secondary GF photons head-on onto the proton beam at the LHC.\footnote{Earlier it was also suggested to use X-ray FEL for that purpose \cite{Serafini-2017}.} With the inverse Compton scattering process, the energy of the resulting photons will be close to the LHC proton-beam energy 6.5\,TeV. However, the cross-section of this process is low and we can expect only several such scattering events per hour with the GF beam intensity \cite{Curatolo2021_TeVLHC_Notebook}. Another promising scenario with a much higher cross-section would be to excite the $\Delta(1232)$ resonance at $\sim$300\,MeV in the rest frame of the proton.
It will be producing $\pi^0$'s and $\pi^+$'s which decay in flight: 
\begin{equation}
    p\ + \ \gamma \to \Delta \to\ p\ + \ \pi^0 \to p\ +2\gamma .
\end{equation}
The incident photon energy must be tuned to $\omega\approx 300\,{\rm MeV}/2\gamma_p\approx22$\,keV for the relativistic factor of the LHC proton beam $\gamma_p=7000$. Such low-energy secondary photons might originate from a lower-energy PSI beam (for example, at the SPS) or by directing the secondary photons on the proton beam at a large relative angle. Either way, the secondary photon beam should be produced not too far from the point at which the collision with the proton beam occurs to have as many secondary photons interacting with the proton beam as possible. 
Note that since the width of the $\Delta(1232)$ is 120\,MeV one can use a fairly broad secondary photon spectrum, say $\omega=260\pm60$\,MeV. 
The $\pi^0$ will be emitted with the energy $\omega_\pi\sim2\gamma_p\,(230\pm60)\,$MeV\,$\sim(3.2\pm0.8)$\,TeV. 
These pions will decay within $\gamma_p\tau_{\pi^0}\approx6\times10^{-13}\,$s 
and all this energy will be shared between two photons going within a small angle $\sim m_\pi/\omega_\pi$.
The cross section for $\pi^0$ production on top of the resonance is $\sigma\approx300\,\mu$b.
The background processes are associated with charged particles, such as pair production (with a cross-section of $\approx 10\,$mb for 300\,MeV photons)  and $\pi^+$ production, where almost equal amount of $\pi^0$'s and $\pi^+$'s are produced (the latter live much longer and do not produce photons). 

Pion production rate can be estimated as follows. Assuming $10^{15}$ photons per second (to account for the divergence of the photon beam) colliding with the proton bunch of $3\times10^{10}$ protons \cite{Citron2018future} in a bunch of $(16\,\mu m)^2$ cross section,
we have 
\begin{equation}
   10^{15}\times3\cdot 10^{10}\times\frac{3\cdot10^{-28}\,\mathrm{cm}^2}{(16\cdot10^{-4}\,\mathrm{cm})^2}\sim4\cdot10^{3}\,\mathrm{s}^{-1}\,.
 \end{equation}
The number of photons is double that number. The photon energy distribution in the lab frame is expected to be broad and monochromatization via collimation may not be possible in this case (this will be studied in more detail in future work). The large width of the $\Delta$ resonance of $\approx 120$\,MeV is of advantage for the number of pions and photons produced: we do not need to tune the secondary photons too precisely and will be integrating over the whole $\Delta$ peak. The highest attainable energy photon beam of 4\,TeV is unprecedented (the fixed-target program COMPASS is limited to below $\approx 100$\,GeV photons). Neutral pions are routinely produced, for example, in $pp$ collisions at the LHC. However, because of the symmetric kinematics, pions and photons are spread over $4\pi$ solid angle. With the asymmetric kinematics considered here, we will obtain a high-energy photon beam with small divergence.

The very high energy gamma rays (VHEGR) are of interest in astrophysics, so having an intense source of VHEGR with energy up to 4\,TeV will help calibrating the detectors for the VHEGR detection and to study VHEGR interactions with matter.

\section{Production of isotopes  and isomers for medicine, dark matter search and gamma lasers}
\label{Sec:Medical_Isotopes}

Isotope and isomer production via photonuclear reactions is among the important applications of gamma sources; see, for example, Refs.\,\cite{Habs2011_Med_Isotopes,pan2020photo} for a detailed discussion. The advantages of the GF are, first of all, high photon flux, but also high monochromaticity allowing one, in some cases, to take advantage of resonant cross section enhancement \cite{Habs2011_Med_Isotopes}.
We consider the production of useful nuclear isotopes and isomers via ($\gamma,n$) or ($\gamma,\gamma'$) reactions from fixed targets containing long-lived nuclei (half-life longer than one year).

\subsection{Production of medical isotopes via (\texorpdfstring{$\gamma,n$}{g,n}) reactions}
\label{subsec:prod_isotopes_g_n}

\begin{table}[t]
    \centering
    \begin{tabular*}{\linewidth}{@{\extracolsep{\fill}} l   c  c  c      c      l      c}
    \hline
    \hline
        Z  &$^{A+1}$X  &$^{A}$X  &$I^{P}$  &$T_{1/2}$  &Decay mode  &$S_n$(keV)\\
    \hline \\[-0.2cm]
        29  &$^{65}$Cu  &$^{64}$Cu  &1$^+$  &12.7\,h  &$\%\varepsilon+\%\beta^+=61.50$  &9910.4\\
        &&&&&$\%\beta^-=38.50$  &\\
        42  &$^{100}$Mo  &$^{99}$Mo  &1/2$^+$  &66.0\,h  &$\%\beta^-=100.00$  &8294.2\\
        46  &$^{104}$Pd  &$^{103}$Pd  &5/2$^+$  &17.0\,d  &$\%\varepsilon=100.00$  &10009.2\\
        68  &$^{170}$Er  &$^{169}$Er  &1/2$^-$  &9.4\,d  &$\%\beta^-=100.00$  &7256.9\\
        75  &$^{187}$Re  &$^{186}$Re  &1$^-$  &3.7\,d  &$\%\beta^-=92.53$  &7360.7\\
        &&&&&$\%\varepsilon=7.47$  &\\
        77  &$^{193}$Ir  &$^{192}$Ir  &4$^+$   &73.8\,d   &$\%\beta^-=95.24$ &7772.0\\
        &&&&&$\%\varepsilon=4.76$  &\\
        79  &$^{197}$Au  &$^{196}$Au  &2$^-$  &6.2\,d  &$\%\varepsilon+\%\beta^+=93.00$  &8072\\
        &&&&&$\%\beta^-=7.00$  &\\

    \hline
    \hline
    \end{tabular*}
    \caption{Examples of medical isotope production via the $^{A+1}_{Z}$X($\gamma, n$)$^{A}_Z$X reaction \cite{Habs2011_Med_Isotopes,SZPUNAR2013_Medical_Isotopes}. For produced medical isotopes $^A_Z$X, their ground-state nuclear spins $I^P$, half-lives $T_{1/2}$ and decay modes are provided.
    Here $\%\varepsilon$ and $\%\beta^+$ represent probabilities of nuclear decay via electron capture $(\varepsilon)$ and 
    $\beta^+$ decay, respectively.
    $S_n$ is the neutron-separation energy of the initial nucleus $^{A+1}_Z$X.
    The data are from Ref.\,\cite{NNDC}.
    }
    \label{tab:Isotope_(g,n)}
\end{table}


Production of medical isotopes via $(\gamma,n)$ can be realized by either resonantly exciting specific nuclear energy levels beyond the neutron separation threshold or exciting the giant dipole resonance (see Sec.\,\ref{Subsect:GDR-multi}) using photons in a broad band. The former approach could be cleaner by taking advantage of photon monochromaticity achievable at the GF and involve less power deposited on the target. The photon bandwidth can be adjusted to match the total width of the excited state predominantly arising from neutron emission.
But the resonant cross section is suppressed due to the small gamma emission branching ratio of the excited state.
The latter approach could be more effective due to a larger peak cross section and a broader width of the GDR (together resulting in a larger integral cross section), allowing for more photons to be absorbed. This was studied in, for example, Ref.\,\cite{SZPUNAR2013_Medical_Isotopes} employing gamma photons produced by laser photons Compton back-scattered off relativistic electrons at the Canadian Light Source (CLS); see also Table\,\ref{tab:gammay_ray_sources}. With orders of magnitude higher photon fluxes at the GF, we can expect a significant improvement of production rates,
thus making $(\gamma,n)$ reaction practical for producing medical isotopes.
Radioisotopes which may have better application in nuclear medicine but are not available due to low production rates accompanied with high cost (see, for example, Ref.\,\cite{Habs2011_Med_Isotopes}) might become accessible with the advent of the GF. A number of isotopes useful for medicine \cite{SZPUNAR2013_Medical_Isotopes} can be produced via  $(\gamma,n)$ reactions; see the examples given in Table\,\ref{tab:Isotope_(g,n)}.

We present an example of producing the $^{99}\textrm{Mo}$ medical isotope at the GF via the GDR. The GDR in heavy nuclei have resonance energies $E_0\approx77\times A^{-1/3}\approx14\,$MeV with widths $\Gamma_{tot}\approx23\times A^{-1/3}\approx5\,$MeV.
Here we can tune the maximal energy of secondary photons at GF to $E_\gamma=E_0+\Gamma_{tot}\approx19\,$MeV and use photons emitted within the $1/\gamma$ cone, i.e., photons with energy between 9.5\,MeV and 19\,MeV (see Sec.\,\ref{subsec:The Gamma Factory}), by suitable collimation.
The average background photon-attenuation cross section
in this energy range is $\sigma_{bg}\approx6\,$b, dominated by electron-positron pair production ($\approx4.4\,$b) and Compton scattering off electrons ($\approx1.7\,$b) \cite{Zyla2020_Particle_Data,Berestetskii_QED}.
The peak cross section of the $^{100}\textrm{Mo}(\gamma,n)^{99}\textrm{Mo}$ reaction is $\sigma_0\approx0.15\,$b \cite{SZPUNAR2013_Medical_Isotopes}.
Therefore, the maximal production rates of the $^{99}\textrm{Mo}$ isotope can be estimated as
\begin{equation}
    p\approx j\frac{\Gamma_{tot}}{E_0+\Gamma_{tot}}\frac{\sigma_0}{\sigma_0+\sigma_{bg}}\approx6\times10^{14}\,\textrm{s}^{-1}.
\end{equation}
This production rate is reached when almost all photons (after collimation) are absorbed in the target. Therefore, the thickness of an enriched $^{100}$Mo target or multiple thin targets should be greater than the absorption length $l\approx1/(\sigma_{bg}n)\approx2.8\,$cm, where $n$ is the number density of $^{100}\textrm{Mo}$ in the target with a density of approximately 10\,g/cm$^3$.
After a week of irradiation, $^{99}\textrm{Mo}$ isotopes with activities of $\approx500$\,TBq can be produced. Such activities are a significant improvement compared to those obtained at other facilities
including some nuclear reactors
\cite{Jang2017_99Mo_elec_accel,Boschi2019_99Mo_Tc}.
The worldwide demand of $^{99}$Mo isotopes is approximately 9000 6-day Ci $^{99}$Mo per week \cite{99Mo_supply}, where 6-day Ci refers to activities measured 6 days after the end of target processing. Assuming one day for the target processing after the one-week irradiation, $\approx2300$ 6-day Ci $^{99}$Mo per week, nearly one quarter of the global supply, can be obtained at the GF.
The $^{99}$Mo isotope having a half-life of $T_{1/2}\approx65.9\,$h decays via $\beta^-$ to $^{99m}$Tc ($T_{1/2}\approx6.0\,$h). Therefore, $^{99}$Mo isotopes are mainly used for the production of the $^{99m}$Tc, the most frequently used isomer in nuclear medicine.
More examples of producing shorter-lived isomers or isotopes from decay of longer-lived mother isotopes, so-called generators, are listed in Ref.\,\cite{Habs2011_Med_Isotopes}.
The worldwide demand for $^{99}$Mo is nowadays mainly satisfied by a small number of nuclear reactors via the $^{235}\textrm{U}(n,f)^{99}\textrm{Mo}$ reaction \cite{Boschi2019_99Mo_Tc}. Some nuclear reactors have suffered shutdown for maintenance or breakdowns, leading to shortages or interruptions in the supply of $^{99m}$Tc. Therefore, looking for alternative methods of producing $^{99}\textrm{Mo}$ isotopes has become important.

Production rates for other medical isotopes such as $^{192}$Ir and $^{196}$Au of $\approx10^{15}$ per second can analogously be derived. Similar numbers of neutrons are also produced from the $(\gamma,n)$ reaction, thus offering tertiary neutron beams; see also Sec.\,\ref{subsusbsec:Neutron_radioact_ion_sorces}.
One convenience of exploiting the GDR is that we may use the same experimental setup for producing many different medical isotopes, since the dependence of resonance energy and width on $A$ is weak, both proportional to $A^{-1/3}$. Multiple thin targets consisting of different atoms can also be used to produce various isotopes simultaneously. 

It is noted that specific activity
is an important quality criteria of medical isotopes. To reach high specific activity via photoproduction, photons with high flux density $\Phi$ (not just high fluxes) are required, which can be estimated from $\sigma\Phi\approx\ln{2}/T_{1/2}$ \cite{Habs2011_Med_Isotopes}, where $\sigma$ is the reaction cross section of transmuting the target isotope into the product isotope and  $T_{1/2}$ is half-life of the product isotope. Since the secondary GF photons with high fluxes are mainly emitted within a small cone (see Fig.\,\ref{Fig:Gamma_Factory_concept}), high flux density can be obtained by putting a target close to the gamma source. When high specific activities are required, one may need to use thin targets to avoid significant photon attenuation.
Besides using photon beams with high flux density, high specific activity can also be acquired via effectively separating and concentrating the product isotope, for example, through the magnetically activated and guided isotope separation method (MAGIS) \cite{Mazur2014dMAGIS}, which can also be applied to preparation of targets containing enriched isotopes.

Not only isotopes, but also nuclear isomers would be copiously produced at the GF via $(\gamma,n)$ reactions, since nuclei will partially decay to isomeric states after neutron emission.
The isomer ratio, i.e., ratio of produced nuclei in the isomeric state to those in the ground state, has been studied with many nuclear reactions including $(\gamma,n)$ via GDR (see, for example, Ref.\,\cite{Thiep2017Isomer_ratio}) using bremsstrahlung photons.
The knowledge of isomer ratios of different isotopes and their energy dependences could help to reveal the nuclear structure and mechanism of nuclear reactions. But currently available data are scarce and have significant discrepancies \cite{Thiep2017Isomer_ratio}. Secondary photons at the GF would offer significant improvement in measuring such data compared with bremsstrahlung photons, as a result of better energy resolution and higher photon fluxes.

\subsection{Production of nuclear isomers via (\texorpdfstring{$\gamma,\gamma'$}{g,g'}) reactions}
\label{subsec:prod_isomers_g_gp}

Radioactive isomers relevant to medicine can be produced via $(\gamma,\gamma')$ reactions. Using the highly monochromatic GF photons, one can selectively excite transitions from the stable or long-lived nuclear ground state to certain higher-energy levels serving as gateway states which will partially decay to the isomeric state directly or by cascade, which is similar to the isomer-depletion process discussed in Sec.\,\ref{sbsec:Low_En_Nucl_Trans}; see Fig.\,\ref{fig:93Mo}.
Table\,\ref{tab:Isomer_(g,g')} lists some isomers relevant for medicine, together with candidate low-lying gateway levels.

To give an example, production of the $^{115m}$In can be realized by selectively exciting the 1078\,keV transition with a radiative width $\Gamma_{rad}\approx4.6\times10^{-4}\,$eV and a branching ratio $f\approx0.16$ of decay to the isomeric state \cite{NNDC}. Doppler broadening of this transition leads to $\Gamma_{tot}\approx1\,$eV at room temperature. The photon-attenuation background is dominated by Compton scattering as $\sigma_{bg}\approx10\,$b is much larger than the effective cross section 
$\sigma_{eff}\approx\sigma_0\Gamma_{rad}/\Gamma_{tot}\approx0.5\,$b with $\sigma_0\approx10^3\,$b being the resonant photon-absorption cross section. The production rate of $^{115m}$In can be approximated as 
\begin{equation}
\label{Eq:isomer_pro_rate}
    p\approx j\frac{\Gamma_{tot}}{E_\gamma}\frac{\sigma_{eff}}{\sigma_{eff}+\sigma_{bg}}f\approx j\frac{\Gamma_{\gamma}}{E_\gamma}\frac{\sigma_0}{\sigma_{bg}}f\approx10^9\,\textrm{s}^{-1}.
\end{equation}
Here, as before, $E_\gamma$ is the maximal energy of secondary photons at the GF; it is tuned to $E_\gamma\approx1078\,$keV in this case.

We note that for exciting narrow resonances in cases where
Doppler width is larger than the radiative width but the effective cross section $\sigma_{eff}$ is still larger than $\sigma_{bg}$, Doppler broadening could allow more photons ($\Gamma_{tot}/E_\gamma\approx10^{-6}$ at room temperature) to be resonantly absorbed leading to higher nuclear isomer production rates as $p\approx jf \Gamma_{tot}/E_\gamma$ [see Eq.\,\eqref{Eq:isomer_pro_rate}].

\begin{table*}[t]
    \centering
    \begin{tabular*}{\textwidth}{@{\extracolsep{\fill}} cccccccccccl}
    \hline
    \hline
        Z  &$^{A}$X  &$^{Am}$X  &$I_g^{P}$  &$E_e(\textrm{keV})$  &$I_e^{P}$  &$E_m(\textrm{keV})$  &$I_m^{P}$  &$T^m_{1/2}$  & $E_{d}(\textrm{keV})$  &
       $\lambda L$  &Decay mode\\
        \hline \\[-0.2cm]
        43  &$^{99}$Tc  &$^{99m}$Tc  &9/2$^+$  &
        1207.26 &(7/2$^-$)  &142.683   &1/2$^-$  &6.0\,h  &2.173  &$E3$  &$\%\textrm{IT} = 99.996$\\
        &&&& &&&  &&142.63  &$M4$  &$\%\beta^-=0.004$\\
        45  &$^{103}$Rh  &$^{103m}$Rh  &1/2$^-$  &357.396  &5/2$^-$  &39.753  &7/2$^+$  &56.1\,min  &39.755  &$E3$  &$\%\textrm{IT} = 100$\\
        &&&&651.716  &(3/2)$^+$  && &&&&\\
        49  &$^{113}$In  &$^{113m}$In  &9/2$^+$  &1024.28  &5/2$^+$  &391.699  &1/2$^-$  &99.5\,min  &391.698  &$M4$  &$\%\textrm{IT} = 100$\\
        &&&&1131.48  &5/2$^+$  && &&&&\\
        49  &$^{115}$In  &$^{115m}$In  &9/2$^+$  &933.780  &7/2$^+$  &336.244  &1/2$^-$  &4.5\,h  &336.241  &$M4$  &$\%\textrm{IT} = 95.0$\\
        &&&&941.424  &5/2$^+$ && &&&&$\%\beta^-=5.0$\\
        &&&&1078.16  &5/2$^+$ && &&&&$\%\beta^-=5.0$\\
        68  &$^{167}$Er  &$^{167m}$Er  &7/2$^+$  &264.874  &3/2$^-$  &207.801  &1/2$^-$  &2.3\,s  &207.801  &$E3$  &$\%\textrm{IT}=100$\\
        &&&&531.54  &3/2$^+$  &&&&&&\\
        &&&&667.900  &(5/2)$^-$  &&&&&&\\
         &&&&745.32  &7/2$^-$  &&&&&&\\
        &&&&810.52  &(5/2)$^+$  &&&&&&\\
        77  &$^{191}$Ir  &$^{191m}$Ir  &3/2$^+$  &658.90  &(3/2$^-$)  &171.29  &11/2$^-$  &4.9\,s  &41.89  &$E3$  &$\%\textrm{IT} = 100$\\

    \hline
    \hline
    \end{tabular*}
    \caption{Examples of medical isomer production in  $^A$X($\gamma$, $\gamma'$)$^{Am}$X reactions that proceed via an intermediate excited state  of energy $E_e$ \cite{Habs2011_Med_Isotopes}. The indices ``g", ``e", ``m" denote the ground, intermediate excited and final isomeric states, respectively, for which the nuclear spin and parity  $I^P$ are provided. The isomeric state is characterized by the energy   $E_m$ and half-life $T^m_{1/2}$, with direct radiative decay via a gamma-ray of energy $E_d$ and multipolarity $\lambda L$. The last column presents decay modes of the isomeric states, which are for the listed cases either isomeric transition (IT), i.e., the transition to a lower-lying level of the same nucleus, or $\beta^-$ decay. The data are from Refs.\,\cite{NNDC,Jain2015}.
   }
    \label{tab:Isomer_(g,g')}
\end{table*}

\begin{table*}[t]
    \centering
    \begin{tabular*}{\textwidth}{@{\extracolsep{\fill}}l c c c c c c c l}
    \hline
    \hline
    Z  &$^{A}$X  &$^{Am}$X  &$E_m$(keV)  &$I_m^{P}$  &$T_{1/2}^m$  &$E_{d}$\,(keV)  &$\lambda L$  &Decay mode\\
    \hline \\[-0.2cm]
        17& $^{34}$Cl & $^{34m}$Cl & 146.36(3) & 3$^+$ &  32.0\,min & 146.36(3) & $M3$  & \%IT = 44.6\\
        &&&&&&&&$\%\varepsilon+\%\beta^+=55.4$\\
        25& $^{52}$Mn & $^{52m}$Mn & 377.749(5) & 2$^+$ &  21.1\,min & 377.748(5) & $E4$ & \%IT = 1.78\\
        &&&&&&&&$\%\varepsilon+\%\beta^+=98.22$\\
        35& $^{80}$Br & $^{80m}$Br & 85.843(4) & 5$^-$ & 4.42\,h & 48.786(5) & $M3$ & \%IT = 100\\
        36& $^{81}$Kr & $^{81m}$Kr & 190.64(4) & 1/2$^-$ & 13.1\,s & 190.46(16) & $E3$ & \%IT = 99.9975\\
        &&&&&&&&$\%\varepsilon=2.5\times10^{-3}$\\
        43& $^{94}$Tc & $^{94m}$Tc & 76(3) & (2)$^+$ &  52.0\,min & 76(3)  &  & $\%\varepsilon+\%\beta^+=100$\\
        &&&&&&&& $\%\textrm{IT}< 0.1$\\
        50& $^{117}$Sn & $^{117m}$Sn & 314.58(4) & 11/2$^-$ & 14.0\,d & 156.02(3)  & $M4$ & $\%\textrm{IT} = 100$\\
        &&&&&&314.3(3)& &\\
        72& $^{178}$Hf & $^{178m}$Hf & 1147.416(6)& 8$^-$ & 4.0\,s & 88.8667(10)  & $E1$ & $\%\textrm{IT} = 100$\\
        78& $^{193}$Pt & $^{193m}$Pt & 149.78(4) & 13/2$^+$ &  4.3\,d & 135.50(3)  & $M4$ & $\%\textrm{IT} = 100$\\
        78& $^{195}$Pt & $^{195m}$Pt & 259.077(23) & 13/2$^+$ &   4.0\,d & 19.8  & $M4$ & $\%\textrm{IT} = 100$\\
        &&&&&&129.5(2)& &\\
        
    \hline
    \hline
    \end{tabular*}
    \caption{Further examples of isomers for medical applications \cite{Dracoulis2016}. In these cases, specific gateway levels for production of these isomers have not yet been identified. However, they may potentially be produced via the excitation of PDR or GDR. See Table\,\ref{tab:Isomer_(g,g')} for explanations of the notations. As additional decay modes, some of these isomers undergo electron capture $(\varepsilon)$ or $\beta^+$ decay.
    }
    \label{tab:Isomers_medicine}
\end{table*}
    
Alternatively, a broad excitation via the PDR or GDR with decay to the isomeric state can be pursued; see examples in Table \ref{tab:Isomers_medicine}. Isomer production via the GDR using $(\gamma,n)$ reactions is briefly discussed at the end of Sec.\,\ref{subsec:prod_isotopes_g_n}.
Photoproduction of isomers via $(\gamma,\gamma')$ reactions was investigated employing photons Compton back-scattered off relativistic electrons, and the
obtained cross sections are larger for photons in the PDR energy range compared to GDR \cite{pan2020photo}.
There exists a maximum cross section between 5 and 10\,MeV on the order of mb with a broad width on the order of MeV, which lies in the PDR energy range; see Sec.\,\ref{Subsect:pygmy}. It is demonstrated in Ref.\,\cite{pan2020photo} that production of $^{99m}$Tc, $^{103m}$Rh, $^{113m}$In, and $^{115m}$In isomers with activities of more than 10\,mCi can be realized by use of gamma rays from laser-electron Compton scattering with photon fluxes of $10^{13}$ per second and 6-hour irradiation. 
The achieved activity is generally sufficient for medical imaging but is still inadequate for therapeutic applications. With nearly four orders of magnitude improvement in photon fluxes at the GF, isomers with considerably higher activities can be produced thus meeting the requirement for therapeutic applications.
We can tune the maximal energy of secondary photons at the GF to a few hundred keV above the neutron-separation threshold to cover the PDR.
Using photons emitted within $\approx1/\gamma$ cone (so roughly 1/2 of all secondary photons, see Sec.\,\ref{subsec:The Gamma Factory}), production rates of these isomers could reach $\sim10^{13}$ per second.

Returning to the example of $^{115m}$In production discussed above, it can also be realized by use of photons in the PDR energy range, about 5 to 10 MeV.
In this energy range, the average background photon-attenuation cross section is $\sigma_{bg}\approx6\,$b and the integrated cross section of the $^{115}\textrm{In}(\gamma,\gamma')^{115m}\textrm{In}$ reaction is around $5\times10^{-3}$\,b\,MeV \cite{pan2020photo}. Following an estimate similar to that for isotope-production rates in Sec.\,\ref{subsec:prod_isomers_g_gp}, we find that production of $\approx10^{13}$ per second of $^{115m}$In can be achieved at the GF, when almost all photons (after collimation) are absorbed in a $^{115}\textrm{In}$ target or multiple thin targets with total thickness greater than the absorption length $l\approx4.4\,$cm.

In this case, the production rate of the isomer of interest is higher when using broadband excitation rather than the narrow resonance, by roughly four orders of magnitude. The isomer-production rate corresponds to activities of produced nuclear isomers after irradiating for a long time comparable to $T_{1/2}$ of the isomeric state.

Besides medical applications, nuclear  isomers  were  recently proposed as detectors for certain  kinds of dark matter which could induce collisional deexcitation of the isomers; see Table\,\ref{tab:Isomer_dark_matter} \cite{Pospelov2020_isomers}.
We note that isomers listed in Table\,\ref{tab:Isomer_dark_matter} are not the only candidates for dark matter search. The GF can produce a range of nuclear isomers with larger quantities accessible, including heavier isomers whose larger nuclear radii allow deexcitation to be induced by smaller momentum exchange during collision with dark matter particles \cite{Pospelov2020_isomers}.
Nuclear isomers have also been suggested for building gamma-ray lasers
(see Sec.\,\ref{Sec:Gamma_Lasers}).

\begin{table*}[t]
    \centering
    \begin{tabular*}{\textwidth}{@{\extracolsep{\fill}}l c c c c c c l}
    \hline
    \hline
    Z  &$^{A}$X  &$E_m$(keV)  &$I_m^{P}$  &$T^m_{1/2}$  &$E_{d}$  &$\lambda L$  &Decay mode\\
    \hline \\[-0.2cm]
        56& $^{137}$Ba & 661.659(3) & 11/2$^-$ & 2.55\,min & 378.0(4) & $E5$ & \%IT = 100\\
    &&&&&  661.657(3)&  $M4$& \\
        71& $^{177}$Lu & 970.175(3) & 23/2$^-$ & 160.4\,d & 115.868(3) & $E3$ & \%$\beta^-=78.6(8)$ \\
        &&&&&&&\%IT = 21.4(8)\\
        72& $^{178}$Hf & 2446.09(8) & 16$^+$ & 31\,y & 12.7(2) & & \%IT = 100\\
    &&&&&  309.50(15)&  $M4(+E5)$&\\
    &&&&&  587.0(1)&  $E5$&\\
    73& $^{180}$Ta & 77.1(8) & 9$^-$ & $>1.2\times10^{15}$\,y & & &\\
    
    \hline
    \hline
    \end{tabular*}
    \caption{Isomers for dark matter detection \cite{Pospelov2020_isomers,Jain2015}. See Table\,\ref{tab:Isomer_(g,g')} for explanations of the notations.
    }
    \label{tab:Isomer_dark_matter}
\end{table*}

Heating due to the high flux of gamma rays is expected to be a tractable problem with above reactions, since there are several applicable heat-dissipation techniques, for example, using a stack of multiple thin targets \cite{Habs2011_Med_Isotopes}.


\section{Induced gamma emission and gamma lasers}
\label{Sec:Gamma_Lasers}

Realization of stimulated photon emission in the gamma range has been an outstanding challenge to the community for many years. Many different proposals have been contemplated \cite{Baldwin_1981_Grasers}; however, only a handful of schemes can be considered realistic (and none have been realized up till now). The advent of the GF motivates an  examination of whether this facility may bring a gamma laser closer to reality.

There are two main groups of gamma-ray laser (graser) proposals. One suggests building a recoilless gamma-ray laser \cite{Baldwin_1981_Grasers,Baldwin1997} using nuclei incorporated in a crystal. The reason to use a crystal is that the M\"ossbauer effect may mitigate the issue of the Doppler broadening of the transition which could significantly suppress the stimulated emission cross section.
But this meets the so-called graser dilemma, where high intensity of pumping required for creating population inversion, i.e., more nuclei in the upper lasing state than in the lower lasing state, could be destructive to other conditions for stimulated emission gain, including the M\"ossbauer and Bormann effects. In order to relax the population inversion requirements, an approach was proposed \cite{Kocharovskaya1999,Kolesov2000Mossb} to suppress resonant absorption for the  M\"ossbauer nuclear transitions of the ions placed in a lattice. This  can  be done via coherent  optical  driving  of  electronic  transitions  of  these  ions and exploiting the hyperfine coupling between the nuclear and electronic degrees of freedom. These concepts are reminiscent of the physics discussed in Sec.\,\ref{Subsec:Interaction_nucl_atom_DF}. 
%
We note that M\"ossbauer effect works for transitions with energy below $\approx180\,$keV.
Candidate low-energy gamma transitions may be selected from a list in Ref.\,\cite{Agda1988}.

We highlight here two examples of graser proposals benefiting from the M\"ossbauer effect.
It has been proposed to build a graser based on collective excitations of $^{57}$Fe M\"ossbauer nuclei, also known as nuclear excitons, in a nuclear forward scattering setup \cite{Brinke2013}.
There is also a proposal for a VUV laser based on  the 8\,eV nuclear transition between the ground and isomeric states of $^{229}$Th ions doped in a VUV transparent crystal \cite{Tkalya2011Th_las}. While pumping of the isomer can be done with a conventional VUV laser, Zeeman splitting induced by an external magnetic field or electric quadrupole splitting in the crystal result in population inversion for the lasing transition between specific hyperfine sublevels. 

\begin{figure}[!htpb]\centering
    \includegraphics[width=1.0\linewidth]{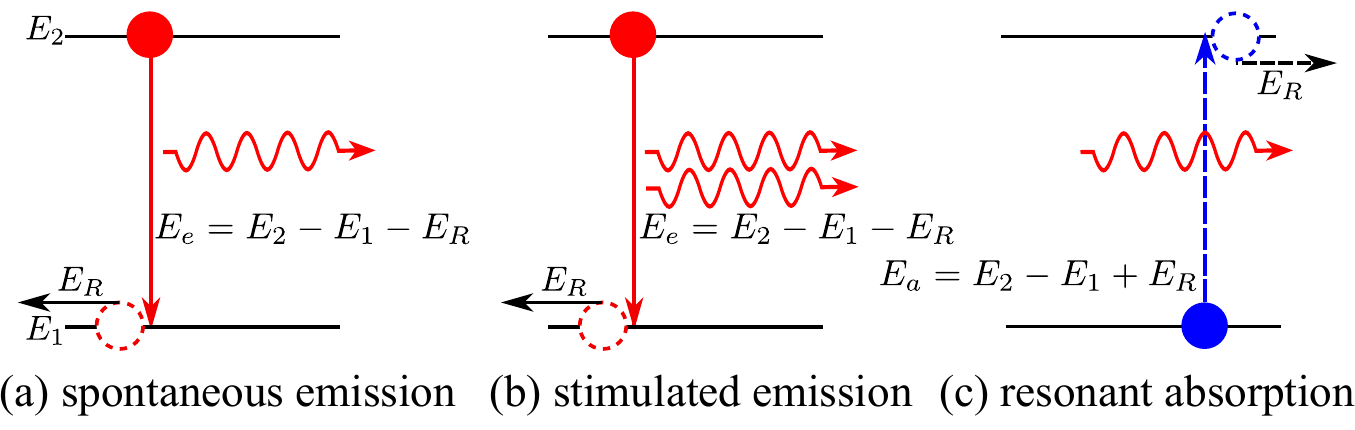}
    \caption{The influence of recoil on stimulated emission and resonant absorption. $E_1$ and $E_2$ are the level energies. The resonant gamma-ray energy for spontaneous and stimulated emission $E_e$ is lower than $E_2 - E_1$ by the recoil energy $E_R$, while the gamma-ray energy required for resonant absorption $E_a$ is higher than the internal energy difference, again, by the recoil energy. Thus, the emitted photons are detuned from resonance and are not readily absorbed.}
    \label{fig:graser_recoil}
\end{figure}

The other group of proposals suggest building a graser with the assistance of recoil \cite{Rivlin1999,Rivlin2007_graser}, which can eliminate the requirement of population inversion due to reduced overlap between the photon-absorption lineshape and photon-emission lineshape. The idea is that if the recoil energy $E_R$ is larger than the Doppler width of the transition, emitted photons would not be absorbed by the nucleus in the ground (initial) state, but they would still stimulate deexcitation of an excited nucleus; see Fig.\,\ref{fig:graser_recoil}. However, implementation of this scheme requires cooling to suppress Doppler broadening
in order to get acceptable stimulated emission gain.
In the two-level pumping scheme with hidden population inversion \cite{Rivlin2007_graser},
the photon intensity required for pumping is estimated to be $\geq10^{30}$\,ph/(cm$^{2}$\,s\,keV) using samples cooled down to 10\,$\mu$K \cite{Zadernovsky2007_graser_hidden_inversion}.

\begin{figure}[!htpb]\centering
    \includegraphics[width=0.95\linewidth]{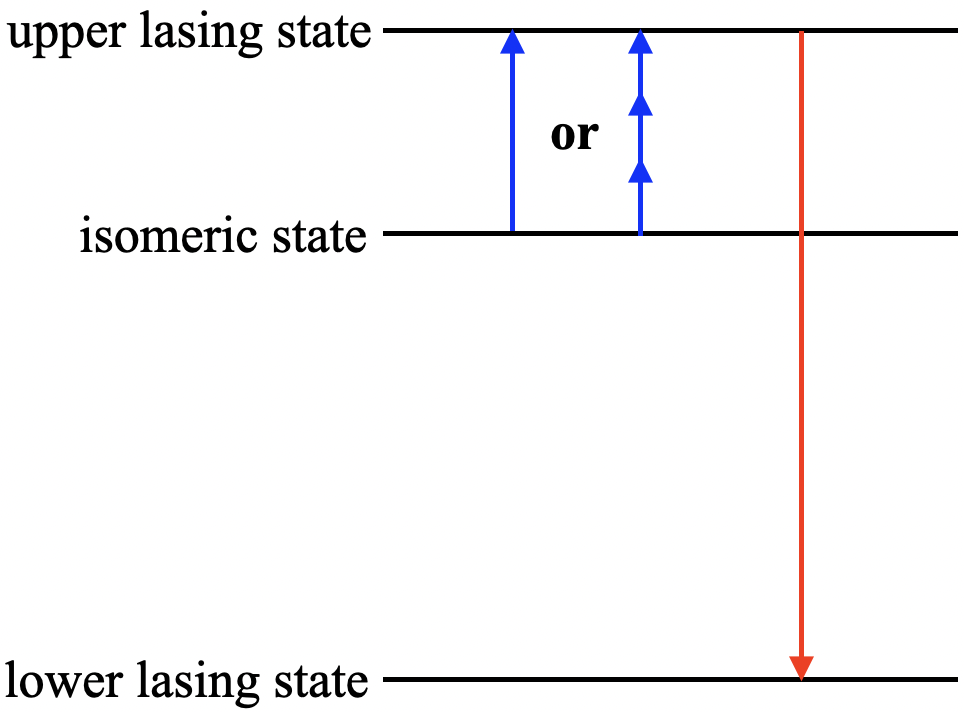}
    \caption{
    A gamma-ray laser scheme \cite{Baldwin1997}. After a large number of isomers are produced, the isomeric state is excited to an upper lasing state via a single-photon or multiphoton process indicated by the blue arrows. Deexcitation from the upper lasing state to the lower lasing state is shown as the red arrow.
    \label{fig:graser_En}
    }
\end{figure}

\begin{figure}[!htpb]\centering
    \includegraphics[width=0.95\linewidth]{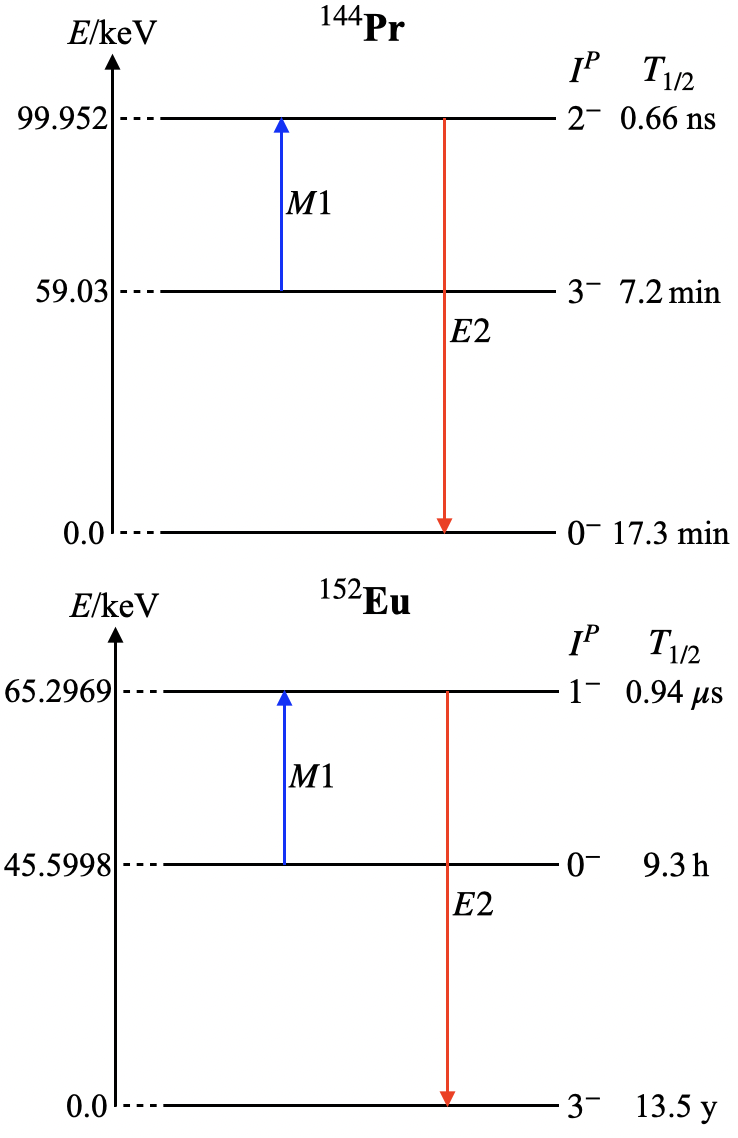}
    \caption{Examples of
    nuclear isomers for
    a prospective gamma-ray laser; see also Fig.\,\ref{fig:graser_En}.}
    \label{fig:graser_eg_PrEu}
\end{figure}

Many graser proposals involve using nuclear isomers; see, for example, Table\,\ref{tab:Isomer_graser}.
A detailed compilation of nuclear isomers with half-lives $\geq$ 10\,ns is provided in Ref.\,\cite{Jain2015}, though possible usages of these isomers are not underlined. Nuclei potentially valuable for realizing a gamma-ray laser based on the concept of hidden population inversion are tabulated in Ref.\,\cite{Zadernovsky2005}.
High photon fluxes from the GF will help produce copious amount of candidate isomers. High tunability and high resolution of the gamma beams could help characterize the samples during the isomer-separation process before incorporating concentrated isomers into a crystal.
The GF may also assist in finding optimal candidate nuclei such as nuclei with small energy difference between the isomeric state and an upper lasing state, where an X-ray or even optical laser can be used for pumping, see the energy level structure suggested in Fig.\,\ref{fig:graser_En}. Here, an isomeric state with a lifetime long enough to enable accumulation of isomers and their incorporation into a host crystal is used. The isomer is then excited via a single-photon or a multiphoton laser-driven transition to an upper lasing state. A likely difficulty with the latter is that multiphoton transitions are strongly suppressed; see Sec.\,\ref{Subsect:GDR-multi}.
Two examples,  $^{144}$Pr and $^{152}$Eu, exhibiting such energy level structures are shown in Fig.\,\ref{fig:graser_eg_PrEu}.
Here, the required excitation energy is relatively low (about 40\,keV for $^{144}$Pr and 20\,keV for $^{152}$Eu); photons at this energy may soon become available at  XFEL facilities. However, the main drawback remains that such transitions have narrow radiative widths of the upper lasing state. This holds particularly true for $^{152}$Eu. Therefore, it may be difficult to realize effective pumping from the isomeric state to the upper lasing state.

For graser schemes where the required transition energy cannot be reached by X-ray lasers,
the GF may be utilized for pumping, such as transferring isomers to an upper lasing state.
We note that secondary photons with total fluxes of $\approx10^{17}$ photons per second, uniformly distributed over the entire energy spectrum, at the GF consist of pulses having fluxes of $10^{10}$ photons per pulse at a repetition rate of $\approx10\,$MHz. The pulse duration is typically $\approx100$\,ps and can be reduced to $30\,$fs limited by the transverse size of the ion beam.
The strong energy-angle correlation of GF photons, see Sec.\,\ref{subsec:The Gamma Factory}, may allow application of the Borrmann effect \cite{Baldwin1997}, i.e., minimized photon absorption when photons are incident at special angles satisfying the Bragg law, to alleviate heating problems.

\begin{table*}[t]
    \centering
    \begin{tabular*}{\textwidth}{@{\extracolsep{\fill}}l c c c c c c l}
    \hline
    \hline
    Z  &$^{A}$X  &$E_m$\,(keV)  &$I_m^{P}$  &$T^m_{1/2}$  &$E_{d}$  &$\lambda L$  &Decay mode\\
    \hline \\[-0.2cm]
        27& $^{60}$Co & 58.59(1) & 2$^+$ & 10.5\,min & 58.603(7) & $M3+(E4)$ & \%IT = 99.75(3)\\
        &&&&&&&\%$\beta^-$ = 0.25(3)\\
        70& $^{169}$Yb & 24.1999(16) & 1/2$^-$ & 46\,s & 24.20(2) & $E3$ & \%IT = 100\\
        72& $^{178}$Hf & 2446.09(8) & 16$^+$ & 31\,y & 12.7(2) & & \%IT = 100\\
    &&&&&309.50(15)&$M4(+E5)$&\\
    &&&&&587.0(1)&$E5$&\\
        72& $^{177}$Hf & 2740.02(15) & 37/2$^-$ & 51.4\,min & 214 & $E3$ &\%IT = 100\\
        72& $^{179}$Hf & 1105.74(16) & 25/2$^-$ & 25.0\,d & 21.01(12) & $M2$ & \%IT = 100\\
        &&&&& 257.37(15) & $E3$ &\\
        73& $^{180}$Ta & 77.1(8) & 9$^-$ & $>1.2\times10^{15}\,$y & & & \\
        77& $^{192}$Ir & 56.720(5) & 1$^-$ & 1.45\,min & 56.71(3) & $E3$ & \%IT = 99.9825\\
        &&&&&&&\%$\beta^-$ = 0.0175\\

    \hline
    \hline
    \end{tabular*}
    \caption{Isomers potentially useful for gamma-ray lasers \cite{Collins97,Balko88}. See Table\,\ref{tab:Isomer_(g,g')} for explanations of the notations.
    }
    \label{tab:Isomer_graser}
\end{table*}

To summarize, the GF can help facilitate research on induced emission and lasing with nuclear transitions, although, a clear path to a graser does not seem to be in sight as yet.

\section{Gamma polarimetry}
\label{Sec:Gamma_pol}

\subsection{Polarimetry with narrow resonances}
\label{subsec:Polarimetry_with_narrow_res}

Narrow--band GF photons in combination with polarized nuclear--target technology open a possibility to perform polarimetry using the target as a polarization filter. To this end, the energy of the GF photons should be tuned in resonance with a gamma transition in the target nucleus.

Suitable transition can be found in the $^{13}$C ($I^P=1/2^-$ for the stable ground state; see also Fig.\,\ref{fig:C13_En}), which can be made into a polarized target \cite{Budker2012PolTar}. For example, there is an $E1$  transition to an $I^P=1/2^+$ state at 3.09\,MeV that has a radiative width of 0.43\,eV and radiatively decays back to the ground state. There is also an $M1$ transition to an excited $I^P=1/2^-$ state at 8.86\,MeV that has a width of 150\,keV. The upper state of this transition lies above the neutron-separation energy for $^{13}$C (4.9\,MeV) and  predominantly decays by neutron emission to the ground state of $^{12}$C (the ground state is the only possible state of the resultant $^{12}$C because the excited states of this nucleus lie sufficiently high).

The resonant cross section for a closed transition is (cf.\,\,Eq.\,\ref{Eq:res_cross_section})
\begin{equation}
    \sigma \simeq \frac{\lambda'^2}{2\pi} ,
 \label{Eq:Resonant_cross section}
 \end{equation}
 where $\lambda'$ is the wavelength in the ion frame.
For the 3.09\,MeV $E1$ transition,
$\sigma\approx$260\,b (1\,b=10$\,^{-24}$\,cm$^2$) is large enough that it is feasible to construct a target comprising multiple absorption lengths for resonant photons. The expected asymmetry for absorption of circularly polarized photons on a $1/2\rightarrow 1/2$ transition is 100\%, and thus such a system can be an efficient circular analyzer, perhaps only limited in precision by the counting statistics of the gamma rays.
One should note in this respect that the  resonance will be narrower than the spectral width of the gamma beam, so only a fraction of the photons will, in fact, be resonant, see Appendix\,\ref{Subsec:Appendix:Gamma_Res_13C}.

This should not be a problem
for the broader 8.86\,MeV $M1$ resonance; however, the resonant cross section is some five orders of magnitude smaller because the width is dominated by
decay via neutron emission rather than gamma transition and because  of the  smaller photon wavelength.

Due to these factors, polarimetry may need to rely, instead of photon-transmission measurements, on the detection of reaction products (neutrons or non-collinear photons).  
\subsection{Other polarimetric techniques}

Polarization measurements of gamma and X-rays have a long tradition in many branches of modern physics. For example, a number of polarization--sensitive studies were performed in astrophysics \cite{WeC76, Ilie2019GammaPol}, atomic, nuclear and plasma physics \cite{ScM94,WeB10,FuK97}. Depending on the energy of the photons, the photoelectric effect, Compton effect, and electron--positron pair production are typically employed in these experiments. In addition, in the recent years, Bragg scattering was successfully used for high--precision polarization measurements in the keV energy range \cite{MaS13,PhysRevResearch.2.023365,Schmitt:2020ttm}.  

During the last two decades, a number of experiments were performed also to study \textit{linear} polarization of photons, emitted in relativistic collisions at ion storage rings \cite{TaS06,WeB10,TaB13}. These measurements usually employ the polarization sensitivity of Compton scattering of light by (quasi--free) electrons. Namely, the angular distribution of Compton--scattered photons is defined by the direction of the linear polarization of incident radiation, as predicted by the well--known Klein--Nishina formula \cite{TaS06}. In order to measure this angular distribution and, hence, to determine the linear polarization one usually employs solid--state position sensitive detectors. These Compton detectors allow (linear) polarization measurements in the energy range of about 10\,keV to 10\,MeV \cite{WeB15}. Above 10\,MeV, pair-production polarimetry can be used \cite{Eingorn2018HEPolarimetry,Chattopadhyay:2021mbb}.

In contrast to linear polarization measurements, less progress has been made in the development of detectors for the measurement of circular polarization of X- and gamma rays. Due to the absence of a polarimetry technique which could be combined with gamma-ray imaging, for example, no circular polarization measurements have been reported in gamma-ray astronomy. In atomic and nuclear physics, however, the experimental approach based on Compton scattering off magnetized solid targets has been applied during the last decades. Moreover, a combined measurement of Compton scattered photons and subsequent bremsstrahlung of the recoiled electron has been recently proposed as a novel approach to circular polarimetry \cite{Tas11}. It is expected that this novel approach will allow accurate studies of circular polarization of X- and gamma rays with energies up to several tens of MeV.

The $\Delta(1232)$ resonance is a dominant feature of the nucleon-excitation spectrum for photon energies $\omega\approx300$\,MeV (in the photon-proton center of momentum frame). It can be used to analyze the circular photon polarization. It is common to describe the photoexcitation of $\Delta(1232)$ by the value of the resonance photoabsorption amplitudes with the parallel and the antiparallel helicity configuration, $A_{3/2,1/2}$. 
    For these, Ref.\,\cite{Zyla2020_Particle_Data} gives the values $A_{3/2}=-0.255(7)$\,GeV$^{-1/2}$ and $A_{1/2}=-0.135(7)$\,GeV$^{-1/2}$~\cite{Zyla2020_Particle_Data}. 
    Note that the uncertainties are relatively large due to averaging over different theoretical analyses which individually have much smaller uncertainties, comfortably within 2-3\% for $A_{1/2}$ and about half that value for $A_{1/2}$. The model dependence stems from the separation of the experimental data into the resonant and the non-resonant parts, a procedure that is model-dependent.
    Their ratio amounts to 
    $|A_{3/2}|/|A_{1/2}|\approx1.89$, resulting in the asymmetry $\frac{|A_{3/2}|^2-|A_{1/2}|^2}{|A_{3/2}|^2+|A_{1/2}|^2}\approx0.56$. The helicity-dependent total cross section integrated over the full spectrum is relevant for the GDH sum rule, see Sec.\,\ref{Subsec:Photopys_protons_light_nucl}.

Development of precision polarimetry and spectropolarimetry (a combination of polarimetry and spectroscopy) at the GF, would open novel possibilities for fundamental-physics experiments such as studying PV in the vicinity of nuclear gamma transitions (Sec.\,\ref{sec:P-violation}) and measurement of vacuum-birefingence effects (Sec.\,\ref{Sec:QuantVacEffects}).


\section{Quantum vacuum effects}
\label{Sec:QuantVacEffects}

\begin{figure}[h]
 \includegraphics[width=0.5\linewidth]{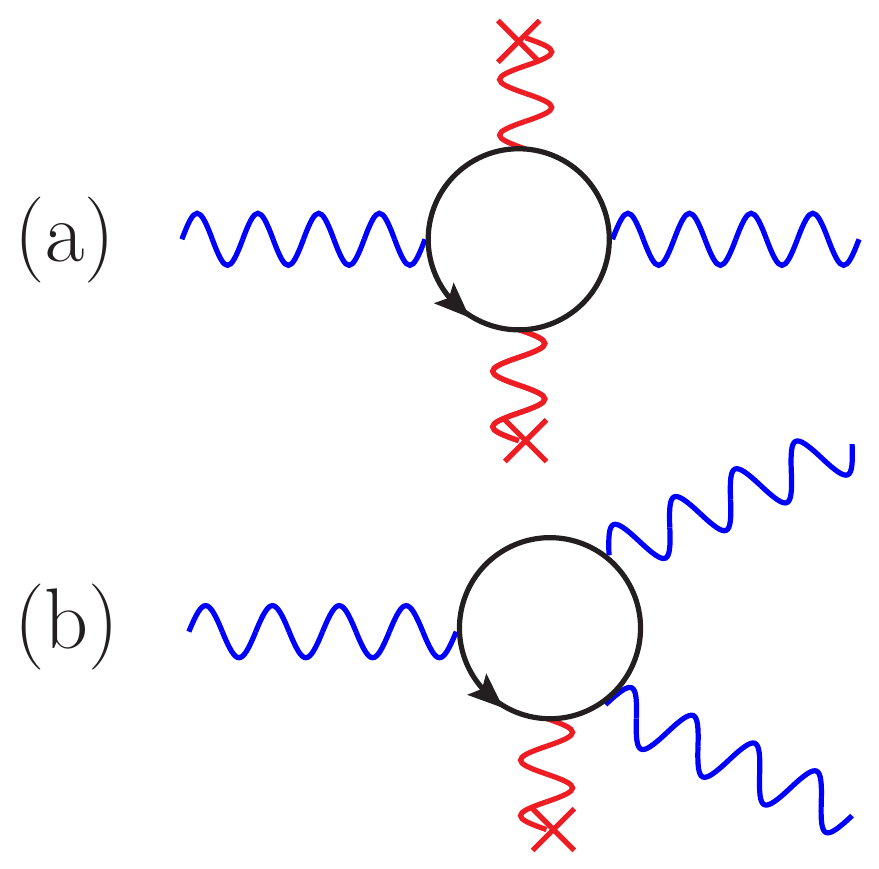}
 \caption{Lowest-order Feynman diagrams of typical quantum vacuum effects, giving rise to (a) vacuum birefringence and (b) photon splitting. 
 The blue wiggly lines are probe photons and the red wiggly lines ending at crosses denote couplings to the prescribed electromagnetic field.}
 \label{fig:vacuum_diags}
\end{figure}

The quantum vacuum amounts to a highly nontrivial state, characterized by the omnipresence of fluctuations of virtual particles.
While the microscopic theory of quantum electrodynamics (QED) does not provide for a direct tree-level interaction among photons, quantum vacuum fluctuations induce effective nonlinear couplings among electromagnetic fields \cite{Halpern:1933dya,Euler:1935zz,Heisenberg:1935qt,Schwinger:1951nm}.
At zero field, the quantum vacuum is characterized by translational invariance and the absence of any preferred direction. 
Conversely, an external electromagnetic field generically introduces a preferred direction for charged particles, and, if inhomogeneous, also breaks translational invariance.
Via the charged particle-antiparticle fluctuations coupling to the external electromagnetic field, this preferred direction can also impact probe photon propagation and give rise to nonlinear QED effects such as vacuum birefringence and photon splitting; see Fig.\,\ref{fig:vacuum_diags} for the corresponding Feynman diagrams at one-loop order. See also the recent reviews~\cite{King:2015tba,Karbstein:2019oej} and references therein.

Here, we briefly discuss the perspectives of studying quantum vacuum effects at the Gamma Factory, using the prominent signatures of vacuum birefringence (in magnetic and laser fields) and photon splitting (in atomic fields) as illustrative examples. Both effects arise from an effective four-photon interaction mediated by a electron-positron fluctuation \cite{Karplus:1950zz,Karplus:1950zza,Costantini:1971cj,DeTollis:1965vna}; cf. Fig.~\ref{fig:vacuum_diags}. The former scales quadratically and the latter linearly with the background field. Besides, we comment on quasi-elastic photon scattering.

\subsection{Vacuum birefringence}
\label{subSec:QuantVacEffects}

Linearly polarized probe photons (energy $\omega$) traversing a strong pump field ($\vec{\cal E}$, $\vec{\cal B}$) can pick up an ellipticity if their polarization vector has a nonvanishing overlap with the two distinct polarization eigenmodes imprinted on the quantum vacuum by the pump field \cite{Toll:1952rq}. See Fig.\,\ref{fig:vacuum_diags}(a) for the corresponding Feynman diagram.

So far, this fundamental effect induced by quantum vacuum fluctuations has never been directly verified in laboratory experiments using macroscopic fields; see the review\,\cite{Battesti:2018bgc} and references therein.
Typical proposals for measuring vacuum birefringence in a laboratory experiment envision the effect to be induced either by (A) quasi-constant static magnetic field, or (B) a counter-propagating high-intensity laser pulse. 

Given that the following conditions hold \cite{Karbstein:2021},
\begin{equation}
\left\{\Bigl(\frac{e{\cal E}}{m_e^2}\Bigr)^2,
\Bigl(\frac{\Omega}{m_e}\Bigr)^2,\frac{\omega\Omega}{m_e^2},
\Bigl(\frac{e{\cal E}\omega}{m_e^3}\Bigr)^2\right\}\ll1,
\label{eq:VBconditions}
\end{equation}
where $\hbar=c=1$, the vacuum-birefringence phenomenon can be studied on the basis of the leading contribution to the renowned Heisenberg-Euler effective Lagrangian \cite{Euler:1935zz,Heisenberg:1935qt}.
Here, ${\cal E}={\rm max}\{|\vec{\cal E}|,|\vec{\cal B}|\}$, $\Omega$ denotes the typical frequency scale of variation of the pump field, $e$ is the elementary charge and $m_e$ is the electron mass.

In the same parameter regime, the polarization-sensitive absorption coefficients associated with the principle possibility of electron-positron pair production in the presence of the background field are exponentially suppressed with $m_e^2/(e{\cal E})\gg1$ for $\omega\ll m_e$ and $m_e^3/(e{\cal E}\omega)\gg1$ for $\omega\gg m_e$; cf., e.g., Refs.~\cite{Baier:2007dw,Karbstein:2013ufa}.

\subsubsection{Quasi-constant magnetic field}\label{subsec:birefringence_B}

First, we consider the effect of vacuum birefringence induced by a quasi-constant static magnetic field $\vec{\cal B}$. In this case, probe photons polarized parallel $\parallel$ (perpendicularly $\perp$) to the plane spanned by their wave vector and the direction of $\vec{{\cal B}}$ experience different refractive indices.
Given that the conditions~\eqref{eq:VBconditions} are met, for probe photons propagating perpendicularly to $\vec{\cal B}$ these refractive indices are given by \cite{Baier:1967zzc}
\begin{equation}
 \left\{\begin{array}{c}
         n_\parallel \\ n_\perp
        \end{array}\right\}
 \simeq 1 + \frac{\alpha}{\pi}\Bigl(\frac{e{\cal B}}{m_e^2}\Bigr)^2\frac{1}{90}
\left\{\begin{array}{c}
         7 \\ 4
        \end{array}\right\}\,, \label{eq:nparallelperp}
\end{equation}
with fine-structure constant $\alpha\simeq1/137$.

\begin{figure}
 \includegraphics[width=1\linewidth]{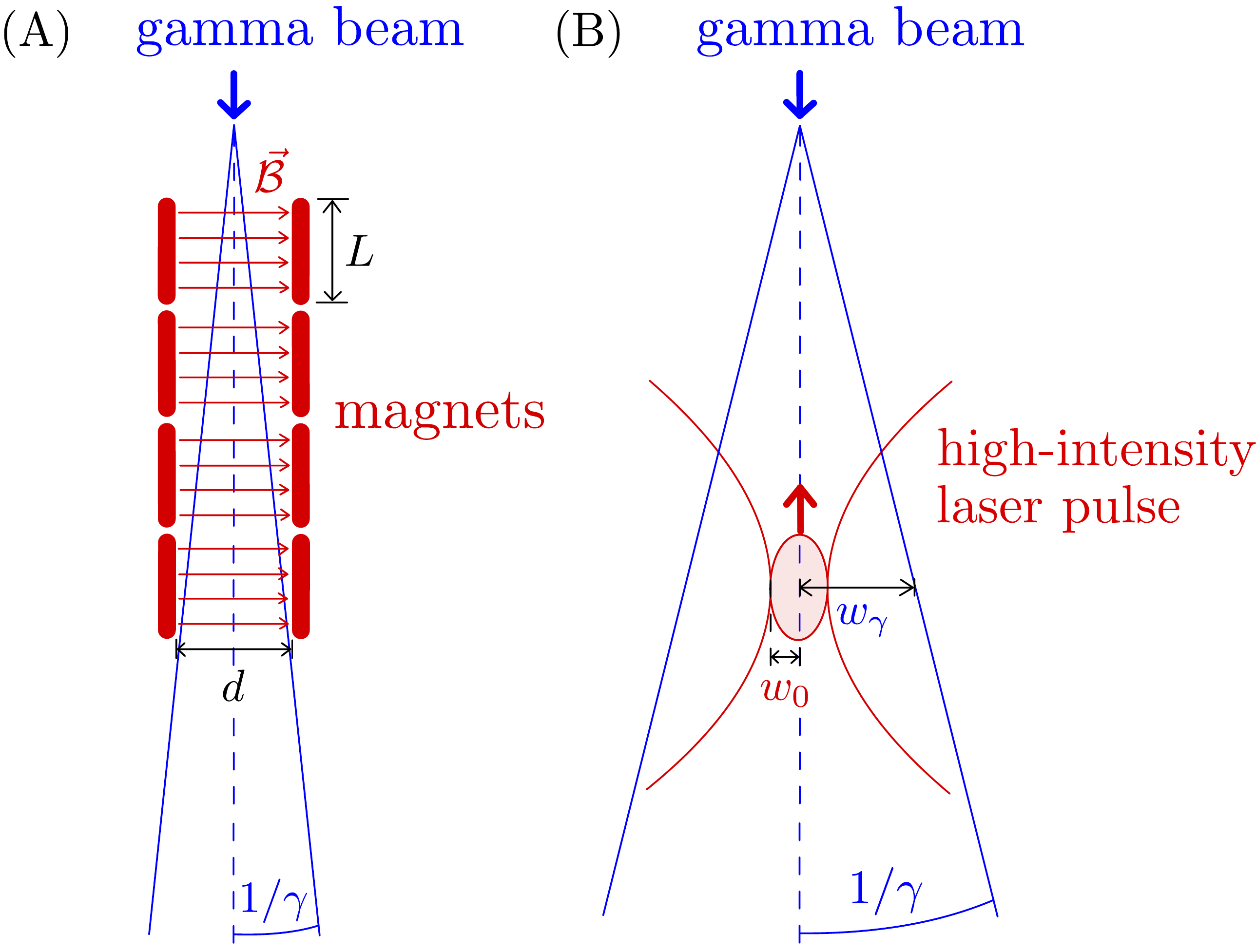}
 \caption{Graphical depiction of the two scenarios envisioned for the detection of QED vacuum birefringence at the Gamma Factory as discussed in the text. In scenario~(A) the birefringence phenomenon is induced by the quasi-constant static magnetic field provided by LHC dipole magnets. In scenario~(B) the effect is driven by a counter-propagating focused high-intensity laser pulse.}
 \label{fig:birefringence_variants}
\end{figure}

Letting initially linearly polarized photons having an equal overlap with both polarization eigenmodes ($\parallel$,$\perp$) traverse a constant magnetic field extending over a length $l$, an ellipticity characterized by the phase difference $\Phi=\omega l(n_\parallel - n_\perp)$ between the two polarization components is induced. 
The explicit expression for the accumulated phase difference is \cite{ERBER1961}
\begin{equation}
\Phi\simeq\omega l\,\frac{\alpha}{30\pi}\Bigl(\frac{e{\cal B}}{m_e^2}\Bigr)^2\,. \label{eq:ellipticity_B}
\end{equation}
Equation~\eqref{eq:ellipticity_B} shows that the effect scales quadratically with the magnetic field strength and linearly with the energy of the probe photons and their propagation distance in the magnetic field. At the same time, the birefringence property of the polarized quantum vacuum results in signal photons scattered into a mode polarized perpendicularly to the incident probe light: for the case of a static quasi-constant magnetic field as considered in this subsection, the number of polarization flipped signal photons is given by $N_\perp\simeq(\Phi/2)^2N$, where $N$ denotes the number of gamma photons available for probing the effect.

The GF will enable such a vacuum-birefringence experiment with a probe-photon energy as high as $\omega\simeq400\,{\rm MeV}$. The driving magnetic field could be provided by a sequence of LHC dipole magnets, delivering a magnetic field of ${\cal B}\simeq8.3\,{\rm T}$ over a length of $L\simeq14.3\,{\rm m}$ each \cite{Savary:2008zz}; the effective diameter $d$ of the bore for traversing light is about $d\simeq45\,{\rm mm}$.
In this case, we have $(\Omega/m_e)^2\simeq\omega\Omega/m_e^2\simeq0$, $(e{\cal B}/m_e^2)^2\simeq 3.54\times10^{-18}$ and $(e{\cal B}\omega/m_e^3)^2\simeq2.17\times10^{-12}$ fully compatible with Eq.\,\eqref{eq:VBconditions}.

To achieve gamma-photon energies up to $\omega\simeq400\,{\rm MeV}$, the Lorentz factor $\gamma$ which effectively governs the generation of the high-energy gamma beam in the GF needs to be as large as $\gamma\approx3000$; cf. Sec.\,\ref{subsec:The Gamma Factory} above. As the opening angle of the gamma beam is given by $\approx1/\gamma$, the bore diameter of the magnet implies a maximum length $l_{\rm max}\approx (d/2)\gamma\simeq67.5\,{\rm m}$ of the magnetic field provided by LHC magnets through which the full gamma beam could travel.
In turn, we can envision the use of up to four LHC dipole magnets resulting in $\omega l\simeq1.16\times10^{17}$, see Fig.~\ref{fig:birefringence_variants}(A) for an illustration.
For these parameters we obtain a small (but potentially measurable, see  Sec.\,\ref{Sec:Gamma_pol}) value of $\Phi\simeq3.18\times10^{-5}$.

We note that this value is about an order of magnitude larger than the one predicted to be accessible in the head-on collision of a state-of-the-art petawatt-class high-intensity-laser and free-electron-laser (FEL) pulses of $\omega\simeq{\cal O}(10)\,{\rm keV}$ \cite{Heinzl:2006xc,Dinu:2013gaa,Dinu:2014tsa,Schlenvoigt_2016,Karbstein:2018omb}. For X-rays of $\omega\simeq{\cal O}(10)\,{\rm keV}$
the possibility of measuring such tiny ellipticities has been demonstrated experimentally \cite{MaS13,PhysRevResearch.2.023365,Schmitt:2020ttm}. While these x-ray techniques cannot be used at $400\,{\rm MeV}$, in the latter parameter regime pair-production polarimetry may be used \cite{Eingorn2018HEPolarimetry,Chattopadhyay:2021mbb}.
See Refs.\cite{Cantatore:1991sq,Wistisen:2013waa} for proposals to measure magnetic-field-induced vacuum birefringence with gamma photons adopting somewhat different experimental parameters.

The advantage of using a static magnetic field to drive the vacuum birefringence phenomenon is that essentially all gamma photons traverse the magnetic field. In turn, the associated number of polarization-flipped signal photons $N_\perp$ scales directly with the total number of photons constituting the gamma beam, and thus is ultimately limited (among other factors) by the repetition rate of the gamma pulses. 
Conversely, in the scenario utilizing a laser pulse to induce the birefringence phenomenon the repetition rate of the high-intensity laser is the limiting factor; see below.

\subsubsection{High-intensity laser pulse}\label{subsec:birefringence_HIT}

Alternatively, the birefringence signal could be driven by a high-intensity laser field \cite{Baier:1975ff,Becker:1974en}; see Fig.~\ref{fig:birefringence_variants}(B).
In such a scenario, the birefringence signal is predominantly induced in the interaction region where the gamma probe collides with the focused high-intensity laser pulse reaching its peak field strength. Outside the focus, the field strength of the high-intensity pump drops rapidly.

State-of-the-art high-intensity lasers of the petawatt-class typically deliver pulses of energy $W={\cal O}(10)\,{\rm J}$ and duration $\tau={\cal O}(10)\,{\rm fs}$ at a wavelength of $\lambda={\cal O}(1)\,\upmu{\rm m}$ and a repetition rate of ${\cal O}(1)\,{\rm Hz}$. These pulses can be focused to a waist radius of $w_0\gtrsim\lambda$.
In turn, the maximum frequency scale of variation of the pump field is given by $\Omega=2\pi/\lambda$.
For our explicit example, we choose the parameters characterizing a commercial 300 TW Titanium Sapphire laser system, such as the one installed at the Helmholtz International Beamline for Extreme Fields (HiBEF) at the European XFEL:
$W=10\,{\rm J}$, $\tau=30\,{\rm fs}$, $\lambda=800\,\upmu{\rm m}$ and a repetition rate of $1\,{\rm Hz}$ focused to $w_0=1\,\upmu{\rm m}$.
The associated Rayleigh range is ${\rm z}_{\rm R}=\pi w_0^2/\lambda$.
Assuming the high-intensity laser field to be well-described as pulsed paraxial fundamental Gaussian beam, the electric peak field strength ${\cal E}_0$ in its focus can be expressed in terms of the pulse energy, pulse duration and waist radius as \cite{Karbstein:2017jgh}
\begin{equation}
 {\cal E}_0^2\simeq 8 \sqrt{\frac{2}{\pi}}\,\frac{W}{\pi w_0^2\tau}\,. \label{eq:peakfield}
\end{equation}

For the laser parameters given above and $\omega\simeq400\,{\rm MeV}$, the dimensionless quantities in Eq.\,\eqref{eq:VBconditions} are $(\Omega/m_e)^2\simeq9.2\times10^{-12}$, $(e{\cal E}_0/m_e^2)^2\simeq 1.46\times10^{-7}$, $\omega\Omega/m_e^2\simeq2.4\times10^{-3}$ and $(e{\cal E}_0\omega/m_e^3)^2\simeq0.09$. These values suggest that in this parameter regime the size of the attainable vacuum birefringence signal can still be reliably estimated from the leading contribution to the Heisenberg-Euler effective Lagrangian.

At the Gamma Factory, the collision point of the gamma beam with the high-intensity laser beam should be sufficiently separated from the source of the gamma photons such that the radius $w_\gamma$ of the gamma beam in the interaction region with the high-intensity laser pulse generically fulfills $w_\gamma\gg w_0$; cf. also Fig.~\ref{fig:birefringence_variants}(B) and Ref.~\cite{Ilderton:2016khs}. 
Even though this implies that only a fraction of the gamma photons will actually see the high-intensity laser focus, focusing the latter less tightly to $w_0\approx w_\gamma$ is not an option: as the ellipticity is proportional to ${\cal E}_0^2\sim1/w_0^2$ [see Fig.~\ref{fig:birefringence_variants}(a) and Eq.\,\eqref{eq:ellipticity_HIT} below], an increase of $w_0$ would immediately reduce the effect.
Moreover, the GF is expected to provide gamma pulses of duration $T\gtrsim160\,{\rm fs}$.

Given that $w_\gamma\gg w_0$ and the pulse duration of the gamma beam meets the criterion $T\gg\{\tau,{\rm z}_{\rm R}\}$, which is true for the parameters of the GF, the phase difference accumulated by the gamma beam can be expressed in a form similar to Eq.\,\eqref{eq:ellipticity_B}, yielding \cite{Karbstein:2021}
\begin{equation}
 \Phi\simeq  \omega{\rm z}_{\rm R}\frac{2^{1/4}\alpha}{30}\Bigl(\frac{e{\cal E}_0}{m_e^2}\Bigr)^2\,{\rm e}^{\frac{1}{2}(\frac{8{\rm z}_{\rm R}}{\tau})^2}\,{\rm erfc}^{1/2}\bigl(\tfrac{8{\rm z}_{\rm R}}{\tau}\bigr)\,,\label{eq:ellipticity_HIT}
\end{equation}
with complementary error function ${\rm erfc}(.)$.
We emphasize that this parameter regime even seems to be particularly beneficial for high-intensity laser driven vacuum-birefringence experiments: given that the conditions $T\gg\tau$ and $w_\gamma\gg w_0$ are met, the experiment is essentially insensitive to the shot-to-shot fluctuations inherent to high-intensity laser systems resulting in spatio-temporal offsets of ${\cal O}(w_0)$ of the position of the high-intensity laser focus. Variations of this order just change the location of the high-intensity laser focus within the forward cone of the gamma probe and thus do not impact the signal; cf. also Fig.\,\ref{fig:birefringence_variants}(B).

Similarly to the result~\eqref{eq:ellipticity_B} for a constant magnetic field, the phase difference~\eqref{eq:ellipticity_HIT} scales quadratically with the field strength of the pump field, and linearly with both $\omega$ and the typical extent of the pump field along the propagation direction of the probe. This extent is typically set by the Rayleigh range ${\rm z}_{\rm R}$ of the high-intensity laser.

Plugging the above parameters into Eq.\,\eqref{eq:ellipticity_HIT}, we obtain $\Phi\simeq0.13$, which is much larger than the analogous value obtained for the case of a static magnetic field discussed in Sec.~\ref{subsec:birefringence_B}.
Assuming in addition that the pulse duration and beam radius of the gamma beam in the interaction region are given by $T\simeq160\,{\rm fs}$ and $w_\gamma\simeq20\,\upmu{\rm m}$, we find $N_\perp/N\simeq2.07\times10^{-6}$ for the ratio of the polarization-flipped signal photons and the number of gamma photons $N$ available for probing the effect;
note that in the present scenario $N_\perp/N=(\tau/T)(w_0/w_\gamma)^2(\Phi/2)^2$ \cite{Karbstein:2021}.

However, we emphasize once again that in contrast to the case of a static magnetic field, for the scenario involving a high-intensity laser pulse, only a fraction of the total number of gamma photons provided by the Gamma Factory is available for probing the vacuum birefringence phenomenon: in this case the repetition rate of the experiment is limited by the repetition rate of ${\cal O}(1)\,{\rm Hz}$ of the high-intensity laser.

Higher laser intensities and/or larger gamma photon energies would even allow for experimental probes of vacuum birefringence in the parameter regime characterized by $(e{\cal E}\omega/m_e^3)^2\gtrsim1$, violating the condition~\eqref{eq:VBconditions}.
The theoretical study of quantum vacuum signatures in this parameter regime requires insights beyond the Heisenberg-Euler effective Lagrangian. Vacuum birefringence in this parameter regime \cite{Kotkin:1996nf,Nakamiya:2015pde,King:2016jnl,Bragin:2017yau} can be reliably studied resorting to the photon-polarization tensor evaluated in the background of the pump field.

Finally, we note that aside from the prospect of directly verifying a fundamental QED prediction for the first time, a precision measurement of the ellipticity constituting the signal of vacuum birefringence in Eqs.~\eqref{eq:ellipticity_B} and \eqref{eq:ellipticity_HIT} at the Gamma Factory would also constitute a sensitive probe for New Physics beyond the standard model of particle physics.
The latter may leave an imprint on the refractive index of the vacuum resulting in deviations from the standard model prediction; see, for example, the recent review \cite{EJLLI20201} and references therein. For a survey of the potential of the Gamma Factory for searches of axion like particles, see Ref.~\cite{Balkin:2021jdr}.

\subsection{Photon Splitting}
\label{subsec:Photon_Splitting}

In contrast to vacuum birefringence, the effect of photon splitting in atomic fields mediated by an electron-positron fluctuation \cite{Li:1997jn,PhysRevA.58.1757,Lee:1998hu,Lee:2001gc} has already been successfully observed in a dedicated laboratory experiment employing probe photons in the energy region of $120-450\,{\rm MeV}$ \cite{Akhmadaliev:2001ik}.
See Fig.~\ref{fig:vacuum_diags}(b) for the corresponding Feynman diagram.  
At present, the experiment and the theory are consistent within the achieved experimental accuracy \cite{Lee:2001gc}.

The high flux of gamma photons at the Gamma Factory as well as the possibility of a precise tuning of their energy will allow for detailed experimental studies of this nonlinear QED process at high statistics and accuracy.
This will provide a sensitive test of theory at unprecedented precision.

For completeness, we note that the photon splitting process can in principle also be triggered by constant electromagnetic fields and laser fields \cite{BialynickaBirula:1970vy,Adler:1971wn,Adler:1970gg,Papanyan:1971cv,Papanyan:1973xa,Stoneham:1979pbh,Baier:1986cv,Baier:1996bq,Adler:1996cja,DiPiazza:2007yx}, but is typically suppressed in these cases.

\subsection{Photon Scattering}
\label{subsec:Photon_Scattering}

In the scenario detailed in Sec.\,\ref{subsec:birefringence_HIT}, also quasi-elastic scattering of gamma photons off the optical high-intensity laser pulse would constitute a signature of quantum vacuum nonlinearity. For the specific scenario considered there, the maximum value for the total number of quasi-elastically scattered gamma photons is given by $N_{\rm tot}=(196/9)N_\perp\simeq21.8 N_\perp$ \cite{Karbstein:2019bhp}. However, for kinematic reasons, these signal photons are scattered into the forward opening angle of the gamma beam \cite{Karbstein:2021}, and thus generically cannot be discerned from the large background of the gamma photons constituting the probe and traversing the interaction region essentially unmodified.
In turn, aiming at measuring the quasi-elastic scattering signal, collimators would be needed to reduce the divergence of the gamma photons before the interaction with the high-intensity laser pulse~\cite{PhysRevA.38.4891,Becker:89,Sangal:2021qeg}.
The same is true for scenarios envisioning the  collision of gamma photons with the photons constituting the initial laser beam.

Experimental bounds on elastic photon-photon scattering from direct searches with optical and X-ray beams are discussed in Refs.\,\cite{Bernard_2000,Inada:2014srv}, while Refs.\,\cite{dEnterria:2013zqi,Aaboud:2017bwk,Sirunyan:2018fhl,Aad:2019ock} present recent experimental evidence of light-by-light scattering with almost real photons in the ATLAS and CMS experiments at CERN.

Finally, we note that for gamma-photon vs. laser-photon collisions, the Gamma Factory will enable center-of-mass energies in the range of $\sqrt{s}\simeq1\ldots120\,{\rm keV}$ at a luminosity of up to $L\simeq10^{40}\,{\rm cm}^{-2}{\rm s}^{-1}$ \cite{Balkin:2021jdr}.
For center-of-mass energies below the electron-positron pair-production threshold, i.e., $\sqrt{s}<2m_e\simeq1\,{\rm MeV}$, no real electron-positron pairs are produced in such collisions.
On the other hand, for prospective gamma-photon vs. gamma-photon collisions, one could in principle
probe center of mass energies in the range $\sqrt{s}\simeq1\ldots800\,{\rm MeV}$ at a luminosity of up to $L\simeq10^{27}\,{\rm cm}^{-2}{\rm s}^{-1}$.
Note however, that in this parameter regime the signature of light-by-light scattering is likely obscured by electron-positron pair production via the Breit-Wheeler process \cite{PhysRev.46.1087} and the corresponding secondary scattering processes; cf. also Ref.\,\cite{Adam:2019mby}.



\section{Nuclear physics with tertiary beams}
\label{Sec:Nucl_phys_tertiary_beams}

\subsection{Tertiary beams at the GF}

The GF beam of gamma rays can be used to produce 
tertiary  beams in collisions with an external stationary target. 
The photon-based scheme represents a change of the present, canonical  
paradigm for the production of such derivative beams
in which the particles are produced in 
strong-interaction-mediated collisions.
The GF tertiary-beam production scheme is based on the peripheral, small-momentum-transfer electromagnetic interactions of the photons with atoms of the target material. 

\begin{figure}[!htpb]\centering
    \includegraphics[width=0.8\linewidth]{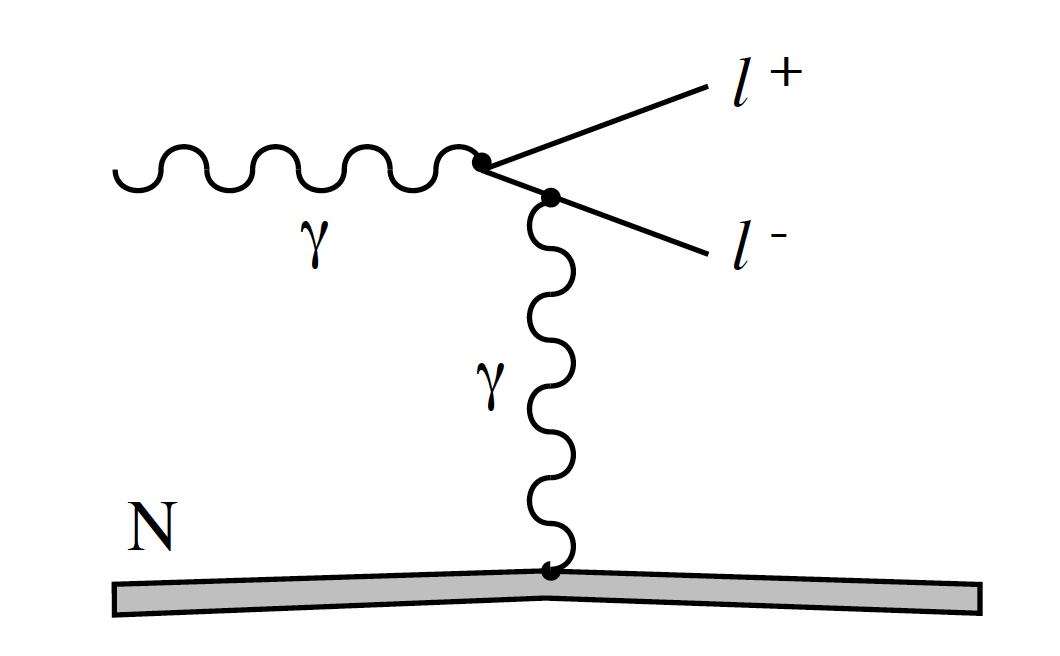}
    \caption{Feynman diagram for lepton-pair production 
    in peripheral photon-nucleus collisions.
    Below the muon-pair production threshold,  only electron-positron pairs 
    can be  produced in this process. The cross section 
    for production of muon pairs rises rapidly with the energy of
    the gamma rays. For example, for a copper target it rises from 0.2\,$\mu$b, to 10\,$\mu$b for the photon energy  
    rising from $\approx$400\,MeV---which is presently accessible at the LHC---to 1600\,MeV, achievable by the high-energy (HE-LHC) upgrade  \cite{Zimmermann:2017bbr}. 
    The cross section also rises rapidly  
    with the atomic number on the target nucleus:
    from 0.016\,$\mu$b for  hydrogen to 30.2\,$\mu$b for lead, 
    for 1\,GeV photons. Muon photoproduction is discussed
    in detail in Ref.\,\cite{Burkhardt:2002vg}.}
    \label{Fig:Feynman_peripheral}
\end{figure}

An example of a Feynman diagram for peripheral production 
of lepton pairs is shown in Fig.\,\ref{Fig:Feynman_peripheral}.

In the case of peripheral electromagnetic processes, 
a large fraction of the wall-plug power 
delivered to the stored PSI beam can be 
transmitted to a chosen type of the tertiary beam as opposed to 
the proton-beam-driver schemes.
A photoproduction scheme may considerably reduce the target heat-load at a fixed intensity of the produced beam, facilitating 
the target design and circumventing the principal technological challenges which limit the intensities of the proton-beam-driven beams.

As an example, a 1\,MW 300\,MeV photon beam, producing a beam of $4 \times 10^{15}$ collected positrons per second deposits 140\,kW of heat power in a one-radiation-length (1X0) graphite target--only 14\% of the wall-plug power dissipated in such a case \cite{Apyan2021}.

%

\subsection{Polarized electron, positron and muon sources} 

The high-intensity beam of gamma rays from the GF can be converted into
a high-intensity beam of positrons and electrons. If the photon-beam energy 
is tuned above the muon-pair-production threshold, such a beam also contains a small 
admixture of $\mu^+$ and $\mu^-$.\footnote{The cross-section of muon pair production is numerically small until the energy of the photon reaches about $4\times m_{\mu}c^2$ \cite{Burkhardt:2002vg}. At the present LHC beam energy, 
crossing this effective muon-production threshold requires laser photons in the EUV region, down to 100\,nm wavelength. While the 
photon pulses in this wavelength-range can be produced--owing 
to the recent 
developments in laser technology--a substantial progress in the 
mirror technology is required to amplify the {\it average} power 
of the laser pulses, by stacking them in a Fabry-Perot cavity,  to the
level of $\approx$10-100\,kW (0.5-5\,mJ pulses with the 20\,MHz repetition rate).
Such an amplification is required to significantly improve the intensity of 
the presently operating muon sources. Two fall-back options can be considered: (1) doubling the LHC beam energy, as proposed in the HE-LHC project \cite{Zimmermann:2017bbr}, and (2) 
using a free electron laser (FEL) rather than a conventional laser
technology.}
 
Beams of different lepton flavors can easily be separated using both their respective and distinct 
kinematic characteristics and the time-of-flight method, since the produced electrons and positrons move with nearly the speed of light while the GF muons are non-relativistic.  

The target intensity  of the GF source of 
electrons/positrons is $10^{17}$ leptons per second, assuming the present CERN accelerator infrastructure and presently available laser technology. 
Such intensity, if achieved, would  be three  orders of magnitude higher than that of the KEK positron source 
\cite{Suwada:2006rp},
and would largely satisfy the positron/electron source  requirements for the proposed  future high-luminosity and high-energy  $ep$ ($eA$)
collider project  
\cite{AbelleiraFernandez:2012cc}.

If the secondary GF photons are circularly polarized--for example, by using 
circularly polarized laser photons  colliding with spin-0, helium-like 
partially stripped ions--the first generation electrons, positrons, or muons, 
are  polarized parallel to the photon-beam direction.

The intensity of the polarized muons which is  reachable at the LHC, using  the GF photon-conversion scheme, is $10^{10}$ muons per second.\footnote{The ratio of cross sections for producing muon pairs and electron-positron pairs scales 
approximately as the square of the ratio of their masses: $m_e^2/m_{\mu}^2$
in the $E_{\gamma} \gg 2 m _{\mu}$ limit.}
If achieved, it would be two  orders of magnitude higher than that of the $\pi E4$ beam at the Paul Scherrer Institute \cite{Prokscha:2008zz}. With the HE-LHC upgrade,  the intensity 
of the polarized muon source is expected to increase  to 
$10^{12}$ muons per second. 

The intensity of the photon-conversion-based GF polarized-muon source
does not satisfy  the muon-beam intensity requirements  for the neutrino factory and muon collider project \cite{Alsharoa2002status}--($10^{13}$ muons of each sign per second).
Therefore, another  production scheme, suitable for generating the requisite 
muon fluxes, is being developed \cite{Apyan2021}. 
Preliminary calculations show that the requisite 
production rate of polarized muons can be reached for 1\,MW photon beams.
The principal advantage of the GF scheme, with respect to proton-beam-driver schemes, is that the product of the muon-source longitudinal and transverse emittances can be improved by more than three  orders of magnitude. 
There are many applications of intense muon beams, including studies of the basic symmetries of nature via searches
for standard-model-forbidden muon-decay modes with unprecedented
precision \cite{Papa2020}, studies of the nuclear properties via spectroscopy of muonic atoms \cite{knecht2020study}, and revisiting the feasibility of the muon catalyzed nuclear fusion  \cite{Alvarez:1957un,Homlid2019}. For the latter two applications, requiring high flux
of negatively charged muons,  the GF  muon source
has an important advantage with respect to proton driven sources
producing predominantly positively charged muons. 
The GF tertiary,  high-brilliance beams of polarized positrons and muons can open the paths to
new type of fixed-target deep inelastic scattering (DIS) experiments requiring high-intensity lepton beams.  

A detailed comparison of production rates of $e^+e^-$ and $\mu^+\mu^-$ pairs is of interest for tests of lepton flavor universality, for example, in the context of the ``proton radius puzzle''~\cite{Pauk:2015oaa}. 
A first observation of di-muonium (i.e., a bound state of $\mu^+\mu^-$) production could also be of interest~\cite{Jentschura:1997tv}.

\subsection{High-purity neutrino beams}

Low-emittance muon beams provided by the GF  source  can be accelerated 
and stored in specially designed storage rings--which preserve 
polarization of stored muons--to produce  muon-neutrino, electron-antineutrino, muon antineutrino and electron neutrino beams 
of precisely controlled fluxes.

The relative fluxes of neutrinos and antineutrinos of each flavor can be precisely controlled by a frequent change of the sign of of the stored muons.

Thanks to the muon polarization and ($V-A$)-type of the weak currents, the relative flux of muon-neutrinos (muon-antineutrinos) coming from negative-muon (positive-muon) decays, and  the flux of   electron-antineutrino (electron-neutrino) can be precisely controlled  on the basis of their respective angular distributions. For more details see, for example, Ref.\,\cite{Blondel:2000vz}. 

The  fluxes of both the neutrino and antineutrino beams should be equal and they can be  predicted to a tenth of a percent accuracy, limited predominantly by the measurement of the muon-beam current in the muon storage ring. 

\subsection{Neutron and radioactive ion sources}
\label{subsusbsec:Neutron_radioact_ion_sorces}

The energy of the GF photons can be tuned to excite the GDR (see Sec.\,\ref{Subsect:GDR-multi}) or fission resonances (Sec.\,\ref{Subsec:Photofission}) of large-$A$ nuclei, providing abundant sources of: 
(1) neutrons with the target intensity reaching $10^{15}$ neutrons per second (first-generation  neutrons),
(2) radioactive, neutron-rich 
ions (coming from fission of heavy nuclei) with the target intensity reaching $10^{14}$ isotopes per second \cite{Nichita:2021iwa} (see also Sec.\,\ref{Sec:Medical_Isotopes}). 

The above fluxes would  approach those of other European projects under construction, such as the European Spallation Source (ESS), FAIR and the future  EURISOL facilities. 
The advantage of the GF sources is their high efficiency -- almost $10\%$ of the LHC RF power can  be  converted into the power of the neutron and radioactive-ion beams. 
The high-flux neutron source could provide new 
opportunities for investigation of  the basic symmetries of nature, for example, via searches for the permanent electric dipole moment (EDM) of the neutron (see Ref.\,\cite{Abel2020_nEDM} and references therein) or neutron-antineutron oscillations \cite{Phillips:2014fgb}.

Finally, the tertiary beams of neutrons and radioactive 
neutron-rich isotopes  could open a wide spectrum of industrial and medical
applications in the domains of:  (1) muon catalyzed cold fusion \cite{Alvarez:1957un,Homlid2019}; (2)
energy-amplifier (EA) research \cite{Carminati:1993zc}; and (3) production of ions for positron emission tomography (PET) and for the selective cancer therapy with alpha emitters
\cite{Mostamand:2020woh}.


\subsection{Production of monoenergetic fast neutrons}
\label{subsusbsec:Production_Monoenerg_neutrons}

Monoenergetic neutrons can be produced in ($\gamma,$n) reactions near narrow resonances like the ones in $^{13}$C (see Appendix\,\ref{Subsec:Appendix:Gamma_Res_13C}). If the gamma beam is polarized, the polarization will be transferred to the neutrons.

As an example, consider neutron production using the 8.86\,MeV resonance of $^{13}$C (see Appendix\,\ref{Subsubsec:8.86MeV_resonance}).
Neutrons will be produced in the reaction \cite{Ajzenberg-Selove91En}
\begin{equation}
    \gamma + {}^{13}\textrm{C}\rightarrow {}^{12}\textrm{C} + n; \quad Q_m=-4.95\,\textrm{MeV}.
\end{equation}
Here the energy release of the reaction is $Q_m\approx T_n-E_{\gamma}$, neglecting the recoil energy of $^{12}$C. With this, we find that the kinetic energy of the neutron $T_n\approx$3.9\,MeV is lower than the energy of the first excited state of $^{12}$C (4.44\,MeV).
Therefore, only ground-state $^{12}$C will be produced and the neutrons will be nearly monoenergetic.

The neutron-production rate is essentially the same as the gamma-absorption rate (Appendix \ref{Subsubsec:8.86MeV_resonance}). For a $1\times1\times1$\,cm$^3$ target containing 1\,g of $^{13}$C, we estimate the neutron-production rate as $5.4\times10^{10}$\,s$^{-1}$.

A potential problem for a source is re-absorption of neutrons: the produced neutrons may further react with the target according to
\begin{equation}
    n + {}^{13}\textrm{C}\rightarrow {}^{14}\textrm{C}+\gamma; \quad Q_m=8.18\,\textrm{MeV} .
\end{equation}
Given $T_{n}\approx3.9$\,MeV, we get $E_{\gamma}\approx12.1$\,MeV. 
The cross section for this neutron-capture process can be estimated as \cite{Krane87}
\begin{equation}
\label{Eq.neutron_cap}
    \sigma_{tot}=2\pi(R+\lambdabar)^2 ,
\end{equation}
where $\lambdabar$ is the reduced wavelength of the neutron, $R=2.5$\,fm,
$\lambdabar=\hbar/p=\hbar c/\sqrt{2mc^2T_n}=2.30$\,fm ($p$ is the neutron momentum), yielding $\sigma_{tot}\approx1.4\,$b. With a $1\times1\times1$cm$^3$ target containing 1\,g of $^{13}$C, fewer than 10$\%$ of the neutrons will be  reabsorbed, so the issue appears tractable.

As for the 7.55\,MeV resonance (Appendix\,\ref{Subsubsec:7.55MeV_resonance}), the production rate of neutrons with $T_n\approx2.6$\,MeV is 
$9.0\times10^{9}$\,s$^{-1}$. The cross section for the neutron-capture process is 2.1\,b. Again, fewer than 10$\%$ of the neutrons will be  reabsorbed.

Narrow gamma resonances in conjunction with the GF, will thus allow producing monoenergetic neutrons at a set of well defined energies (e.g., 2.6\,MeV, 3.9\,MeV,... for the case of a $^{13}$C target). Such neutrons may be useful for measuring cross sections of processes (e.g., neutron capture) relevant to astrophysics. 

It is important to note that the angular distribution of the neutrons will be nearly isotropic, suggesting a geometry for the target to study the interaction of the produced neutrons: it could be a ``ball" surrounding the neutron-production target.


\subsection{Metrology with keV neutrons} 

Tertiary neutrons produced in $(\gamma,n)$ reactions can be used to improve the operation quality of the GF. The gamma-beam energy above the neutron threshold (typically, around 7\,MeV) can be
measured with high accuracy by measuring the time-of-flight (TOF) of the neutrons. By subtracting the large fixed neutron binding energy, resulting in rather slow neutrons with keV energies together with a sharp start
signal for the TOF measurement, a high resolution at the $10^{-7}$
level can be achieved \cite{ELI-NP_gamma_n}. This enables a fast measurement of the average
gamma-beam energy and the width of the gamma beam without perturbing the
GF. This could provide feedback for adjusting the ion-beam energy, thus facilitating experiments requiring
precise control of the ion-beam or gamma-ray energy.

\section{Nuclear physics opportunities at the SPS}
\label{Section:NP_PoP}

The Proof-of-Principle (PoP) GF experiment is proposed at the SPS \cite{Krasny2019PoP}. Here, we have a much lower relativistic factor, $\gamma\lesssim220$.
Since it is nontrivial to get low-energy gamma rays with $\gamma\geq200$ at the LHC, the operation at the SPS extends the available energy range of the secondary photons. Note that for X-rays, there exist intense coherent sources, for example, the European XFEL \cite{EuropeanXFELBLs} producing X-rays with energies $\lesssim30\,$keV.

At the SPS, with $\lesssim10$\,eV primary laser photons, secondary photon energies of up to 1.6\,MeV will be available. The use of an FEL primary-photon source (primary photon energies of up to a few hundred eV) can extend the energy range, partially overlapping with the range that should be available at the LHC with conventional lasers.
Already with the secondary-photon energies available with the currently planned PoP experiment \cite{Krasny2019PoP}, $\lesssim$44\,keV, interesting new results could be obtained. Table\,\ref{tab:Low_En_Gamma_Trans} contains information on nuclear transitions in the range $0.008-60$\,keV. We have also discussed nuclear Raman transitions accessible with the secondary photon beams (or even with primary beams if an FEL is used) at the SPS in Sec.\,\ref{Subsec:Nuclear_Raman_Transitions}.

The SPS experiments could also serve as a platform for development of laser cooling of the PSI and ion sources, for example, $^{201}$Hg and investigation of its 1.5\,keV gamma resonance (see Table\,\ref{tab:Low_En_Gamma_Trans}).

\section{Speculative ideas and open questions}
\label{Sec:More_ideas_open_Q}

\subsection{Applying the Gamma Factory ideas at other facilities}
\label{Subec:GF_at_other_accell}

The ideas of the Gamma Factory can also be implemented at other facilities. In fact, back-scattering of laser photons from hydrogen-like ions and laser cooling of PSI were considered for RHIC  \cite{Zolotorev1997}. A survey of the existing and future accelerator facilities where GF concepts may be implemented and the projected parameters of the corresponding gamma sources are surveyed in Table\,\ref{tab:prospective_GF_facilities}.  

\begin{table}[htpb]
    \centering
    \begin{tabular*}{\linewidth}{@{\extracolsep{\fill}} llll}
    \hline 
    \hline
    Facility & Lab. & p$^+$ energy & Max. photon energy \\
    \hline \\[-0.2cm]
    LHC       & CERN & 6.5\,TeV  & Pb$^{81+}+12.6~\mathrm{eV} \rightarrow 373$\,MeV \\
    SPS       & CERN & 450\,GeV    & Ti$^{21+}+11.7\,\mathrm{eV} \rightarrow 2.1$\,MeV\\
    RHIC      & BNL  & 255\,GeV    & Cl$^{16+}+11.8\,\mathrm{eV} \rightarrow 0.74$\,MeV\\
    \hline \\[-0.2cm]
    NICA      & JINR & 12.6\,GeV      & Li$^{2+}+10.2\,\mathrm{eV} \rightarrow 0.83$\,keV\\
    SIS 100   & GSI  & 29\,GeV        & B$^{4+}+10.2\,\mathrm{eV} \rightarrow 6.4$\,keV\\
    SIS 300   & GSI  & 87\,GeV        & Ne$^{9+}+12.1\,\mathrm{eV} \rightarrow 86$\,keV\\
    \hline \\[-0.2cm]
    SC-SPS    & CERN & 1.3\,TeV      & Kr$^{35+}+11.4\,\mathrm{eV} \rightarrow 15$\,MeV\\
    HE-LHC    & CERN & 13.5\,TeV      & U$^{91+}+7.8\,\mathrm{eV} \rightarrow 0.96$\,GeV\\
    FCC-hh    & CERN & 50\,TeV        &  U$^{91+}+2.1\,\mathrm{eV} \rightarrow 3.5$\,GeV \\
    \hline
    \hline
    \end{tabular*}
    \caption{Survey of the existing (first three lines) and future relativistic heavy-ion facilities. The maximum photon energy is given assuming hydrogen-like ions interacting with a primary laser beam in the optical range (down to 100~nm). With a dedicated FEL used as a primary source of higher energy photons (+~heavier ion) the maximum secondary photon energy can be increased further.}
    \label{tab:prospective_GF_facilities}
\end{table}


\subsection{Nuclear waste transmutation}
\label{Subec:Nuclear_Waste}

Nuclear waste characterization and transmutation is an important topic on the world-wide scale. Nuclear waste usually contains both stable and unstable isotopes including long-lived fission products (LLFPs) that are particularly troublesome as they usually require secure storage for thousands of years. A possible route toward a solution is transmutation of the dangerous isotopes \cite{Perot:2017un}. To effectively transmute LLFPs into stable isotopes or short-lived radioactive products, one needs to avoid newly producing dangerous isotopes in the process. A selective isotope-transmutation method for LLFPs with neutron-separation thresholds lower than those of other isotopes using quasi-monochromatic gamma-ray beams was proposed in Ref.\,\cite{Hayakawa2016_Transmut}. To realize this,
narrow-band gamma rays with tunable energy and high photon fluxes 
beyond the capabilities of existing facilities are needed.\footnote{Even with the GF running non-stop for four months, we will produce $\approx 10^{24}$ gammas in total. Even if \textbf{each} of these photons transmutes a $^{93}$Zr, we are talking on the order of 100\,g of material processed; in practice, this will be many orders of magnitude less. This does not seem to be a practical approach.} While the secondary photons at the Gamma Factory with expected total photon fluxes $j\approx10^{17}$\,ph/s will not bring this proposal into practice and transmute nuclear waste efficiently, the role of the GF could be to enable proof-of-principle experiments and to measure ($\gamma$,$n$) cross sections of unstable isotopes, most of which have not been measured.

\subsection{Laser polarization of PSI}
\label{subsec:Laser_polarization_PSI}

The interaction of polarized primary photons with the PSI generally leads to polarization of both electron and nuclear spins of the PSI (assuming the ground-state electronic angular momentum and nuclear spin are nonzero). It is currently unclear if there is a viable scheme to preserve the electron polarization on a round trip in the storage ring. As for the nuclear spins, the task should presumably be easier on the account of much smaller values of the nuclear magnetic moments.  Techniques for preserving proton polarization using so-called ``Siberian snakes'' \cite{Derbenev2021siberian} have been implemented, for instance, at RHIC, and could, presumably, be adopted to the PSI at the GF. 
There is an additional question of whether nuclear polarization will survive a round trip in a non-bare ion \cite{Nortershauser-2021}.


As mentioned in Ref.\,\cite{Budker2020_AdP_GF}, regardless of whether the polarization survives a round trip, there are interesting physics opportunities with polarized PSI beams, for example, parity-violating structure-function studies (Sec.\,\ref{subsec:Photoabs_structure_finctions}) or fixed-target experiments with polarized PSI.

\subsection{Quark-gluon plasma with polarized PSI}

The availability of nuclear spin-polarized PSI produced at the GF opens a possibility, briefly discussed in Ref.\,\cite{Budker2020_AdP_GF}, to study collisions of such polarized ions and the resulting quark-gluon plasma.

Indeed, collisions of polarized nuclei, particularly deformed ones, is an active area of research in heavy-ion collisions. Here, optical polarization control may offer an alternative to the currently employed inference of the polarization of initial deformed nuclei based on the final particle distribution using two observables-—multiplicity and ellipticity--as proposed \cite{Shuryak2000} and experimentally realized at RHIC \cite{Huang2013}.

\subsection{Ground-state hyperfine-structure transitions in PSI}

The ground-state hyperfine intervals of heavy PSI with nonzero-spin nuclei can be in the eV range. Knowledge of these interval is important, for example, for the study of QED effects and hyperfine anomalies in strong magnetic fields as well as for determination of nuclear moments \cite{Skripnikov2018Bi}. A few of such hyperfine intervals were measured in electron-beam ion traps (EBIT) via emission spectroscopy (see, for example, Refs.\,\cite{Crespo1996,Crespo1998,Beiersdorfer2001}). Measurements were also performed via laser-spectroscopy in low-energy storage rings, where one can benefit from laser cooling of the PSI, see, for example, Refs.\,\cite{Nortershauser_2013,Ullmann_2015,Ullmann-2017}.

The GF may enable a somewhat different approach to precision spectroscopy of hyperfine intervals of heavy PSI. Optical pumping (Sec.\,\ref{subsec:Laser_polarization_PSI}) easily creates hyperfine polarization. (Hyperfine polarization will also naturally result from the radiative decay of the upper state.) Conversely, the rate of production of secondary photons by polarized primary photons depends on the hyperfine state, and so monitoring this rate can be used for ``optical probing'' of the hyperfine sate. 
The experiment would consist in measuring the flux of secondary photons produced by optically pumped PSI as a function of the frequency of an applied magnetic field that, when resonant with the hyperfine interval, will drive the M1 transition between the hyperfine states and redistribute their population. The matrix element of the M1 transition between the hyperfine states is on the order of a Bohr magneton, and the transition can thus be driven with a relatively modest magnetic field. To produce such a magnetic field in the frame of the PSI, one can pass the ions through an undulator (e.g., a Halbach array made up of permanent magnets similar to the common ``refrigerator magnets'') or a microwave source. Transverse magnetic fields as seen by the PSI are scaled up in magnitude by the relativistic factor compared to their value in the laboratory frame. Scanning of the frequency can be done by adjusting the relativistic factor of the PSI. 

To get a feeling for some of the parameters, let us consider the 1.2\,eV hyperfine interval in the ground state of hydrogen-like
$^{207}$Pb$^{81+}$ ions. The required microwave frequency is around 100\,GHz, and the required undulator period is about 3\,mm. We note also that spontaneous emission on the hyperfine transition results in forward-directed radiation with maximum photon energy of $\approx 7$\,keV in the laboratory frame. However, due to a relatively long lifetime of the upper ground-state hyperfine level, which is about 50\,ms in the ion rest frame, the radiation will not be localized to the interaction point.

  
  

\subsection{Detection of gravitational waves} 

Laser cooling of the PSI beam in general and extreme laser cooling with nuclear transitions  (Sec.\,\ref{subsubsec:laser_cooling_nuc_trans}) in particular can help turn the LHC into a gravitational wave detector \cite{Orava_2017}. This possibility was discussed for many years, including a dedicated workshop  \cite{SRGW2021_workshop}. The general idea is that gravitational waves can resonantly couple to specific modes of the ion motion in the storage ring and the low emittance that may be achievable with laser cooling combined with precise beam-position monitors (BPM) may result in a competitive scheme for gravitational-wave detection, complementing other Earth-based and space-borne facilities. The practicality and ultimate sensitivity of this method will need to be evaluated in future work.

\section{Conclusions and optimistic outlook}
\label{Section:Conclusions}

In compiling this review of nuclear physics opportunities associated with the Gamma Factory, the interdisciplinary team of authors, including both theorists and experimentalists, had an opportunity to take a fresh look at this vast field as well as at a range of related fields. We are firmly convinced that the new technologies associated with the GF will lead to significant progress on many ``fronts,'' and will likely lead to disruptive breakthroughs, although exactly where these will occur is difficult to predict.\footnote{The famous quote that ``predictions are difficult, especially about the future'' is often attributed to Yogi Berra but, apparently, Niels Bohr said that long before him and Mark Twain---even earlier than that.}

One of the interesting aspects of the GF science program is its complementarity to other experimental approaches, for instance, the physics done with electron-beam facilities. Many of the opportunities discussed in this review are unique to the GF, the advent of which will lift the limitations of the hitherto available photon sources. But even in the cases where the same underlying physics can be accessed with multiple approaches (for instance, measurement of neutron distributions within nuclei), consistency of the results obtained with these different probes would serve as robust check for the different experiments and supporting theory.

As with the earlier review of atomic-physics opportunities with the GF \cite{Budker2020_AdP_GF}, it is likely that, due to the rapid progress of ideas, the paper will be outdated even before it is published. For example, the ongoing optimization of the LHC operation parameters dedicated to the GF operation may lead to more optimistic performance estimates across the board, especially with laser cooling. Nevertheless, it is our hope that these overviews will serve as suitable departure points for generating further ideas, conducting more in-depth studies, and, importantly, will become the bases for planning for specific experiments.

\section*{Acknowledgments}

The authors are grateful to Hartmuth Arenh\"ovel, Sonia Bacca, Hendrik Bekker, Carlos Bertulani, Carsten Brandau, Camilla Curatolo, Catalina Curceanu, Alejandro Garcia, Dieter Habs, Roy Holt, Magdalena Kowalska, Gerda Neyens, Jorge Piekarewicz, Szymon Pustelny, Mark Raizen, Concettina Sfienti, Jennifer Shusterman, Evgeny V.\,Tkalya, Edward Shuryak, and Bogdan Wojtsekhowski for inspiring discussions, and to the Mainz Institute for Theoretical Physics (MITP) for hosting a workshop on Physics Opportunities with the Gamma Factory that catalyzed this review. This work was supported in part by the DFG Project ID 390831469:  EXC 2118 (PRISMA+ Cluster of Excellence).
FK has been funded by the Deutsche Forschungsgemeinschaft (DFG) under Grant No. 416607684 within the Research Unit FOR2783/1. A.~P\'alffy gratefully acknowledges funding from the DFG in the framework of the Heisenberg Program. The work of  VF is supported by the Australian Research Council grants DP190100974 and  DP200100150 and the JGU Gutenberg Fellowship.
AS acknowledges support by the Deutsche Forschungsgemeinschaft (DFG, German Research Foundation) under Germany's Excellence Strategy EXC-2123 QuantumFrontiers-390837967.
MG's work was supported by EU Horizon 2020 research and innovation programme, STRONG-2020 project under grant agreement No 824093 and by the German-Mexican research collaboration Grant No. 278017 (CONACyT) and No. SP 778/4-1 (DFG).
YAL acknowledges the European Research Council (ERC) under the European
Union's Horizon 2020 research and innovation programme (grant agreement No
682841 ``ASTRUm''). The work of A.~Petrenko is supported by the Foundation for the Advancement of Theoretical Physics and Mathematics ``BASIS''.

\section*{Appendix}
\appendix

\section{Survey of existing and forthcoming gamma facilities}
\label{Appendix:gamma_facilities}
%
\begin{table*}[t]
    \centering
    \begin{tabular*}{\textwidth}{@{\extracolsep{\fill}}l      c      c      c      c}
    \hline 
    \hline
    Facility name     & MAMI A2    & JLab Hall D     & ELSA      & MAX IV \\
    \hline \\[-0.2cm]
     Location & Mainz & Newport News  & Bonn & Lund \\
     Electron energy (GeV) & 1.6 & 12 & 4.68 & 220 \\
     Max $\gamma$ energy (MeV)  & 1600 & 9200 & 2400 & 180 \\
     Energy resolution (MeV) & 2--4\,MeV & 30 & 12.5 & 0.3 \\
     Photon polarization & $\le$ 0.8 & $\le$ 0.4 &  & -- \\
     Max on--target flux ($\gamma$/s)  & 10$^8$ & 10$^8$ & 5 $\times$ 10$^6$ & 4 $\times$ 10$^6$\\
     Reference & \cite{Kaiser:2008zza} & \cite{GLUEX} & \cite{Hillert:2006yb} &  \cite{Adler:2013} \\
    \hline
    \hline
    \end{tabular*}
    \caption{Bremsstrahlung tagged-photon facilities around the world.}
    \label{tab:tagged-photon facilities}
\end{table*}
\begin{table*}[t]
    \centering
    \begin{tabular*}{\textwidth}{@{\extracolsep{\fill}}lcccccccc}
    \hline 
    \hline
    Facility name     & ROKK-1M    & GRAAL     & LEPS      & HI$\gamma$S & ELI-NP & SLEGS &CLS\footnote{Parameters of this facility are from Ref.\,\cite{SZPUNAR2013_Medical_Isotopes}.}   &GF \\
     Location & Novosibirsk & Grenoble  & Harima & Duke &  Bucharest & Shanghai  &Saskatoon    &CERN\\
     Storage ring & VEPP-4M & ESRF & SPring--8 & Duke--SR & linac  & SSRF &2.9\,GeV  &LHC\\
        \hline \\[-0.2cm]
        Laser--photon energy (eV) & 1.17--4.68 & 2.41--3.53 & 2.41--4.68 & 1.17--6.53 & 1.50--1.52   & 0.117\,(CO$_2$)  &0.117\,(CO$_2$)  &multiple\\
     $\gamma$--beam energy (MeV)  & 100--1600 & 550--1500 & 1500--2400 & 1--100 (158) &  0.2--20    & $<$22  &$\leq15$   &$\leq400$\footnote{For possibility of achieving higher energy, see Sec.\,\ref{subsubsec:higher_en_gammas} and \ref{Sec:Colliding-beam_opportunities}.}\\
     $\Delta E / E$  & 0.01 -- 0.03 & 0.011 & 0.0125 & 0.008 -- 0.1 & 0.005 &  &$\sim0.0011$\footnote{energy spread of 2.9\,GeV electrons}  &$\sim 10^{-4}$ -- $10^{-6}$\\
     Max on-target flux ($\gamma$/s)  & 10$^6$ & 3$\times$10$^6$ & 5$\times10^6$ & 10$^4-5\times$10$^8$ & 8$\times 10^8$   & 10$^9-10^{10}$  &$10^{10}$\footnote{\label{total_flux}the total photon flux}   &$10^{17}$\footref{total_flux}\\
    \hline
    \hline
    \end{tabular*}
    \caption{Parameters of existing  and forthcoming Compton back-scattering $\gamma$-ray sources around the world, from Refs.\,\cite{WELLER2009257,Ur2015_ELI-NP,Pan2009_SLEGS}. These sources are based on photon scattering from beams of relativistic electrons circulating in storage rings.
    }
    \label{tab:gammay_ray_sources}
\end{table*}

Here we briefly discuss the currently available sources of photons (partially) overlapping in energy with the reach of the GF.

Tagged-photon facilities (summarized in Table\,\ref{tab:tagged-photon facilities}) can be exemplified by The Electron Stretcher Accelerator (ELSA) facility in Bonn \cite{Schroder2014}. Here, a pulsed electron beam from a linac system is stretched in a storage ring before it hits a an aluminium foil in which bremsstrahlung is produced. A magnetic spectrometer is used to determine the energy of the post-bremsstrahlung electrons. The ``tagged photons'' range in energy from 10 to 180\,MeV. The energy resolution of about 300\,keV; the photon rates of up to $4\times10^6$\,photons/s/MeV have been achieved. 

Compton-backscattering facilities using relativistic electron beams are summarized in Table\,\ref{tab:gammay_ray_sources}.

\section{Gamma resonances in \texorpdfstring{\textsuperscript{13}}{13}C}
\label{Subsec:Appendix:Gamma_Res_13C}

The relevant nuclear levels of $^{13}$C are depicted in Fig.\,\ref{fig:C13_En}. Here we discuss several specific resonances.

\subsection{The 8.86\hspace{0.5mm}MeV \texorpdfstring{$M1$}{M1} resonance}
\label{Subsubsec:8.86MeV_resonance}

On-resonance cross section for the gamma transition is \cite{Krane87}
\begin{equation}
\label{Eq:res_cross_section}
    \sigma_0=2\pi\left (\frac{\hbar c}{E_\gamma}\right )^2\frac{2I_e+1}{2I_g+1}\frac{\Gamma_{\gamma}}{\Gamma_{tot}} ,
\end{equation}
where the quantity in parentheses is the reduced wavelength of gamma rays and $I_g$ and $I_e$ are the spin quantum numbers of the nuclear ground and excited states, respectively. $\Gamma_{tot}$ is the total width of the excited state and $\Gamma_{\gamma}$ is the 
partial width corresponding to the direct gamma transition from the excited state to the ground state.
In this case, $I_g^P=1/2^-$,$I_e^P=1/2^-$, $\Gamma_{\gamma}$=3.36\,eV, $\Gamma_{tot}$=150\,keV \cite{Ajzenberg-Selove91En}, yielding  $\sigma_0=7.0\times10^{-4}$\,b.

By proper collimation, we can tune the energy-spread width of GF photons $\Gamma_{ph}$ to be $\Gamma_{ph}=\Gamma_{tot}=150$\,keV. The effective photon flux is approximately $j_{eff}= j\Gamma_{ph}/E_{\gamma}=1.7\times10^{15}$\,s$^{-1}$, where $j=10^{17}$\,s$^{-1}$ is the total photon flux expected at the GF before collimation.


For a $^{13}$C target with a volume of $1\times1\times1$\,cm$^3$ and $^{13}$C density of 1\,g/cm$^3$, the number of gamma-absorption events is
\begin{equation}
    N=j_{eff}\frac{N(^{13}\textrm{C})\sigma_0}{1\,\textrm{cm}^2}= 5.4\times10^{10}\,\textrm{s}^{-1}\,.
\end{equation}

\subsection{The 7.55\hspace{0.5mm}MeV \texorpdfstring{$E2$}{E2} resonance}
\label{Subsubsec:7.55MeV_resonance}

For the 7.55\,MeV energy level, we have $I_g^P=1/2^-$, $I_e^P=5/2^-$, $\Gamma_{\gamma}$=0.115\,eV, $\Gamma_{tot}$=1.2\,keV \cite{Ajzenberg-Selove91En}. Then, from Eq.\,\eqref{Eq:res_cross_section} we get $\sigma_0=1.2\times10^{-2}$\,b ($\sigma_{7.55}=17.7\times\sigma_{8.86}$).

Following a similar procedure as in the case of the 8.86\,MeV transition above,
we get the effective photon flux $j_{eff}=1.6\times10^{13}$\,s$^{-1}$ and using a $^{13}$C target with the same volume and density, the number of gamma-absorption events is $N=9.0\times 10^9$\,s$^{-1}$.

\subsection{The 3.09\hspace{0.5mm}MeV \texorpdfstring{$E1$}{E1} resonance}
\label{Subsubsec:3.09MeV_resonance}

The first excited state of $^{13}$C has an energy of 3.09\,MeV with a linewidth of $\Gamma_{tot}=\Gamma_{\gamma}=0.43$\,eV \cite{NNDC} and $I^P=1/2^+$. The on-resonance cross section (Eq.\,\ref{Eq:res_cross_section}) is $\sigma_0$=260\,b.
We note that Doppler broadening at room temperature $T\approx 300$\,K leads to a Doppler width $\Gamma_D=2\sqrt{\ln{2}}E_{\gamma}\sqrt{2k_B T/(Mc^2)}\approx10.6\,$eV$\gg\Gamma_{tot}$, yielding the effective resonance photon-absorption cross section $\sigma_{eff}\approx\sigma_0\Gamma_\gamma/\Gamma_{tot}=10\,$b.
For the interaction of $\approx 3$\,MeV photons with a ${}^{13}$C target, the photon-attenuation background is dominated by Compton scattering off electrons \cite{Zyla2020_Particle_Data}, which has a cross-section $\sigma_{\textrm{Compton}}\approx0.7$\,b $\ll \sigma_{eff}$.
The maximal resonant photon-absorption rate is
$p_{max}
\approx j \Gamma_D/E_{\gamma}\approx 3.4\times 10^{11}$\,s$^{-1}$, giving $1/\sqrt{p_{max}\times100\,\textrm{s}}\approx1.7\times10^{-7}$.
Here we chose a 100\,s measurement time for the purpose of an example. In order to reach such an absorption rate, the thickness of the $^{13}$C target should be greater than the absorption length $l\approx2\,$cm.

\section{Other reviews and nuclear databases}
An early review of photonuclear experiments with Compton-backscattered gamma beams is given in Ref.\,\cite{Nedorezov_2004}. Throughout the present paper, we frequently refer to a comprehensive review of nuclear photophysics \cite{ELI-NP-WB} conducted in the context of the Extreme Light Infrastructure (ELI).

A comprehensive database of giant dipole resonances (GDR) for many nuclei is maintained by the Russia Lomonosov Moscow State University Skobeltsyn Institute of Nuclear Physics
Center for Photonuclear Experiments Data. 
The database, ``Chart of Giant Dipole Resonance Main Parameters,'' can be accessed at
 \url{http://cdfe.sinp.msu.ru/saladin/gdrmain.html}. 
 
The US National Nuclear Data Center maintained by the Brookhaven National Laboratory  provides databases for gamma transitions in nuclei, for instance NuDat 2.8 or the Evaluated Nuclear Structure Data Files \url{https://www.nndc.bnl.gov/ensdf}. 

A database called \textit{BrIcc} \cite{Kibedi2008IC_coeff} provides theoretical values of internal conversion coefficients: \url{http://bricc.anu.edu.au/index.php}.

\bibliographystyle{mybib}
\bibliography{GF_Nuclear_References}

\begin{thebibliography}{100}
\bibitem{Note_GT}
The definition of a Gamow-Teller transition is the same as in beta decay: it is
  a pseudovector transition, no parity change and a spin change of 0 or $\pm
  1$; 0 to 0 is not allowed.
\end{thebibliography}

\end{document}